\documentclass[aps, prc,twocolumn, superscriptaddress, showkeys, nofootinbib]{revtex4-1}  

\usepackage[utf8]{inputenc}

\usepackage{graphicx,color}
\usepackage{amsmath,amssymb,amsfonts}
\usepackage{hyperref}

\usepackage{xspace}	% Include xspace

\def  \etas       {\mbox{$\eta / \textit{s}$ }\xspace}

\def \cnm2   {\mbox{$\mathrm{C}_{nm}\lbrace 2 \rbrace$ }\xspace}
\def \cnm4   {\mbox{$\mathrm{C}_{nm}\lbrace 4 \rbrace$ }\xspace}

\def \sc23  {\mbox{$\mathrm{SC}(2,3)$ }\xspace}
\def \sc24  {\mbox{$\mathrm{SC}(2,4)$ }\xspace}

\def \nsc23 {\mbox{$\mathrm{NSC}(2,3)$}\xspace}
\def \nsc24 {\mbox{$\mathrm{NSC}(2,4)$}\xspace}

\newcommand{\psirp}{\Psi^{\rm  RP}}

\begin{document}
\title{Measuring differential particle correlations in relativistic nuclear collisions}
%----------------------------------------------------------------------------
\medskip
%----------------------------------------------------------------------------
%----------------------------------------------------------------------------
\author{Niseem~Magdy} 
\email{niseemm@gmail.com}
\affiliation{Department of Physics, University of Tennessee, Knoxville, TN 37996, USA}
%----------------------------------------------------------------------------

%----------------------------------------------------------------------------
%\date{\today}
%----------------------------------------------------------------------------
\begin{abstract}
%----------------------------------------------------------------------------
This study explores the transverse momentum ($p_T$) dependencies of Symmetric and Asymmetric Correlations (SC and ASC) with one and two particles of interest in Au+Au collisions at 200 GeV. Leveraging the AMPT model, the investigation delves into the sensitivity of these correlations to the final state effects, providing valuable insights into their potential for constraining the final state effects' $p_T$ dependencies.
The HIJING model is employed as a benchmark for non-flow correlations, shedding light on their impact on interpreting SC and ASC data. Moreover, the study points out that differential SC and ASC with one and two particles of interest (POIs) typically incorporate contributions from event-plane angle fluctuations. 
Consequently, this work highlights the significance of SC and ASC with one and two POIs as valuable tools for investigating the $p_T$ nature of the final state effects and advocates for comprehensive experimental measurements across various beam energies and system sizes to enhance our understanding and provide additional constraints for theoretical models.
%----------------------------------------------------------------------
\end{abstract}
%----------------------------------------------------------------------
%\pacs{25.75.-q, 25.75.Gz, 25.75.Ld}% PACS, the Physics and Astronomy
                             % Classification Scheme.
%----------------------------------------------------------------------
%Use showkeys class option if keyword
%----------------------------------------------------------------------
\maketitle
%----------------------------------------------------------------------
%\linenumbers
%----------------------------------------------------------------------
\section{Introduction}
%----------------------------------------------------------------------
%----------------------------------------------------------------------
Studies conducted at prominent facilities like the Relativistic Heavy Ion Collider (RHIC) and the Large Hadron Collider (LHC) are dedicated to unraveling the properties of quark-gluon plasma (QGP), a state of deconfined nuclear matter created in the relativistic nuclear collisions~\cite{Shuryak:1978ij, Shuryak:1980tp, Muller:2012zq}. 
In the realm of relativistic nuclear collisions, the evolution of QGP is governed by hydrodynamics. The dynamic interplay of participant distribution, subject to event-by-event fluctuations, dictates the initial geometry of the collision. Both experimental and theoretical efforts, spanning past and present, are deeply invested in comprehending the transport properties of the QGP. 
Specifically, the focus lies on elucidating fundamental characteristics such as specific shear viscosity (\etas)~\cite{Shuryak:2003xe, Romatschke:2007mq, Luzum:2008cw, Bozek:2009dw, ALICE:2019zfl, Parkkila:2021yha, ALICE:2020sup, Adam:2020ymj}, which play key roles in shaping the behavior and evolution of the QGP.

%----------------------------------------------------------------------
Particle correlations and fluctuations, as manifested in anisotropic flow measurements, represent a crucial avenue for estimating specific shear viscosity (\etas)~\cite{Bozek:2020drh, Alver:2008zza, Giacalone:2020byk, Schenke:2014tga, Staig:2010pn, Heinz:2001xi, Hirano:2005xf, Huovinen:2001cy, Hirano:2002ds, Romatschke:2007mq, Luzum:2011mm, Song:2010mg, Qian:2016fpi, Magdy:2017ohf, Magdy:2017kji, Bilandzic:2021voo, Schenke:2011tv, Teaney:2012ke, Gardim:2012yp, Lacey:2013eia, Bhatta:2022dml}. These measurements provide a way to understand the viscous hydrodynamic response elicited by the anisotropy inherent in the initial state energy density, which is illustrated by the complex eccentricity vectors $\mathcal{E}_{n}$. The complex eccentricity vectors $\mathcal{E}_{n}$~\cite{Alver:2010dn, Petersen:2010cw, Lacey:2010hw, Teaney:2010vd, Qiu:2011iv} can be given as:
%----------------------------------------------------------------------
\begin{eqnarray}
\mathcal{E}_{n} & \equiv &  \varepsilon_{n} e^{i {\textit{n}} \Phi_{n} }  \\  \nonumber
&\equiv  &
  - \frac{\int dx^{'}\,dy^{'}\,\textit{r}^{n}\,e^{i {\textit{n}} \phi}\, \textit{E}(r,\phi)}
           {\int dx^{'}\,dy^{'}\,\textit{r}^{n}\,\textit{E}(r,\phi)}, ~(\textit{n} ~>~ 1),
\label{epsdef1}
\end{eqnarray}
%----------------------------------------------------------------------
where $\varepsilon_{n}$ and $\mathrm{\Phi_{n}}$ denote the magnitude and angle direction, respectively, of the n$^{\rm th}$ eccentricity vector. Here, $r$ represents the radial coordinate, while $\phi$ signifies the spatial azimuthal angle. The initial energy density profile, denoted by ${\textit{E}}(r,\phi)$, plays a key role in characterizing the geometric features of the collision system and is fundamental to understanding the subsequent hydrodynamic evolution~\cite{Teaney:2010vd, Bhalerao:2014xra, Yan:2015jma}.
%----------------------------------------------------------------------

%----------------------------------------------------------------------
In the context of heavy ion collisions, the QGP medium undergoes a pressure gradient, transforming the spatial anisotropies inherent in the initial state into momentum anisotropies in the final state~\cite{Voloshin:1994mz}. This transformation process underscores the significance of analyzing the azimuthal distribution of particles produced in these collisions. Such analysis is commonly approached through a Fourier expansion of the azimuthal particle production~\cite{Voloshin:1994mz, Poskanzer:1998yz}, given as:
%----------------------------------------------------------------------
\begin{eqnarray}
\dfrac{dN}{d\varphi} &\propto & 1 + \sum_{n=1}^{\infty} 2 v_{n} \cos\left[  n(\varphi -\psirp) \right] ,
\label{equ:Fourier_expansion}   
\end{eqnarray}
%----------------------------------------------------------------------
where $\varphi$ is the particle's azimuthal angle, $v_n$ represents the n$^{th}$-order flow harmonic, and $\psirp$ gives the reaction plane of the collision. This methodology provides a systematic framework for quantifying the azimuthal asymmetries in particle distributions, offering valuable insights into the underlying dynamics and evolution of the collision system.
%----------------------------------------------------------------------

%----------------------------------------------------------------------
While the flow harmonics of order n$^{th}$ exhibit a linear relationship with the corresponding initial state anisotropies $\varepsilon_{{{n}}}$, higher-order flow harmonics (n$>$3) exhibit a more complex behavior. Specifically, they encompass a non-linear response to the lower-order eccentricities $\varepsilon_{{{n=2, 3}}}$. Both linear and non-linear relationships encode the medium's response to the QGP transport properties (i.e., \etas, $\zeta/s$, etc)~\cite{Song:2010mg, Niemi:2012aj, Gardim:2014tya, Fu:2015wba, Holopainen:2010gz, Qin:2010pf, Qiu:2011iv, Gale:2012rq, Liu:2018hjh, Teaney:2012ke, Bhalerao:2014xra, Yan:2015jma, Gardim:2011xv}. Accordingly, achieving an accurate estimation of the QGP transport properties requires strict model constraints to the initial state eccentricities and their associated fluctuations and correlations across a broad spectrum of beam energies and collision systems~\cite{Schenke:2019ruo, Alba:2017hhe}.
%----------------------------------------------------------------------

%----------------------------------------------------------------------
To this end, comprehensive data-model comparisons of various correlation types, including Symmetric Correlations (SC), Asymmetric Correlations (ASC), Normalized Symmetric Correlations (NSC), and Normalized Asymmetric Correlations (NASC), are essential. These correlations have been experimentally and theoretically investigated in Pb--Pb collisions at LHC energies and in Au--Au at RHIC energies~\cite{Mordasini:2019hut, Moravcova:2020wnf, Bilandzic:2020csw, Taghavi:2020gcy, Li:2021nas, Magdy:2024ooh, Magdy:2022ize, Magdy:2021sba, Magdy:2018itt, ALICE:2017fcd, CMS:2019nct, CMS:2013wjq, ALICE:2011ab, Yan:2015jma, Bilandzic:2021rgb, STAR:2022gki, STAR:2022vkx}. These efforts collectively contribute to a deeper understanding of the interplay between initial state conditions, medium response, and the resulting flow observables, paving the way for refined characterizations of the QGP properties.
%----------------------------------------------------------------------

%----------------------------------------------------------------------
Recent investigations have prompted inquiries into the interplay between the soft and hard properties of the QGP, raising questions about how to effectively measure such a relationship~\cite{Holtermann:2023vwr, Holtermann:2024vdw, Plumberg:2024leb}. It is anticipated that measurements of the $p_{T}$ differential flow could provide valuable insights into the energy loss mechanism of jets while also serving as indicators of the high $p_{T}$ nature of flow fluctuations and correlations. Consequently, addressing such inquiries requires a detailed investigation of the differential particle correlations and fluctuations. Moreover, several studies have highlighted potential limitations of simple differential flow measurements involving only one particle of interest (POI), such as fluctuations in the event-plane angle ($\psi_{n}$)~\cite{Gardim:2012im, Magdy:2022jai, Zhu:2024tns}. This underscores the importance of considering a more comprehensive approach in analyzing differential flow measurements to capture the full spectrum of underlying dynamics.
%----------------------------------------------------------------------

%----------------------------------------------------------------------
In this study, I extend previous investigations of integrated SC, ASC~\cite{Taghavi:2020gcy, Bilandzic:2021rgb, Bilandzic:2020csw, Mordasini:2019hut, Li:2021nas, Moravcova:2020wnf} to encompass a comprehensive differential analysis. Utilizing the HIJING and AMPT models, I conduct a detailed examination of SC and ASC employing one and two Particles of Interest (POIs). Additionally, I compare the AMPT calculations with available experimental data. 
%----------------------------------------------------------------------
In addition, the present study examines the influence of both (i) long-range correlations, such as those arising from dijet correlations, and (ii) short-range correlations, including resonance decays, Bose-Einstein correlations, and fragments of individual jets, on the presented quantities, utilizing the HIJING model~\cite{Wang:1991hta, Gyulassy:1994ew}.
%----------------------------------------------------------------------

%----------------------------------------------------------------------
The structure of this paper is outlined as follows: Section~\ref{sec:2} provides an overview of the theoretical models utilized for examining SC and ASC, which is given in section~\ref{sec:3}. The findings derived from the conducted studies are delineated in Section~\ref{sec:4}, followed by the summary in Section~\ref{sec:5}.
%--------------------------------------------------------------------

%--------------------------------------------------------------------
%--------------------------------------------------------------------
%\section{Methodology}
%--------------------------------------------------------------------
\section{Models} \label{sec:2}
%--------------------------------------------------------------------
The analysis is performed utilizing events simulated by the AMPT (v2.26t9b)~\cite{Lin:2004en} and HIJING (v1.411)~\cite{Wang:1991hta, Gyulassy:1994ew} models for Au+Au collisions at $\sqrt{s_{NN}}$ = 200 GeV. Charged particles within the transverse momentum range of $0.0 < p_T < 5.0$ GeV/$c$ and pseudorapidity $|\eta| < 1.0$ are selected for analysis as POI and of $0.0 < p_T < 2.0$ GeV/$c$ are selected for analysis as Particles of Reference (POR). The HIJING model elucidates non-flow effects, whereas the AMPT model scrutinizes the dependence of flow correlators on final state effects.

The HIJING model~\cite{Wang:1991hta, Gyulassy:1994ew} is a Monte Carlo event generator designed to simulate parton and particle production in high-energy heavy ion collisions. It employs Glauber geometry to simulate heavy-ion collisions via binary nucleon-nucleon interactions. It is utilized for studying jet and mini-jet production alongside associated particle production in high-energy p+p, p+A, and A+A collisions. Furthermore, the HIJING model integrates PYTHIA~\cite{Sjostrand:1986hx} for determining kinematic variables of scattered partons in each hard- or semihard-interaction, and Lund string fragmentation~\cite{Andersson:1983ia} for the hadronization process.

The AMPT model is widely employed in exploring heavy-ion collision physics~\cite{Lin:2004en, Ma:2016fve, Ma:2013gga, Ma:2013uqa, Bzdak:2014dia, Nie:2018xog, Haque:2019vgi, Zhao:2019kyk, Bhaduri:2010wi, Nasim:2010hw, Xu:2010du, Magdy:2020bhd, Guo:2019joy, Magdy:2020gxf, Magdy:2022cvt}. It encompasses several key components: (i) an initial partonic state provided by the HIJING model~\cite{Wang:1991hta, Gyulassy:1994ew}, with parameters specified in the Lund string fragmentation function~\cite{Ferreres-Sole:2018vgo} denoted by $f(z) \propto z^{-1} (1-z)^a\exp (-b~m_{\perp}^2/z)$, where $a=0.55$ and $b=0.15$ GeV$^{-2}$, as detailed in Ref~\cite{Xu:2011fi}. Here, $z$ represents the light-cone momentum fraction of the produced hadron with transverse mass $m_\perp$ around that of the fragmenting string. (ii) Partonic scattering characterized by a cross-section,
%--------------------------------------------------------------------
\begin{eqnarray}\label{eq:21}
\sigma_{pp} &=& \dfrac{9 \pi \alpha^{2}_{s}}{2 \mu^{2}}.
\end{eqnarray}
%--------------------------------------------------------------------
In this context, $\mu$ denotes the screening mass, while $\alpha_{s}$ represents the QCD coupling constant, typically characterizing the expansion dynamics of A--A collisions~\cite{Zhang:1997ej}. Moreover, the model incorporates (iii) the hadronization process through coalescence followed by hadronic interactions~\cite{Li:1995pra}.

{
Within the present AMPT model framework, only the initial value of \etas evaluated at the beginning of the heavy-ion collision can be assessed via a proper choice of the $\sigma_{pp}$ by varying $\mu$ and/or $\alpha_s$. This choice assumes an initial temperature of $T_{i}$=378~MeV for the system of massless quarks and gluons~\cite{Xu:2011fi, Nasim:2016rfv, Solanki:2012ne}:
\begin{eqnarray} \label{eq:22}
 \dfrac{\eta}{s} &=& \dfrac{3 \pi}{40 \alpha^{2}_{s}}  \dfrac{1}{ \left(  9 +  \dfrac{\mu^2}{T_{i}^2} \right)  \ln\left(\dfrac{18 + \mu^2/T_{i}^2}{ \mu^2/T_{i}^2 } \right) - 18}.
\end{eqnarray}
}
%--------------------------------------------------------------------

%--------------------------------------------------------------------
\begin{table}[h!]
\begin{center}
\caption{The summary of the AMPT sets used in this work.\label{tab:1}}
 \begin{tabular}{|c|c|c|c|}
 \hline 
 AMPT-set      &  $\mu$~($fm^{-1}$) &   $\sigma_{pp}$   &  \etas     \\
  \hline
  Set-1        &      2.265       &       3.0         &   0.232    \\    
  \hline
  Set-2        &      1.602       &       6.0         &   0.156   \\ 
 \hline
\end{tabular} 
\end{center}
\end{table}
%--------------------------------------------------------------------
%--------------------------------------------------------------------
This study simulated Au+Au collisions at $\sqrt{s_{\rm NN}}=$ 200~GeV using AMPT and Hijing models. The AMPT model with fixed values of $\alpha_{s}$ = 0.33 and $\mu$ =1.602 and 2.265 is used, see tab.\ref{tab:1}. The choice of these model parameters is motivated by previous works~\cite{Xu:2011fi, Nasim:2016rfv, Magdy:2021sba, Magdy:2020fma}. The analysis focuses on minimum bias Au+Au collisions at $\sqrt{s_{\rm NN}}=$~200~GeV, with approximately 20 million events generated for each case considered.

%--------------------------------------------------------------------
\section{Correlators}~\label{sec:3}
%--------------------------------------------------------------------
%--------------------------------------------------------------------
The events generated were analyzed using the two- and multi-particle correlations given via the use of the traditional and two-subevents cumulant methods~\cite{Bilandzic:2010jr, Jia:2017hbm, Huo:2017nms, Zhang:2018lls, Magdy:2020bhd} see Appendix~\ref{App:A}. The observables discussed in this work can be presented in terms of the integrated/differential flow vectors;
%--------------------------------------------------------------------
\begin{eqnarray}\label{eq:2-1}
    Q_{n,P} &=& \sum_{j} e^{\mathit{i}n\phi_{j,P} },
\end{eqnarray}
where $\phi_j$ is the azimuthal angle of the $\mathit{j}^{th}$ particle, $P$ will represent the particle states (i.e., zero for integrated and non-zero for differential particle).
%--------------------------------------------------------------------

%--------------------------------------------------------------------
%--------------------------------------------------------------------
\begin{widetext}
%--------------------------------------------------------------------
\subsection{Symmetric Correlations (SC)}
%--------------------------------------------------------------------
The two-, four-, and six-particle symmetric correlations $\langle V_{n_{1}}... V_{n_{1}} \rangle_{P}$ can be given as;
%--------------------------------------------------------------------
{\small
\begin{eqnarray}\label{eq:2-2}
    \langle  V_{n_{1}} V_{n_{2}} \rangle_{P} &=& e^{\mathit{i}  (n_{1} \phi_{1,P} + n_{2} \phi_{2,P}) } \equiv v_{n_{1},P}~v_{n_{2},P}  ~cos(n_{1} \psi_{n_{1},P} + n_{2} \psi_{n_{2},P}),
\end{eqnarray}
}
{\small
\begin{eqnarray}\label{eq:2-4}
\langle V_{n_{1}} V_{n_{2}} V_{n_{3}} V_{n_{4}} \rangle_{P} &=& e^{\mathit{i}  (n_1 \phi_{1,P} + n_2 \phi_{2,P} + n_3 \phi_{3,P} + n_4 \phi_{4,P}) } \nonumber \\
&\equiv& v_{n_1,P}~v_{n_2,P}~v_{n_3,P} ~v_{n_4,P} ~cos(n_1 \psi_{n_1,P} + n_2 \psi_{n_2,P} + n_3 \psi_{n_3,P} + n_4 \psi_{n_4,P}), 
\end{eqnarray}
\begin{eqnarray}
\langle V_{n_{1}} V_{n_{2}} V_{n_{3}} V_{n_{4}} V_{n_{5}} V_{n_{6}} \rangle_{P}  &=& e^{\mathit{i}  (n_1 \phi_{1,P} + n_2 \phi_{2,P} + n_3 \phi_{3,P} + n_4 \phi_{4,P} + n_5 \phi_{5,P} + n_6 \phi_{6,P}) } \nonumber \\
&\equiv& v_{n_1,P}~v_{n_2,P}~v_{n_3,P} ~v_{n_4,P} ~v_{n_5,P} ~v_{n_6,P}  \nonumber \\
& & cos(n_1 \psi_{n_1,P} + n_2 \psi_{n_2,P} + n_3 \psi_{n_3,P}  + n_4 \psi_{n_4,P} + n_5 \psi_{n_5,P} + n_6 \psi_{n_6,P}).
\end{eqnarray}
}
%--------------------------------------------------------------------
For harmonic orders of the particle of interest (POI), denoted as $P$, the values can be $P=00$, $P=000$, $P=0000$, $P=00000$, and $P=000000$, corresponding to no particle of interest in two, three, four, five, and six particle correlations, respectively. When there are one or two POIs, the value of $P$ can range from 1 to $k$, where $k$ represents the flow harmonic. For example, in two-particle correlations, $n_1$ or $n_2$ representing the POI can be denoted as $\langle V_{n_{1}} V_{n_{2}} \rangle_{10}$ or $\langle V_{n_{1}} V_{n_{2}} \rangle_{02}$, respectively. Additionally, two-particle correlations with two POIs are represented as $\langle V_{n_{1}} V_{n_{2}} \rangle_{12}$.

\subsection{Asymmetric Correlations (ASC)}
%--------------------------------------------------------------------
The three-, four-, and five-particle asymmetric correlations $\langle V_{n_{1}}...V_{n_{5}} \rangle_{P}$ can be given as;
%--------------------------------------------------------------------
{\small
\begin{eqnarray}
\langle  V_{n_{1}} V_{n_{2}} V_{n_{3}}  \rangle_{P} &=& e^{\mathit{i}  (n_1 \phi_{1,P} + m \phi_{2,P} + k \phi_{3,P}) } \equiv v_{n_1,P}~v_{n_2,P}~v_{n_3,P}  ~cos(n_1 \psi_{n_1,P} + n_2 \psi_{n_2,P} + n_3 \psi_{n_3,P}),
\end{eqnarray}
}
%--------------------------------------------------------------------
\small{
\begin{eqnarray}
\langle V_{n_{1}} V_{n_{2}} V_{n_{3}} V_{n_{4}} V_{n_{5}} \rangle_{P} &=& e^{\mathit{i}  (n_1 \phi_{1,P} + n_2 \phi_{2,P} + n_3 \phi_{3,P} + n_4 \phi_{4,P} + n_5 \phi_{5,P}) } \nonumber \\
&\equiv& v_{n_1,P}~v_{n_2,P}~v_{n_3,P} ~v_{n_4,P} ~v_{n_5,P} \nonumber \\
& & cos(n_1 \psi_{n_1,P} + n_2 \psi_{n_2,P} + n_3 \psi_{n_3,P} +  n_4 \psi_{n_4,P} + n_5 \psi_{n_5,P}),   \nonumber \\
\end{eqnarray}
}

The current work contains two assumptions; first, the integrated (differential) event planes are equivalent (i.e., we do not consider the longitudinal $\eta$ decorrelations). Second, non-flow effects are small for the two-particle correlations with a large $\eta$ gap. The second assumption is validated in this work using the HIJING model. The k-particle correlations using the traditional cumulant method will be used for the five- and six-particle correlations, while the two-subevent cumulant method with $\eta$ gap ($|\Delta\eta|> 0.7$) will be used for the two-, three-, and four-particle correlations, see Appendix~\ref{App:A}.

In this work, the discussions are limited to introducing the differential two- to six-particle correlations and discussing their sensitivity to (i) the non-flow effects given by the HIJING model and (ii) the final state effects given by the $\sigma_{pp}$ variation in the AMPT model.
%--------------------------------------------------------------------
%--------------------------------------------------------------------
\end{widetext}

%--------------------------------------------------------------------
%--------------------------------------------------------------------
\section{Results and discussion}\label{sec:4}
%--------------------------------------------------------------------
This section will explore the impact of final state effects on symmetric and asymmetric correlations and the potential influence of non-flow effects on these presented observables. Therefore, I found it important to compare the AMPT elliptic and triangular flow, $v_2$ and $v_3$, with the experimental measurements~\cite{PHENIX:2011yyh}. Such comparison can help us understand to what degree the experimental measurements can help constrain the theoretical model. 
%--------------------------------------------------------------------

%--------------------------------------------------------------------
\begin{figure}[ht] 
\includegraphics[width=0.90 \linewidth, angle=-0,keepaspectratio=true,clip=true]{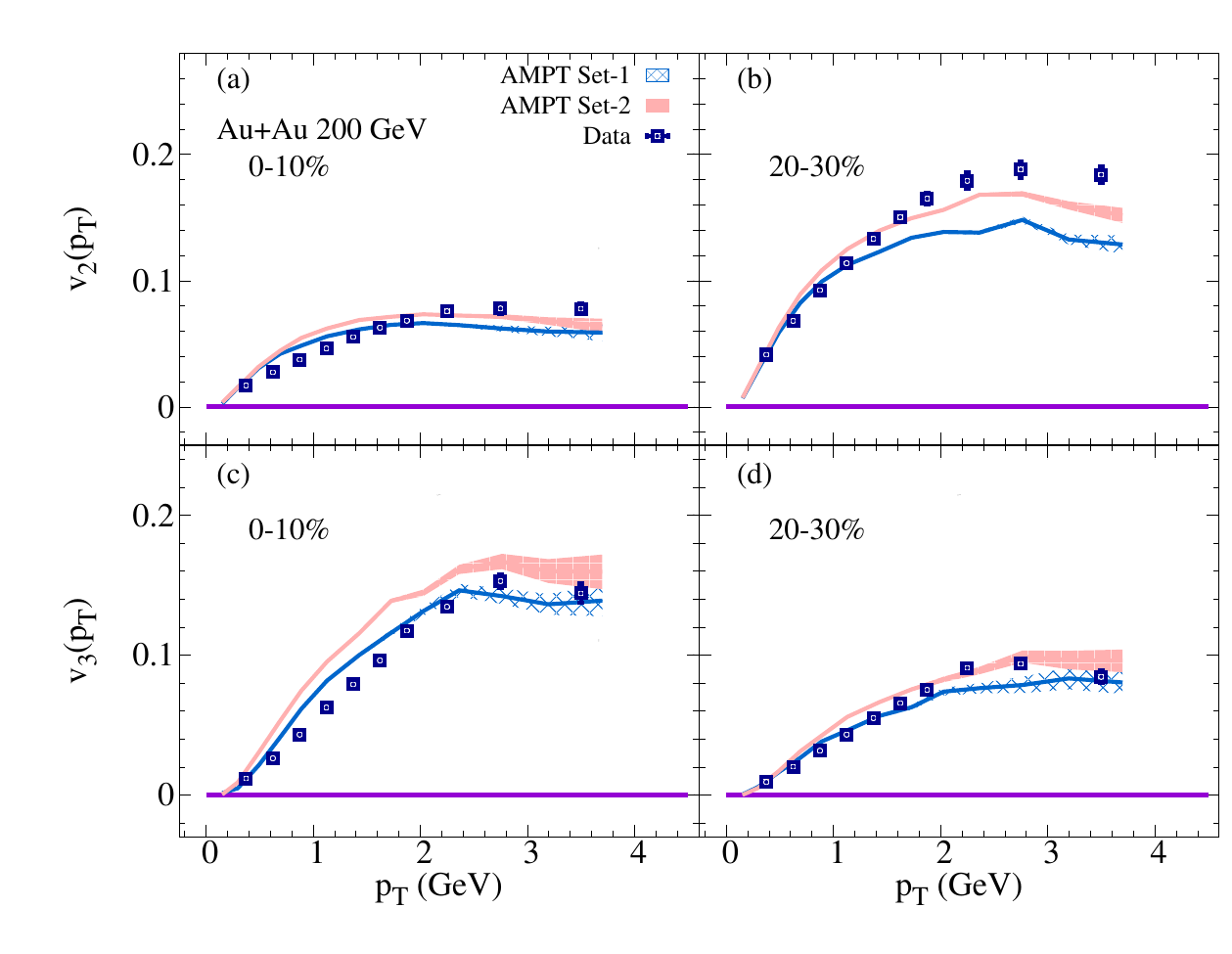}
\vskip -0.4cm
\caption{
The $p_{T}$ dependence of the $v_{2}$ and $v_{3}$, using the two-subevents method for Au+Au at $\sqrt{\textit{s}_{NN}}~=$ 200~GeV from the AMPT compared to experimental data~\cite{PHENIX:2011yyh} given by open symbols.
\label{fig:0-0comp}}
\vskip -0.1cm
\end{figure}
%--------------------------------------------------------------------
Figure~\ref{fig:0-0comp} shows a comparison of the $p_{T}$ dependence of the $v_{2} (p_{T})$ (a)-(b) and $v_{3} (p_{T})$ (c)-(d)  for Au+Au collisions at 200~GeV from the AMPT model.  The presented $v_{n} (p_{T})$ from the AMPT exhibit a sensitivity to $\sigma_{pp}$ variations. 
The calculations indicate a reasonable agreement with the experimental measurements Ref.~\cite{PHENIX:2011yyh} (open symbols). Comparisons between the  AMPT calculations and the available experimental measurements of $v_n$ reveal the ability of the flow harmonic correlations to constrain the AMPT final state effects.

%--------------------------------------------------------------------
\subsection{Symmetric correlations with one POI}
%--------------------------------------------------------------------

\begin{figure}[!h] 
\includegraphics[width=0.99 \linewidth, angle=-0,keepaspectratio=true,clip=true]{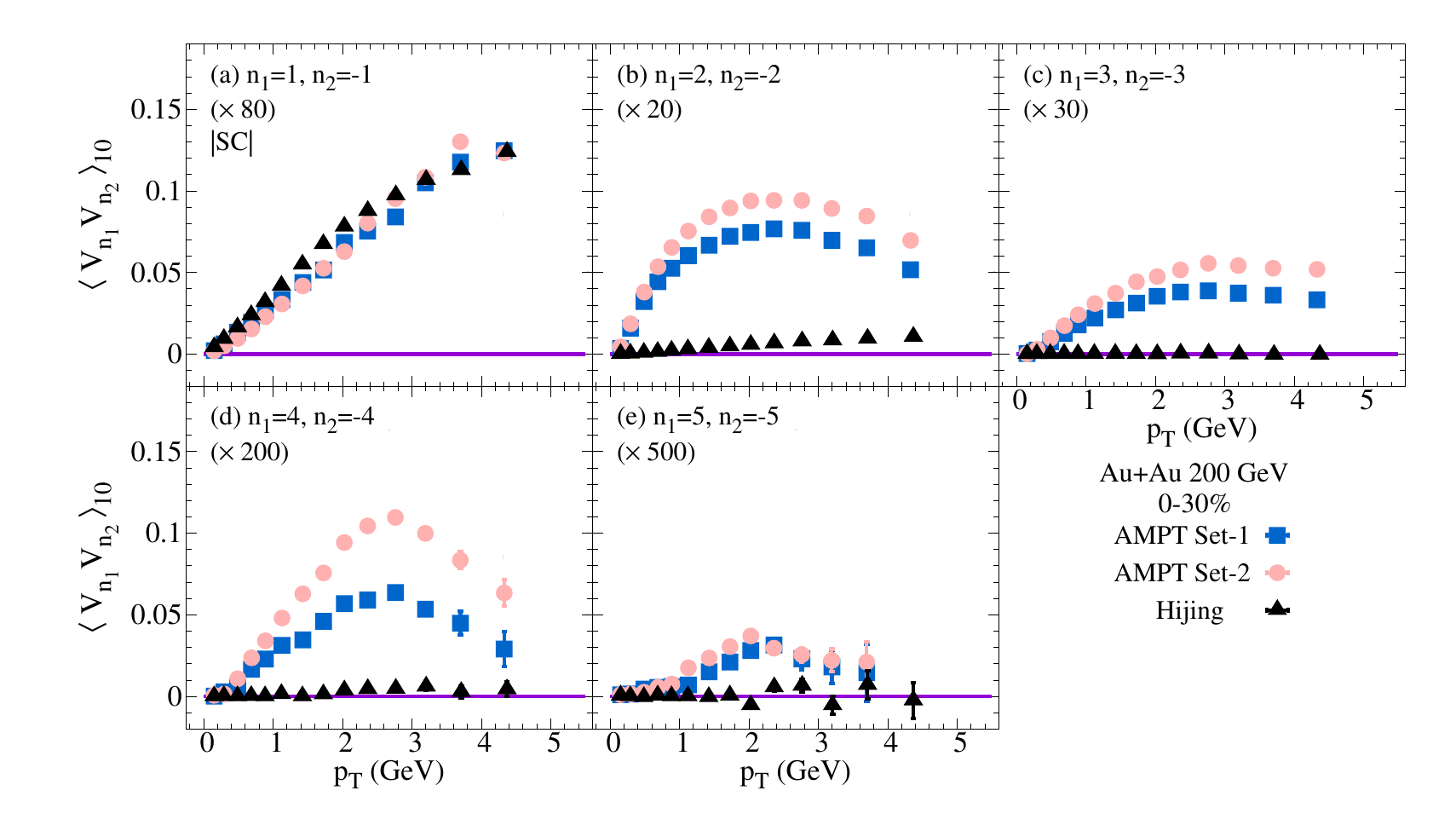}
\vskip -0.4cm
\caption{
The $p_{T}$ dependence of the two-particle flow harmonics with one POS, $v^{2}_{n}(p_{T}) = \langle V_{n_1} V_{n_2}\rangle_{10}$, using the two-subevents method for Au+Au at $\sqrt{\textit{s}_{NN}}~=$ 200~GeV from the AMPT and HIJING models. 
\label{fig:1-1psc}}
\vskip -0.1cm
\end{figure}
%--------------------------------------------------------------------
\begin{figure}[!h] 
\includegraphics[width=0.90 \linewidth, angle=-0,keepaspectratio=true,clip=true]{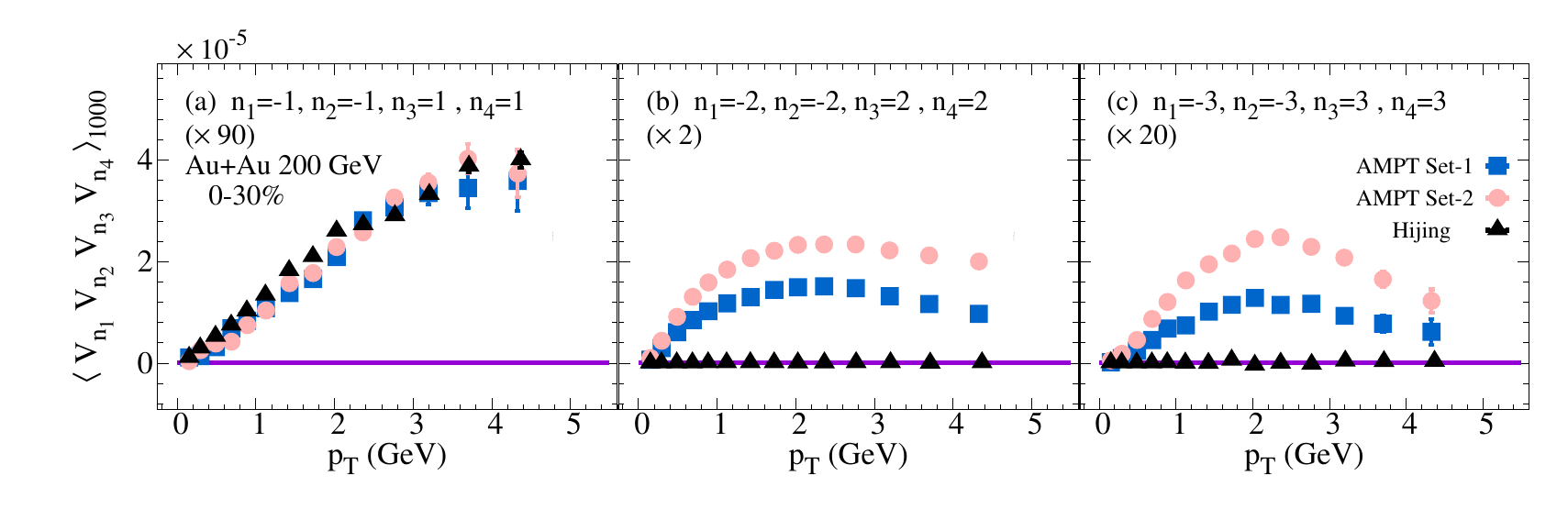}
\vskip -0.4cm
\caption{
Same as in Fig.~\ref{fig:1-1psc} but for same harmonic correlations $\langle V_{n_1} V_{n_2} V_{n_3} V_{n_4}\rangle_{1000}$.
\label{fig:2-1psc}}
\vskip -0.1cm
\end{figure}
%--------------------------------------------------------------------
\begin{figure}[!h] 
\includegraphics[width=0.90 \linewidth, angle=-0,keepaspectratio=true,clip=true]{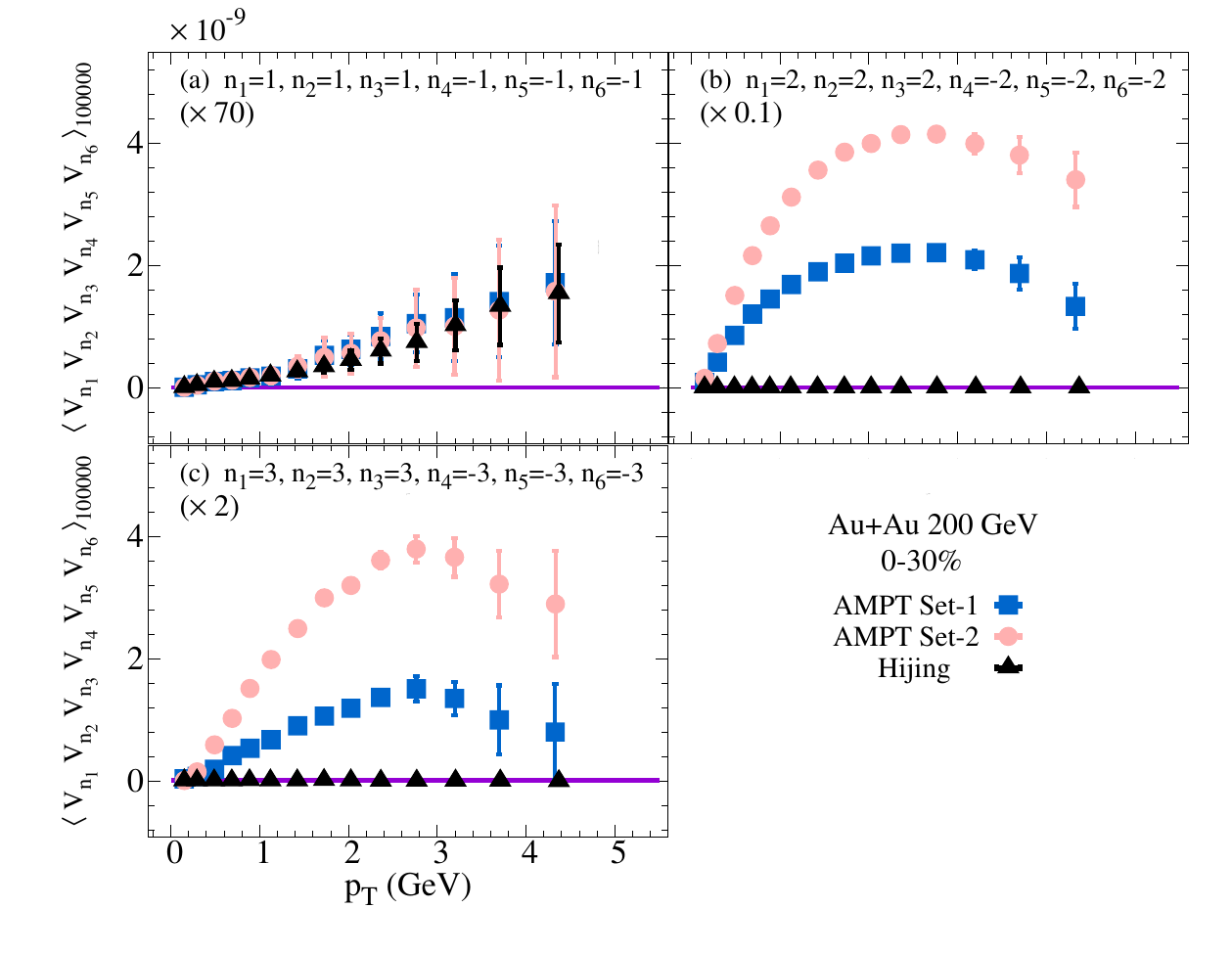}
\vskip -0.4cm
\caption{
Same as in Fig.~\ref{fig:1-1psc} but for same harmonic correlations $\langle V_{n_1} V_{n_2} V_{n_3} V_{n_4} V_{n_5} V_{n_6}\rangle_{100000}$ using the traditional cumulant  method.
\label{fig:3-1psc}}
\vskip -0.1cm
\end{figure}
%--------------------------------------------------------------------
%--------------------------------------------------------------------
In Figs.~\ref{fig:1-1psc}--\ref{fig:3-1psc}, we present our AMPT and HIJING models predictions for the symmetric correlations involving two, four, and six particles as a function of transverse momentum ($p_{T}$) and final state effects ($\sigma_{pp}$) for Au+Au collisions at 200~GeV. The four- and six-particle correlations are anticipated to contain contributions from flow and flow fluctuations (i.e., initial state density fluctuations and hydrodynamic evolution fluctuations). 
The results reveal a discernible decrease in the same harmonic correlations (n$>$1) with the $\sigma_{pp}$, underscoring their sensitivity to the final state effects delineated by the AMPT model.  The observed differences in sensitivity to $\sigma_{pp}$ between two-particle, four-particle, and six-particle SC likely arise due to the number of particles that are correlated~\cite{Niemi:2015qia, Sorensen:2009cz, Voloshin:2007pc}. Further investigation into these differences in sensitivity could be an interesting direction for future work.
Moreover, the results obtained from the HIJING model for $n > 1$ suggest small non-flow contributions in the two-, four-, and six-particle symmetric correlations. 
%--------------------------------------------------------------------

%--------------------------------------------------------------------
The two-particle SC $\langle V_{1} V_{-1}\rangle_{10}$ shown in Fig.~\ref{fig:1-1psc} (a) demonstrates similar effects between the HIJING and AMPT models, albeit with a larger magnitude from the HIJING model. This can be attributed to the contribution of the global momentum conservation (GMC) effect on such two-particle SC. The GMC is expected to impact $\langle V_{1} V_{-1}\rangle$ as~\cite{Borghini:2000cm, Borghini:2002mv, Retinskaya:2012ky, ATLAS:2012at}:
\begin{eqnarray}\label{corrv1}
\langle V_{1} V_{-1}\rangle(p_{T}^{a},p_{T}^{b})  &=& v_{1}(p_{T}^{a})v_{1}(p_{T}^{b}) - K p_{T}^{a} p_{T}^{b},
\end{eqnarray}
where $K$ is the GMC parameter that depends on the average multiplicity $\langle N_{ch} \rangle$ and the variance of $p_T$ (i.e., $K \propto 1/(\langle N_{ch} \rangle \langle p_{T}^{2}\rangle)$). The GMC effect is expected to have a more complicated contribution to higher-order correlations~\cite{STAR:2018gji, STAR:2019zaf}. Investigating this effect on higher-order correlations is beyond the scope of this study.
%--------------------------------------------------------------------

%--------------------------------------------------------------------
\begin{figure}[!h]
\includegraphics[width=0.90 \linewidth, angle=-0,keepaspectratio=true,clip=true]{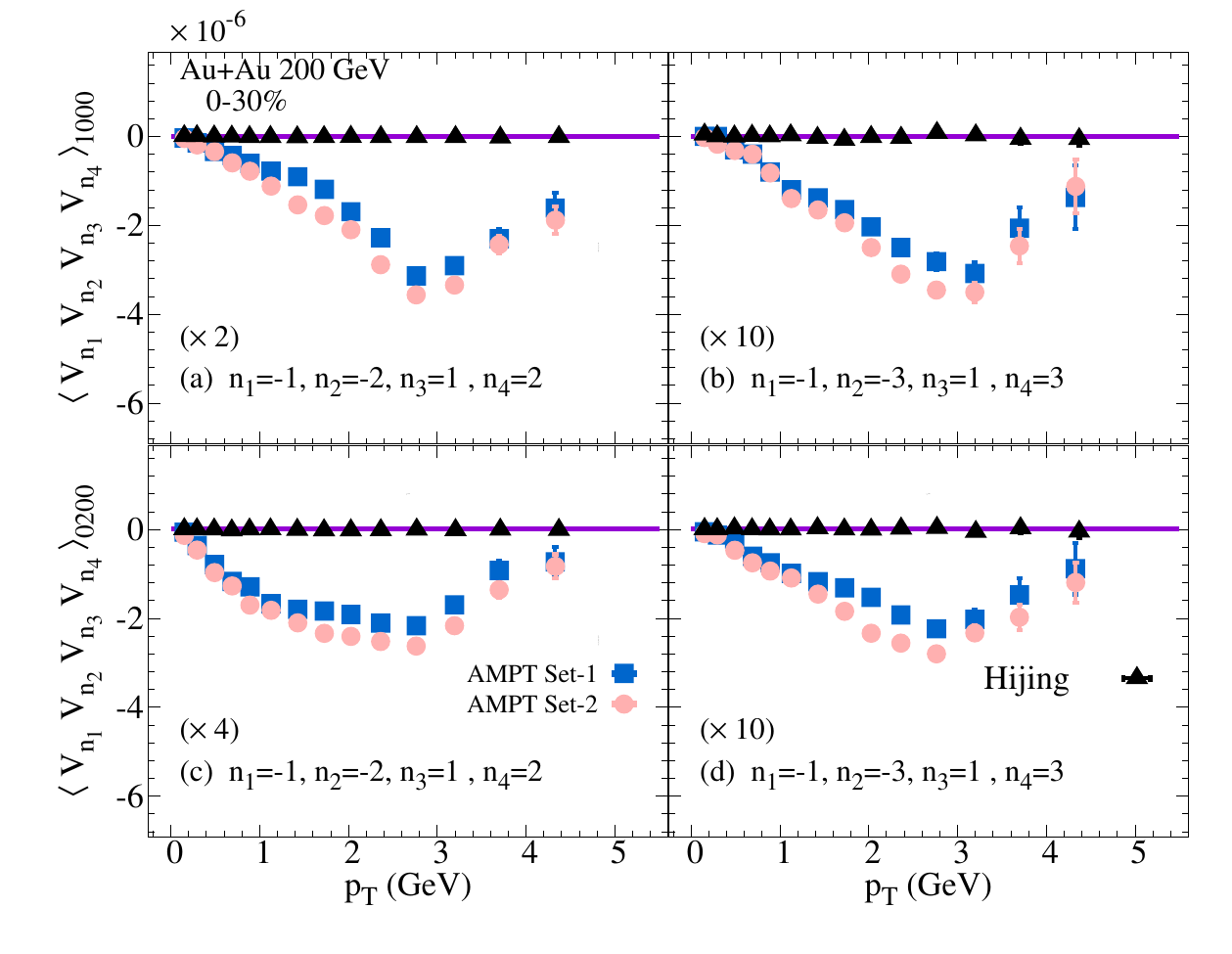}
\vskip -0.4cm
\caption{
Same as in Fig.~\ref{fig:1-1psc} but for mixed harmonic correlations $\langle V_{n_1} V_{n_2} V_{n_3} V_{n_4}\rangle_{1000}$ and $\langle V_{n_1} V_{n_2} V_{n_3} V_{n_4}\rangle_{0200}$.
\label{fig:4-1psc}}
\vskip -0.1cm
\end{figure}

\begin{figure}[!h] 
\includegraphics[width=0.90 \linewidth, angle=-0,keepaspectratio=true,clip=true]{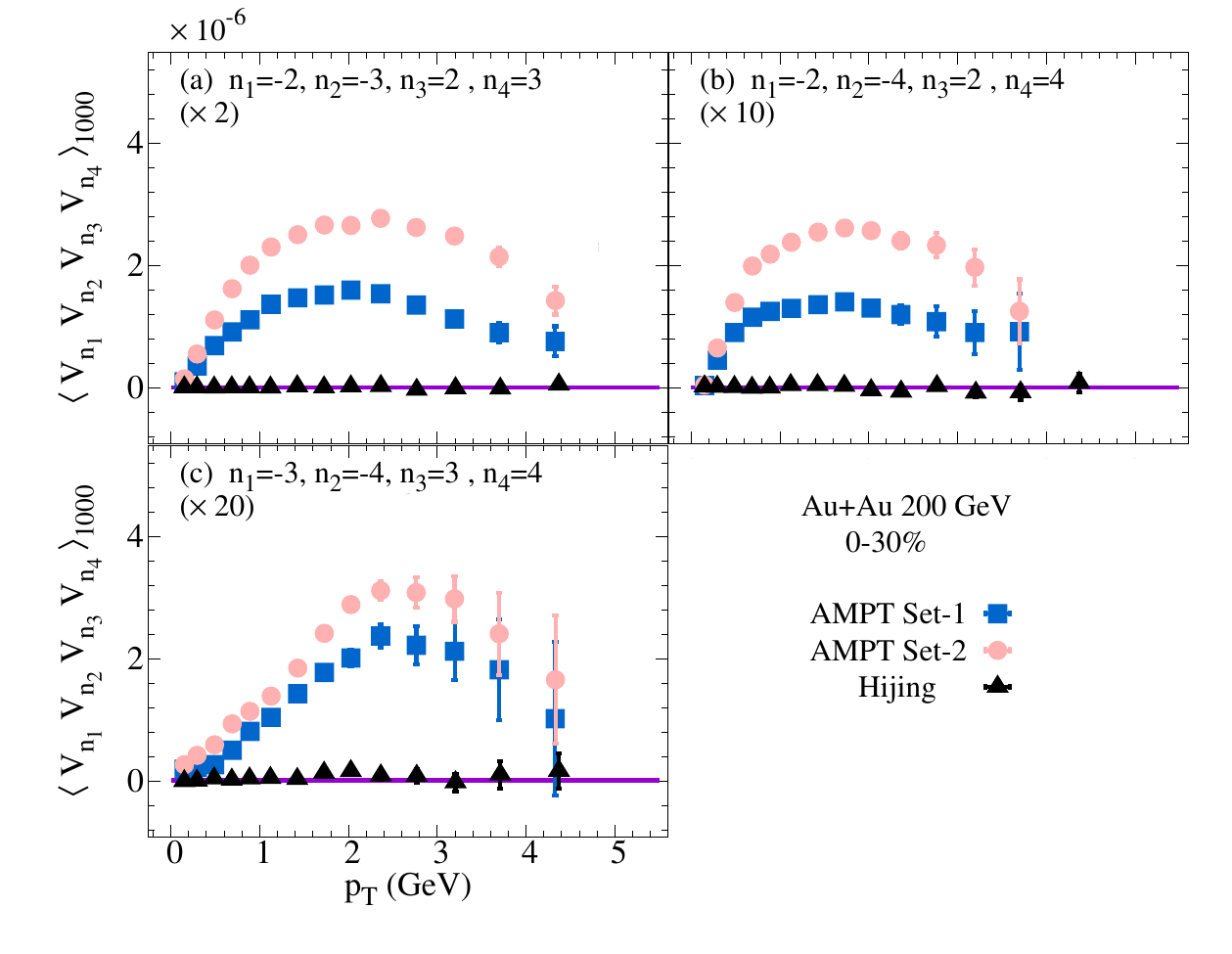}
\vskip -0.4cm
\caption{
Same as in Fig.~\ref{fig:1-1psc} but for mixed harmonic correlations $\langle V_{n_1} V_{n_2} V_{n_3} V_{n_4}\rangle_{1000}$.
\label{fig:5-1psc}}
\vskip -0.1cm
\end{figure}

\begin{figure}[!h]
\includegraphics[width=0.90 \linewidth, angle=-0,keepaspectratio=true,clip=true]{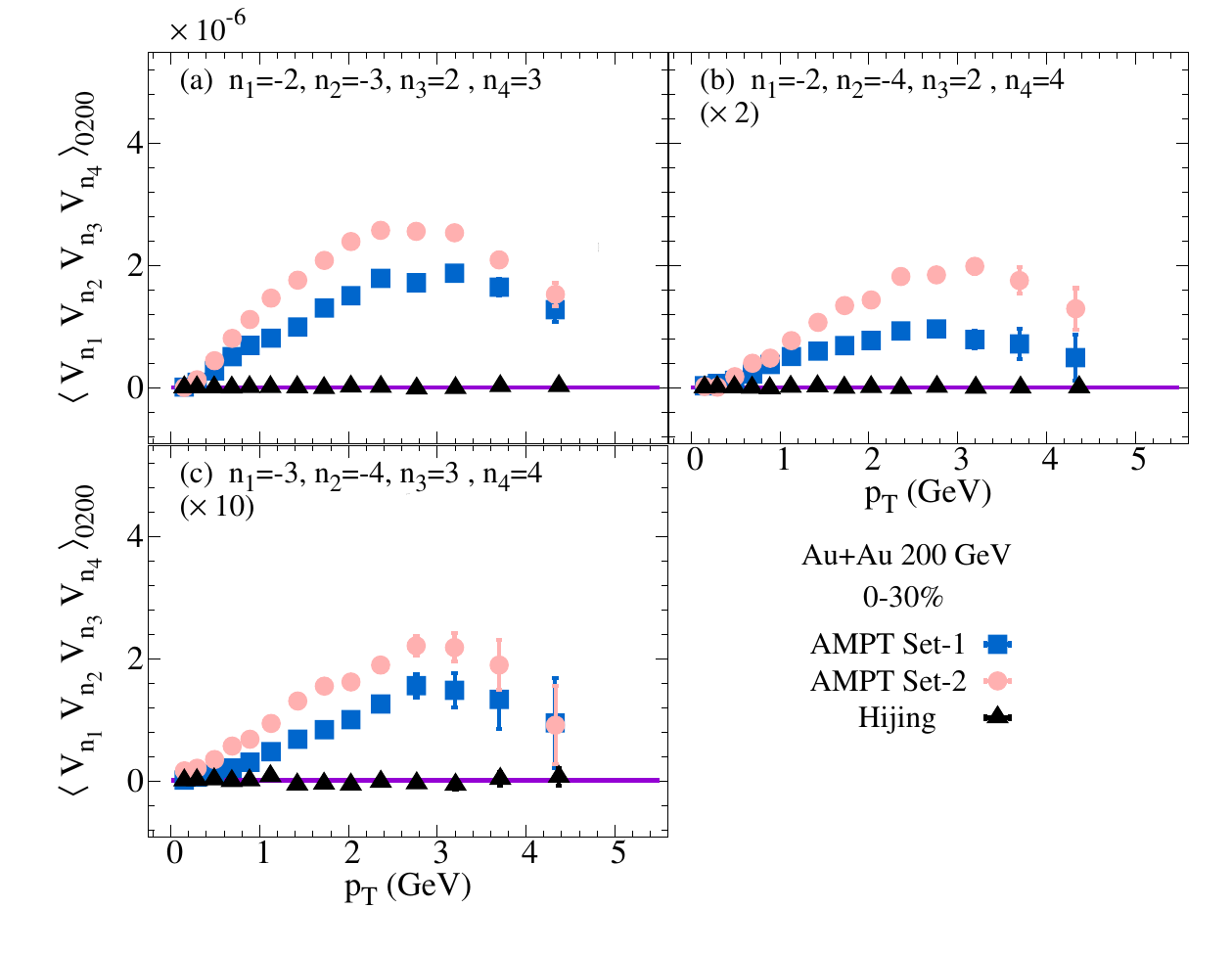}
\vskip -0.4cm
\caption{
Same as in Fig.~\ref{fig:1-1psc} but for mixed harmonic correlations $\langle V_{n_1} V_{n_2} V_{n_3} V_{n_4}\rangle_{0200}$.
\label{fig:6-1psc}
}
\vskip -0.1cm
\end{figure}

\begin{figure}[!h] 
\includegraphics[width=0.90 \linewidth, angle=-0,keepaspectratio=true,clip=true]{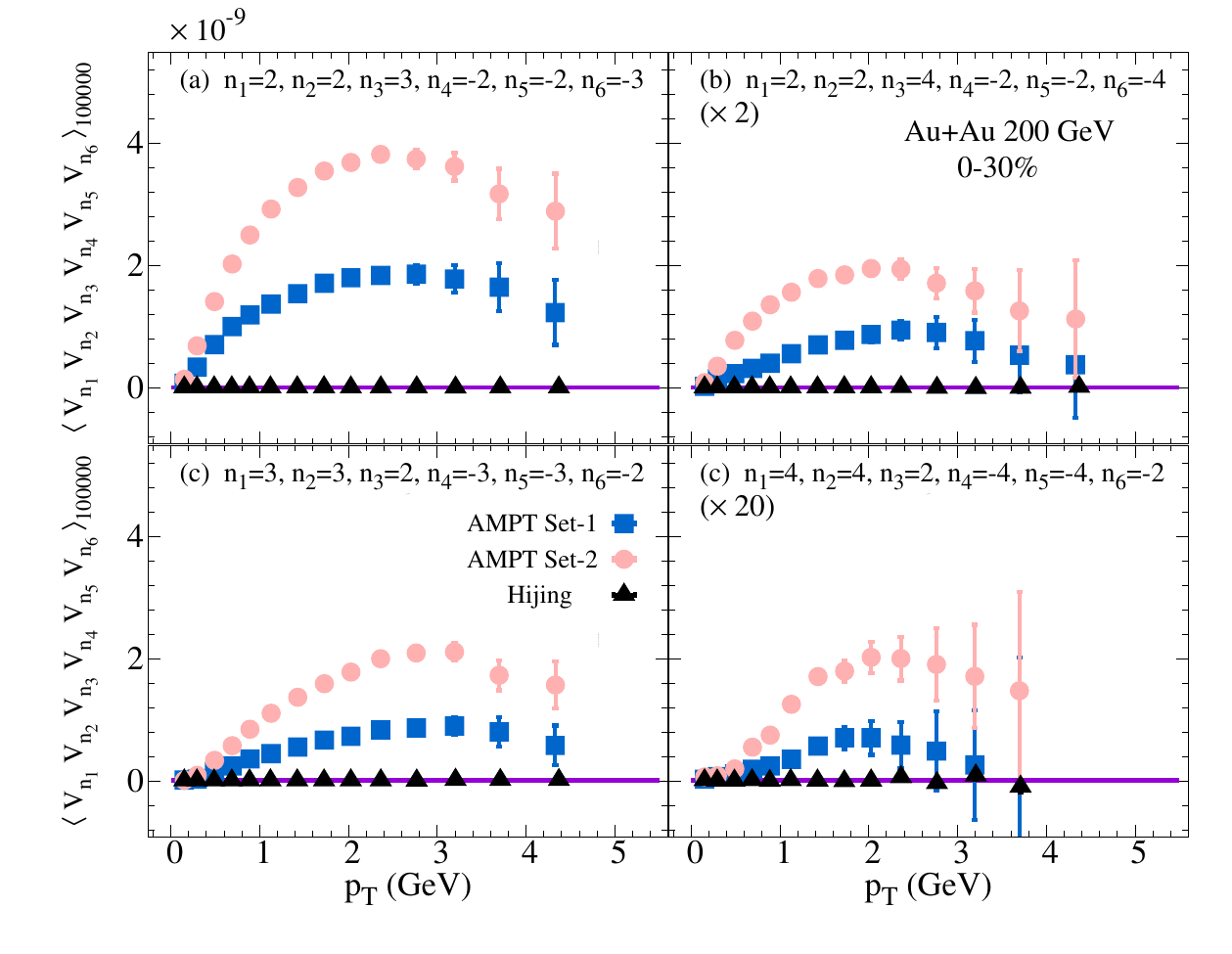}
\vskip -0.4cm
\caption{
Same as in Fig.~\ref{fig:1-1psc} but for mixed harmonic correlations $\langle V_{n_1} V_{n_2} V_{n_3} V_{n_4} V_{n_5} V_{n_6}\rangle_{100000}$  using the traditional cumulant  method.
\label{fig:7-1psc}}
\vskip -0.1cm
\end{figure}

\begin{figure}[!h] 
\includegraphics[width=0.90 \linewidth, angle=-0,keepaspectratio=true,clip=true]{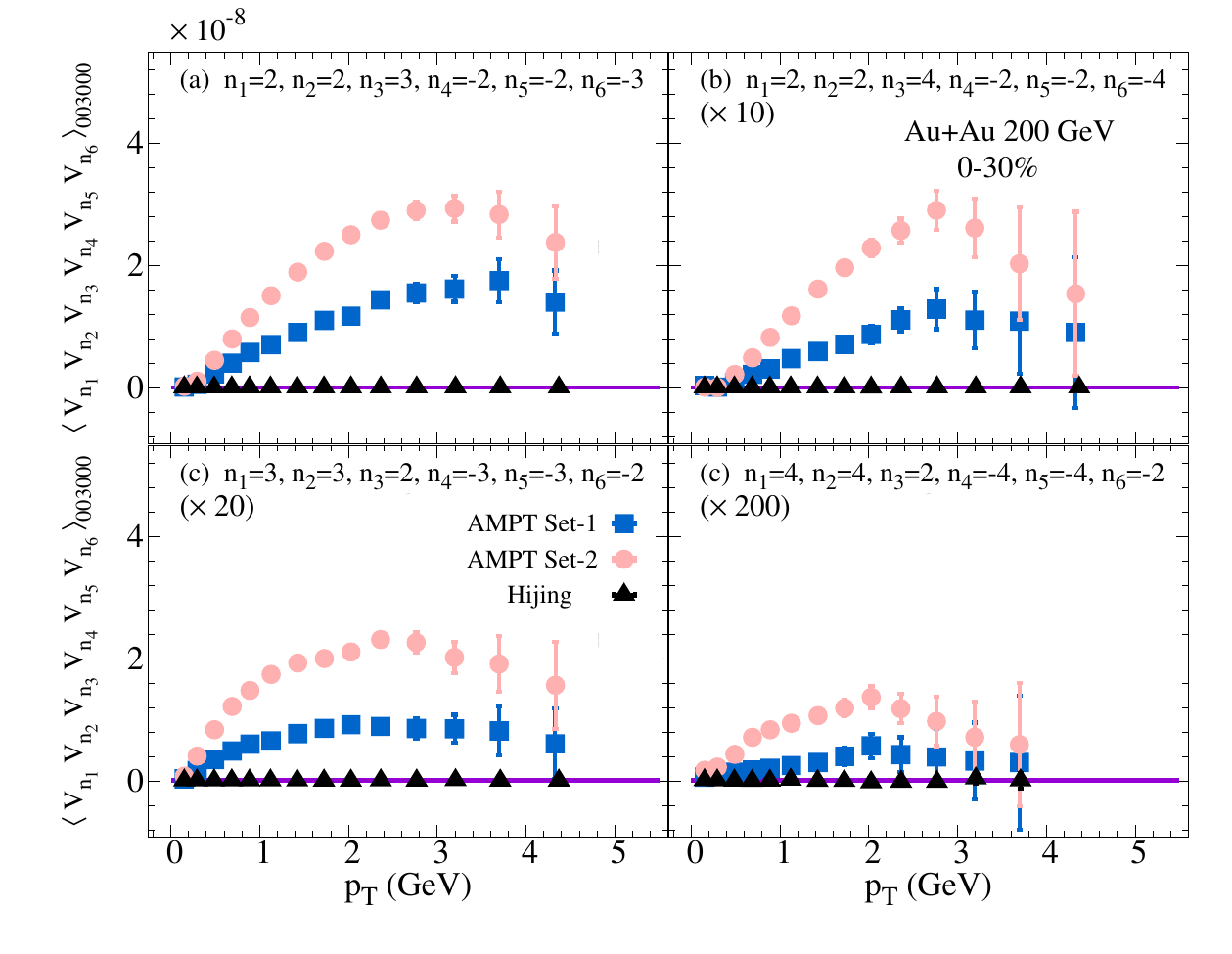}
\vskip -0.4cm
\caption{
Same as in Fig.~\ref{fig:1-1psc} but for mixed harmonic correlations $\langle V_{n_1} V_{n_2} V_{n_3} V_{n_4} V_{n_5} V_{n_6}\rangle_{003000}$  using the traditional cumulant  method.
\label{fig:8-1psc}}
\vskip -0.1cm
\end{figure}

\begin{figure}[!h] 
\includegraphics[width=0.90 \linewidth, angle=-0,keepaspectratio=true,clip=true]{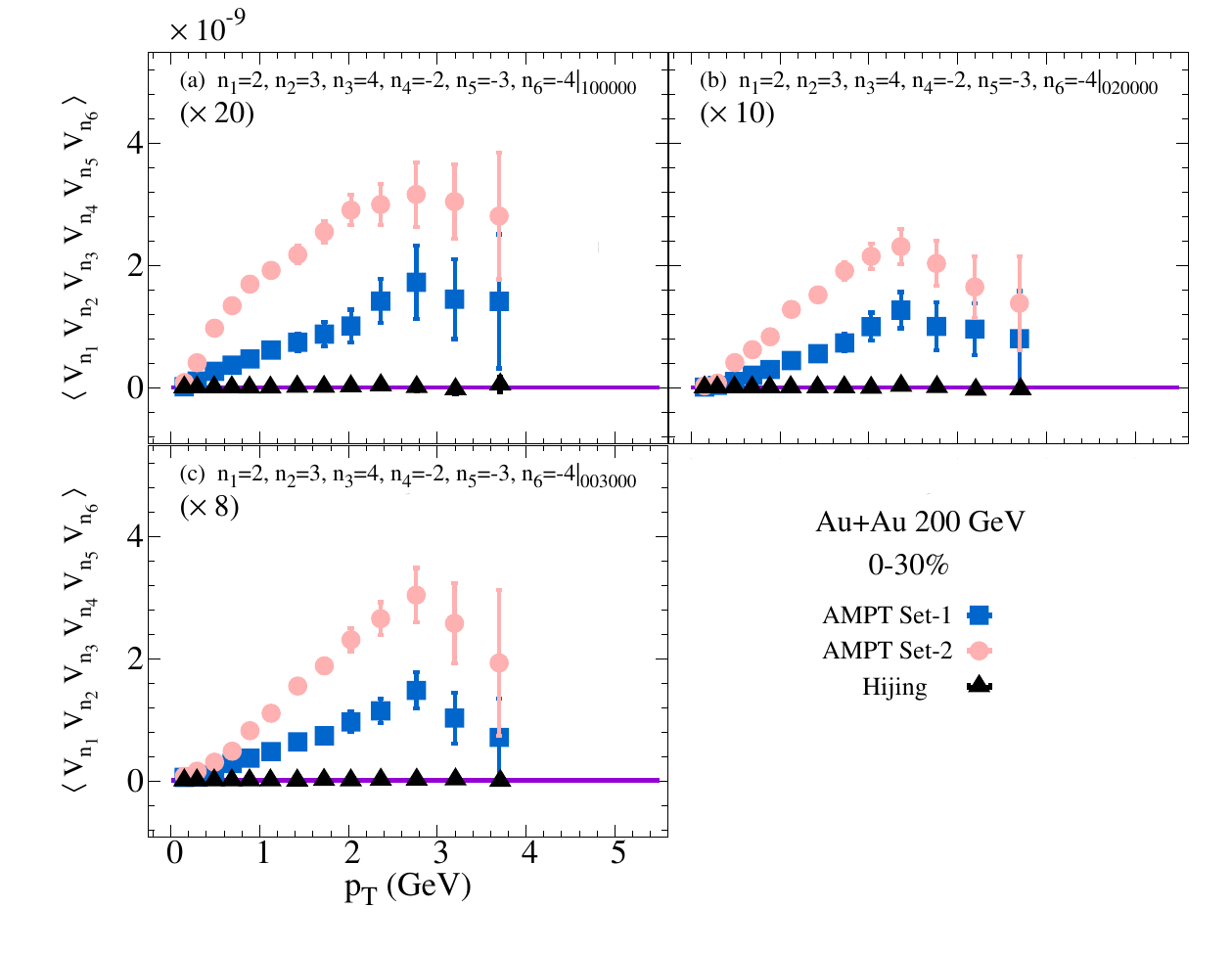}
\vskip -0.4cm
\caption{
Same as in Fig.~\ref{fig:1-1psc} but for mixed harmonic correlations $\langle V_{n_1} V_{n_2} V_{n_3} V_{n_4} V_{n_5} V_{n_6}\rangle_{100000}$,  $\langle V_{n_1} V_{n_2} V_{n_3} V_{n_4} V_{n_5} V_{n_6}\rangle_{020000}$ and  $\langle V_{n_1} V_{n_2} V_{n_3} V_{n_4} V_{n_5} V_{n_6}\rangle_{003000}$  using the traditional cumulant  method.
\label{fig:9-1psc}}
\vskip -0.1cm
\end{figure}
%--------------------------------------------------------------------
The mixed harmonics four- and six-particle symmetric correlations are given in Figs.~~\ref{fig:7-1psc}--~\ref{fig:9-1psc}, as a function of $p_{T}$ for Au+Au collisions at 200~GeV from HIJING and the AMPT models. The mixed harmonics four- and six-particle correlations are expected to give the flow harmonics correlations induced by initial and final state effects. 
The AMPT calculations indicate that mixed harmonic correlations had a characteristic dependence on the $p_T$, and they decrease with the $\sigma_{pp}$. Such dependence on $p_T$ and $\sigma_{pp}$ illustrates the mixed harmonics correlations' sensitivity to the final state effects given by the AMPT model.  In addition, the HIJING model calculations of the mixed harmonics symmetric correlations indicated that the results have little (if any) non-flow contribution.

Here, it's very essential to mention that the results presented with one POI in Figs.~\ref{fig:1-1psc}--\ref{fig:9-1psc} will contain a nonvanishing event-plane angle fluctuation term, $\langle cos(n \psi_{n}(p_{T}) - n \psi_{n}) \rangle$ (see Refs~\cite{Magdy:2022jai, Nielsen:2022jms}). Such an effect can complicate the interpolation of the experimental measurements of such observables. Thus, it is important to consider such an effect spatially for small collision systems and high $p_T$ differential flow messmates~\cite{ALICE:2022dtx}. In addition, in future work, when presenting the NSC and NASC, one needs to discuss the event-plane angle fluctuation effect in more detail.

%--------------------------------------------------------------------
%--------------------------------------------------------------------
%--------------------------------------------------------------------
\subsection{Asymmetric correlations with one POI}

\begin{figure}[!h] 
\includegraphics[width=0.90 \linewidth, angle=-0,keepaspectratio=true,clip=true]{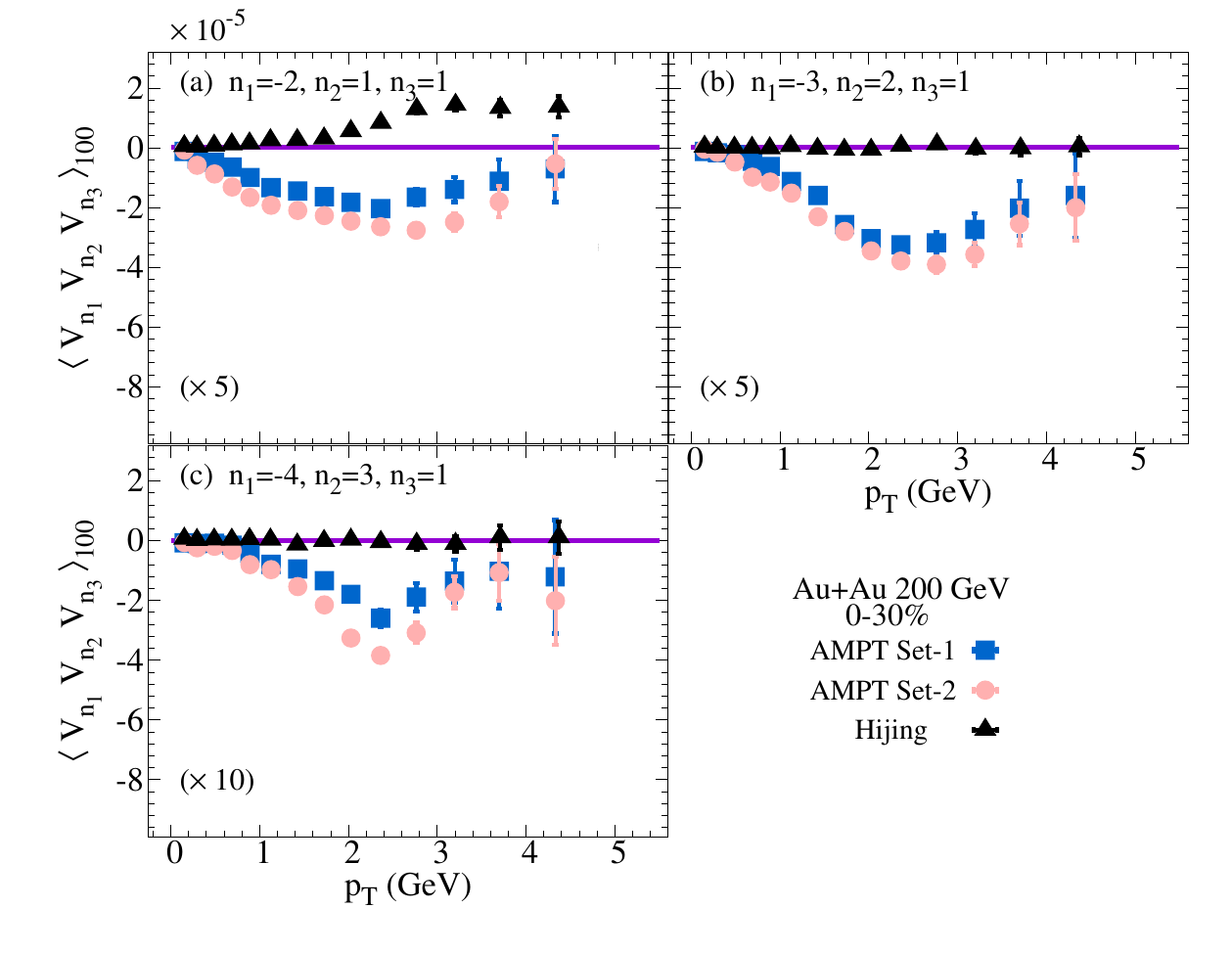}
\vskip -0.4cm
\caption{
The $p_{T}$ dependence of the three-particle asymmetric correlations $\langle V_{n_1} V_{n_2} V_{n_3}\rangle_{100}$ with one POS using the two-subevents method for Au+Au at $\sqrt{\textit{s}_{NN}}~=$ 200~GeV from the AMPT and HIJING models.
\label{fig:1-1pasc}}
\vskip -0.1cm
\end{figure}

\begin{figure}[!h] 
\includegraphics[width=0.90 \linewidth, angle=-0,keepaspectratio=true,clip=true]{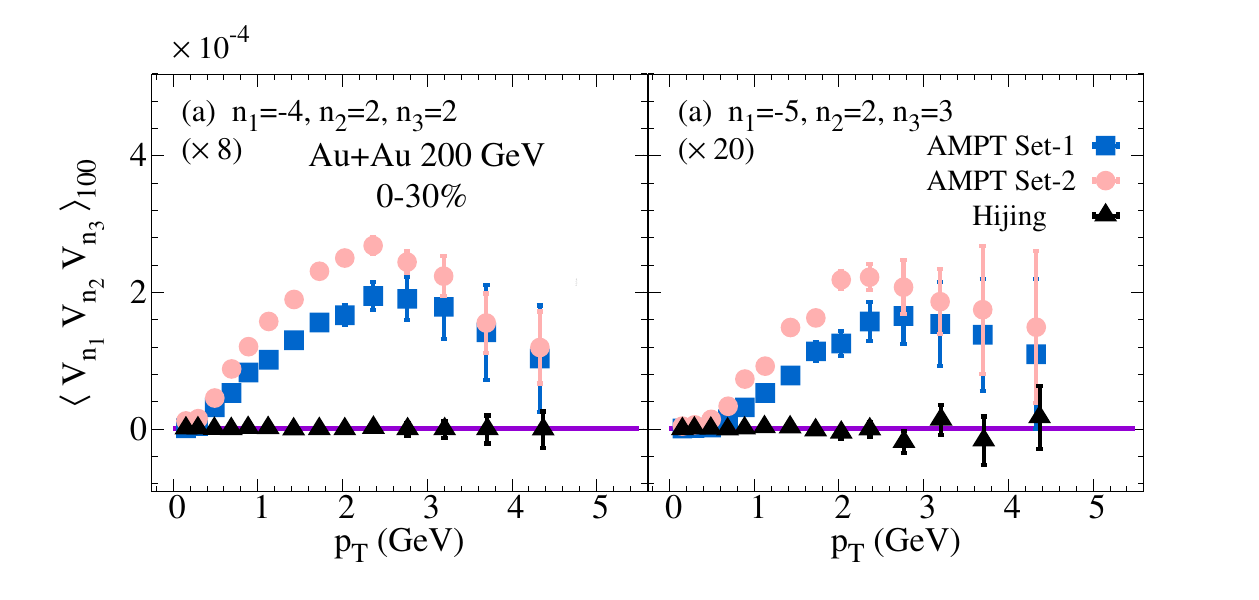}
\vskip -0.4cm
\caption{
Same as in Fig.~\ref{fig:1-1pasc} but for asymmetric correlations $\langle V_{n_1} V_{n_2} V_{n_3}\rangle_{100}$.
\label{fig:2-1pasc}}
\vskip -0.1cm
\end{figure}

\begin{figure}[!h] 
\includegraphics[width=0.90 \linewidth, angle=-0,keepaspectratio=true,clip=true]{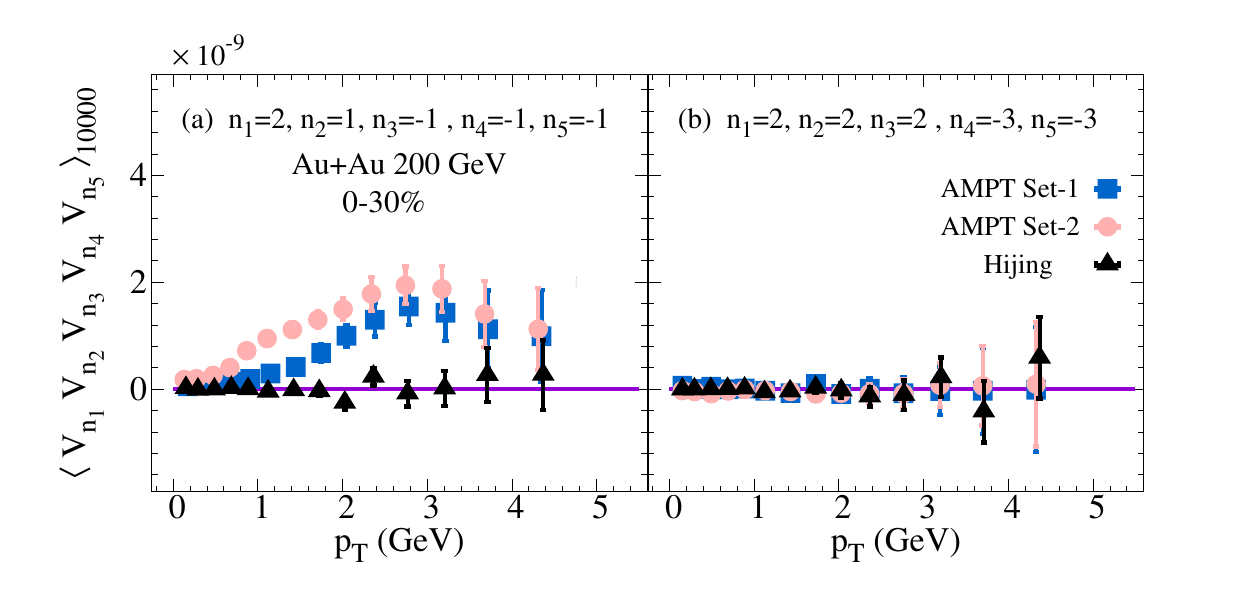}
\vskip -0.4cm
\caption{
Same as in Fig.~\ref{fig:1-1pasc} but for asymmetric correlations $\langle V_{n_1} V_{n_2} V_{n_3} V_{n_4}  V_{n_5}\rangle_{10000}$ using the traditional cumulant method.
\label{fig:3-1pasc}}
\vskip -0.1cm
\end{figure}

\begin{figure}[!h] 
\includegraphics[width=0.90 \linewidth, angle=-0,keepaspectratio=true,clip=true]{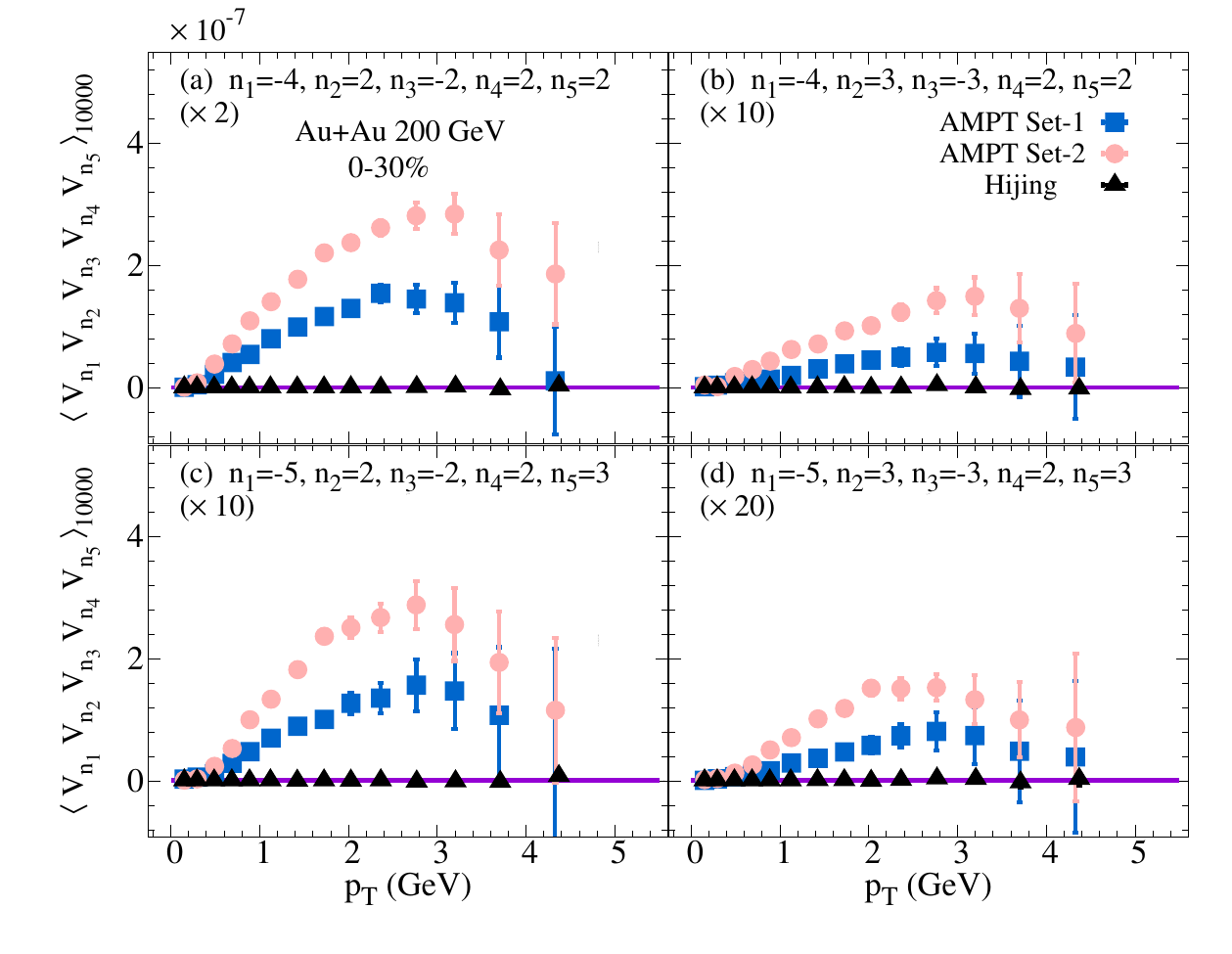}
\vskip -0.4cm
\caption{
Same as in Fig.~\ref{fig:1-1pasc} but for asymmetric correlations $\langle V_{n_1} V_{n_2} V_{n_3} V_{n_4}  V_{n_5}\rangle_{10000}$ using the traditional cumulant method.
\label{fig:4-1pasc}}
\vskip -0.1cm
\end{figure}
%--------------------------------------------------------------------
%--------------------------------------------------------------------
The Figs.~\ref{fig:1-1pasc}--~\ref{fig:4-1pasc} depict the three- and five-particle asymmetric correlations as functions of transverse momentum ($p_{T}$) for Au+Au collisions at 200 GeV, simulated using the HIJING and AMPT models. These correlations are anticipated to capture the flow angle correlations arising from the initial state $\psi_{n}$ correlations. The presented calculations reveal a characteristic dependence on $p_{T}$ alongside a correlation strength reduction with the $\sigma_{pp}$ value. This behavior reflects the sensitivity of the three- and five-particle asymmetric correlations to final state effects as modeled by AMPT. 
The calculations presented in Fig.~\ref{fig:1-1pasc} (a) highlight that only $\langle V_{-2} V_{1} V_{1} \rangle_{100}$ exhibits a noticeable impact due to the non-flow effects within the HIJING model. Such an effect at $p_T$ $>$ 2.0~GeV is expected to be caused by the residual non-flow effect on $v_{2}(p_T)$, as one can observe in Fig.~\ref{fig:1-1psc}  (b). Additionally, $\langle V_{-2} V_{1} V_{1} \rangle_{100}$ will have a constant ($p_T$-independent) contribution from the GMC (see Eq.\ref{corrv1})~\cite{STAR:2018gji, STAR:2019zaf}.
%-------------------------------------------------------------------

%-------------------------------------------------------------------
Moreover, the asymmetric correlations $\langle V_{2} V_{2} V_{2} V_{-3} V_{-3} \rangle_{10000}$ in panel (b) of Fig.~\ref{fig:3-1pasc} consistently display values around zero as a function of $p_{T}$, indicating the weak nature of correlations between $\psi_{2}$ and $\psi_3$. The HIJING model calculations in Figs.~\ref{fig:1-1pasc}--~\ref{fig:4-1pasc}, barring $\langle V_{-2} V_{1} V_{1} \rangle_{100}$, yield values consistent with zero within the uncertainties, suggesting minimal (if any) non-flow contribution to the presented three- and five-particle asymmetric correlations.
%-------------------------------------------------------------------

%-------------------------------------------------------------------
Similarly to the symmetric correlations, the asymmetric correlations computed using one POI displayed in Figs.~\ref{fig:1-1pasc}--~\ref{fig:4-1pasc} will incorporate a nonvanishing event-plane angle fluctuation effect. This effect plays a crucial role in complicating the interpolation of experimental measurements of such observables. Therefore, it is imperative to account for this effect. Moreover, in future analyses, a more detailed discussion of the event-plane angle fluctuation effect will be considered when discussing the normalization of the asymmetric correlations.

%\clearpage
%--------------------------------------------------------------------
%--------------------------------------------------------------------
%--------------------------------------------------------------------
\subsection{Symmetric correlations with two POIs}
%--------------------------------------------------------------------
\begin{figure}[!h] 
\includegraphics[width=0.90 \linewidth, angle=-0,keepaspectratio=true,clip=true]{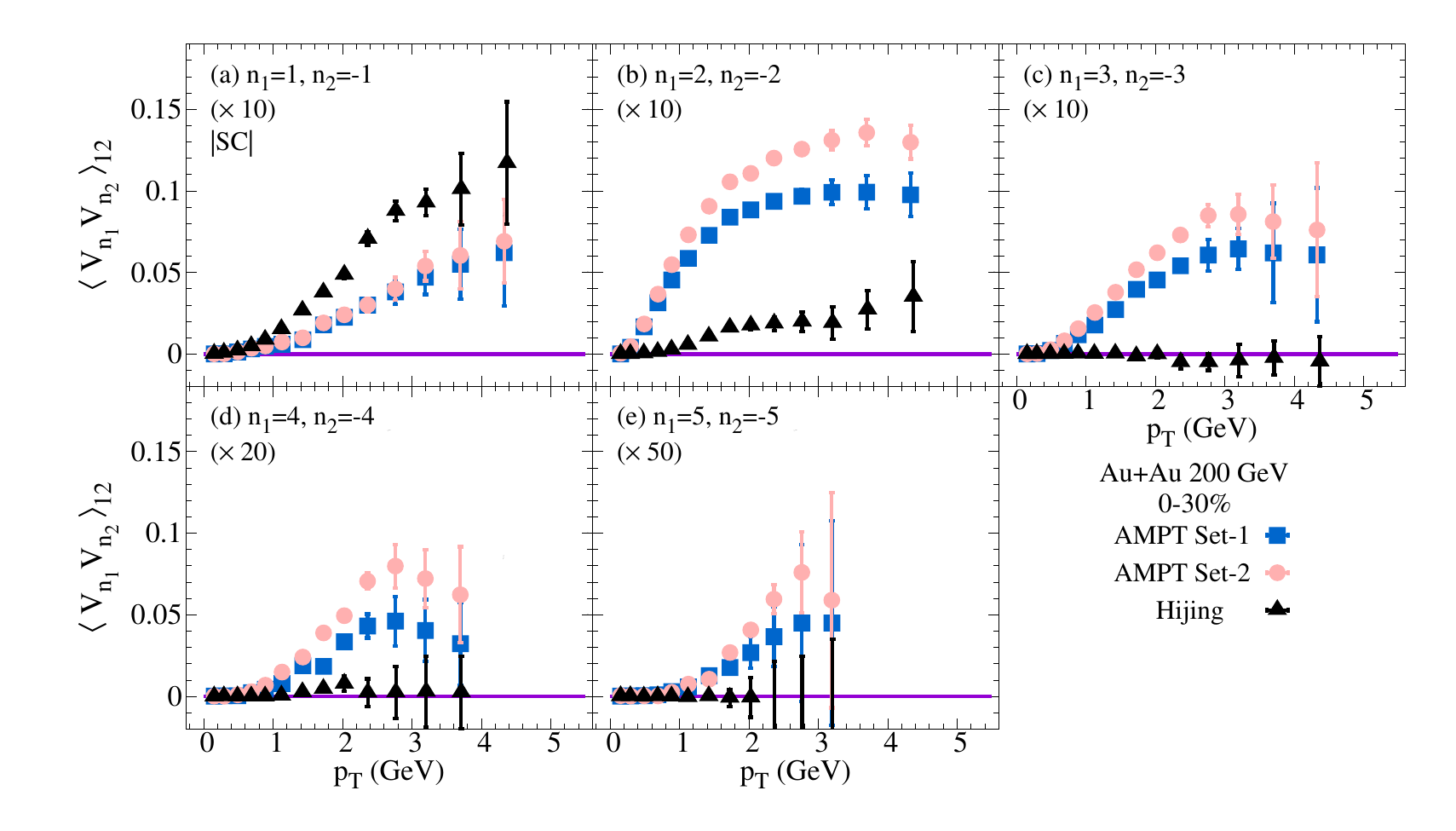}
\vskip -0.4cm
\caption{
The $p_{T}$ dependence of the two-particle flow harmonics with two POS, $v^{2}_{n}(p_{T}) = \langle V_{n_1} V_{n_2}\rangle_{12}$, using the two-subevents method for Au+Au at $\sqrt{\textit{s}_{NN}}~=$ 200~GeV from the AMPT and HIJING models.
\label{fig:1-2psc}}
\vskip -0.1cm
\end{figure}
%--------------------------------------------------------------------
\begin{figure}[!h] 
\includegraphics[width=0.90 \linewidth, angle=-0,keepaspectratio=true,clip=true]{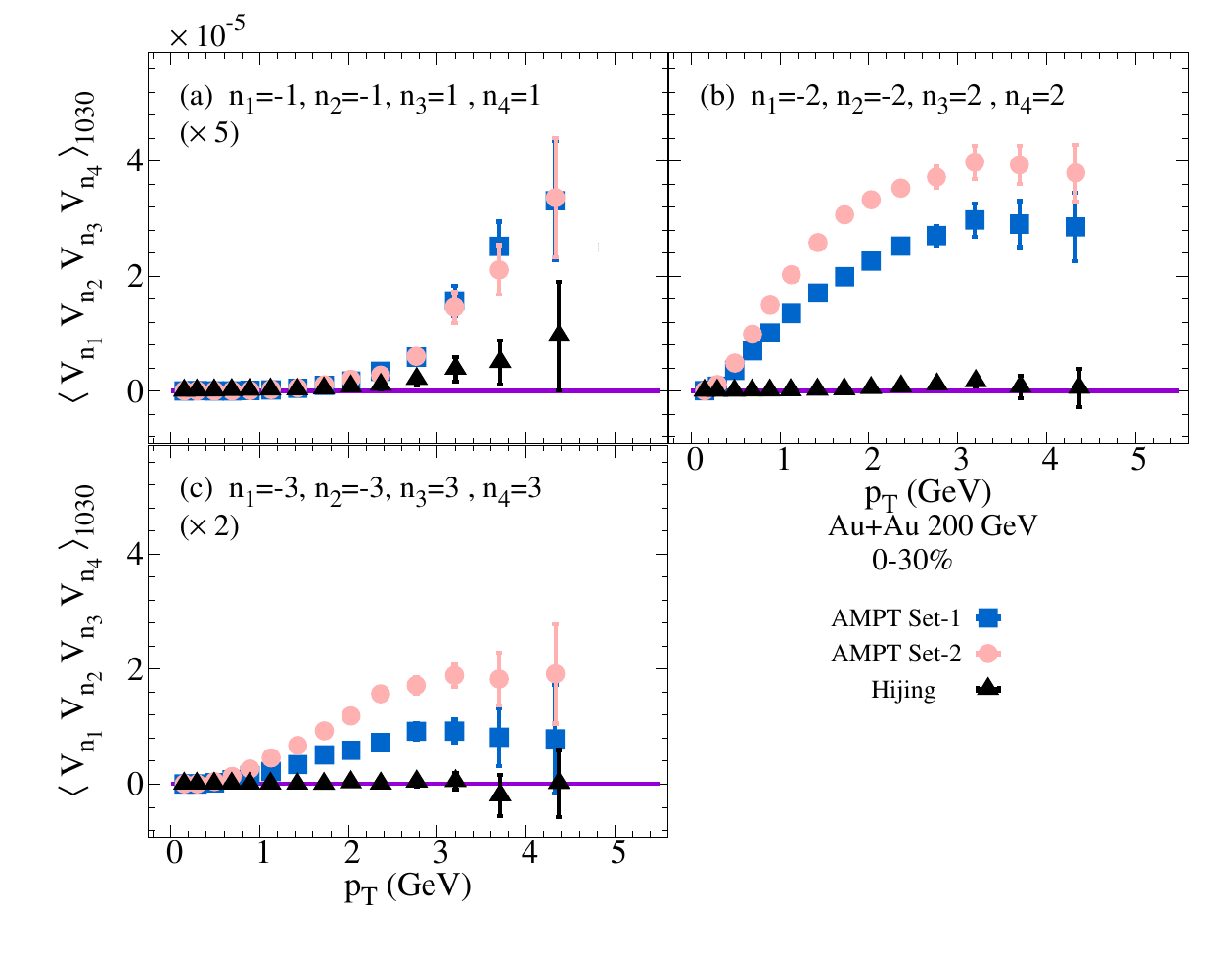}
\vskip -0.4cm
\caption{
Same as in Fig.~\ref{fig:1-2psc} but for symmetric same harmonic correlations $\langle V_{n_1} V_{n_2} V_{n_3} V_{n_4}\rangle_{1030}$.
\label{fig:2-2psc}}
\vskip -0.1cm
\end{figure}
%--------------------------------------------------------------------
\begin{figure}[!h] 
\includegraphics[width=0.90 \linewidth, angle=-0,keepaspectratio=true,clip=true]{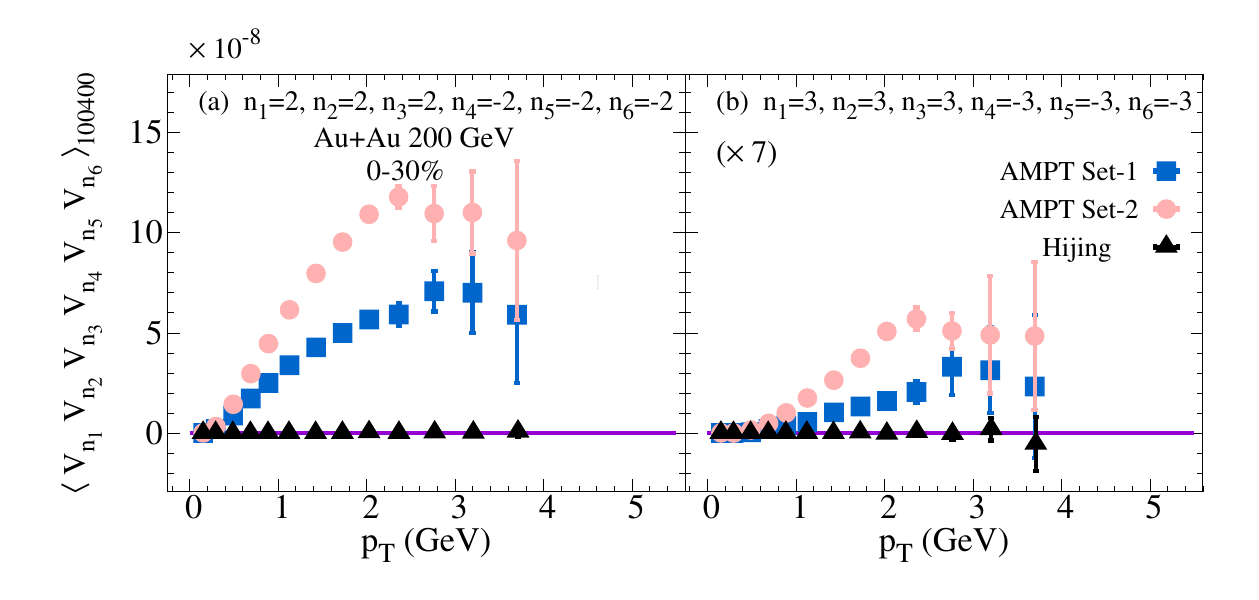}
\vskip -0.4cm
\caption{
Same as in Fig.~\ref{fig:1-2psc} but for symmetric same harmonic correlations $\langle V_{2} V_{2} V_{2} V_{-2} V_{-2} V_{-2}\rangle_{100400}$ and $\langle V_{3} V_{3} V_{3} V_{-3} V_{-3} V_{-3}\rangle_{100400}$ using the traditional cumulant method.
\label{fig:3-2psc}}
\vskip -0.1cm
\end{figure}
%--------------------------------------------------------------------
Figures~\ref{fig:1-2psc}--\ref{fig:3-2psc} illustrate the two-, four-, and six-particle same harmonic symmetric correlations with two particles off interests (POIs) as functions of $p_{T}$ and $\sigma_{pp}$ for Au+Au collisions at 200 GeV. The particle correlations depicted in these figures, with two POIs, are expected to contain no event-plane angle fluctuation.  Furthermore, the absence of event-plane angle fluctuations renders these observables particularly well-suited for discerning the $p_T$ nature of the anisotropic flow fluctuations through ratios between four- and two-particle correlations\cite{STAR:2022gki}. In addition, the presented calculations elucidate the dependence of the two-, four-, and six-particle symmetric correlations on $p_{T}$ and underscore their sensitivity to variations of the $\sigma_{pp}$.
%--------------------------------------------------------------------

%--------------------------------------------------------------------
The two-particle SC $\langle V_{1} V_{-1}\rangle_{12}$ with two POIs shown in Fig.~\ref{fig:1-2psc} (a) indicates similar behavior but different magnitudes between the HIJING and AMPT models. This observation can be understood in light of Eq.~\ref{corrv1}, which contains two terms: (i) $v_{1}(p_{T}^{a})v_{1}(p_{T}^{b})$, which stems from the interplay of initial-state fluctuations and the hydrodynamic expansion~\cite{Luzum:2010fb, Teaney:2010vd, Gardim:2011qn, Magdy:2018whk, STAR:2018gji}. (ii) $K p_{T}^{a} p_{T}^{b}$, reflecting the GMC effect on $\langle V_{1} V_{-1}\rangle$. The first term will have no contribution in the case of the HIJING model, and $\langle V_{1} V_{-1}\rangle$ will be dominated by the GMC term. In contrast, the AMPT model will have contributions from both terms as given in Eq.~\ref{corrv1}. Consequently, one expects the HIJING model to have larger values of $\langle V_{1} V_{-1}\rangle_{12}$ than the AMPT model spatially in the case of having two POIs. It is also important to emphasize that the GMC effect is expected to have a more complicated contribution to higher-order correlations with two POIs.
%--------------------------------------------------------------------

%--------------------------------------------------------------------
\begin{figure}[!h] 
\includegraphics[width=0.90 \linewidth, angle=-0,keepaspectratio=true,clip=true]{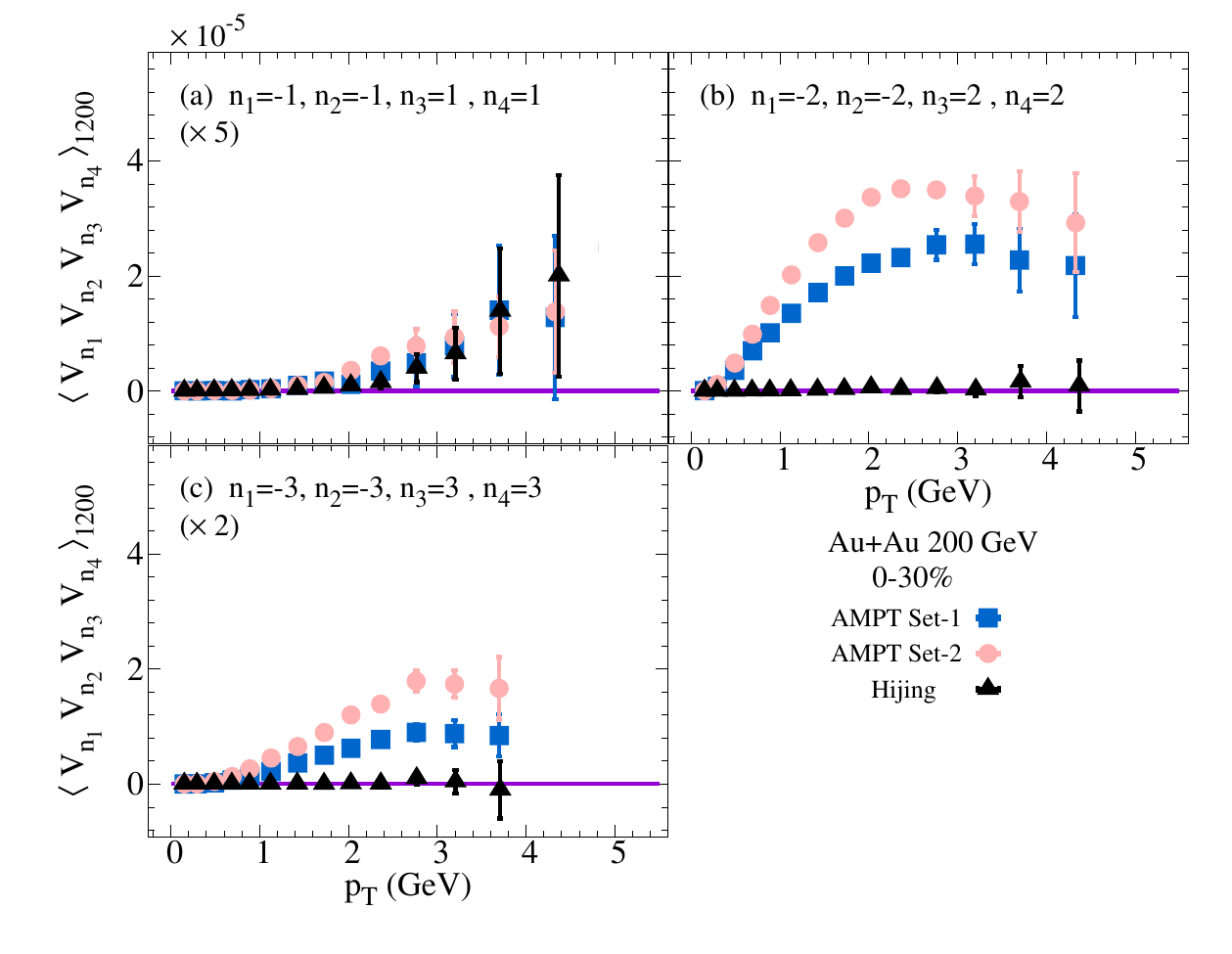}
\vskip -0.4cm
\caption{
Same as in Fig.~\ref{fig:1-2psc} but for symmetric same harmonic correlations $\langle V_{n_1} V_{n_2} V_{n_3} V_{n_4}\rangle_{1200}$.
\label{fig:4-2psc}}
\vskip -0.1cm
\end{figure}
%--------------------------------------------------------------------
\begin{figure}[!h] 
\includegraphics[width=0.90 \linewidth, angle=-0,keepaspectratio=true,clip=true]{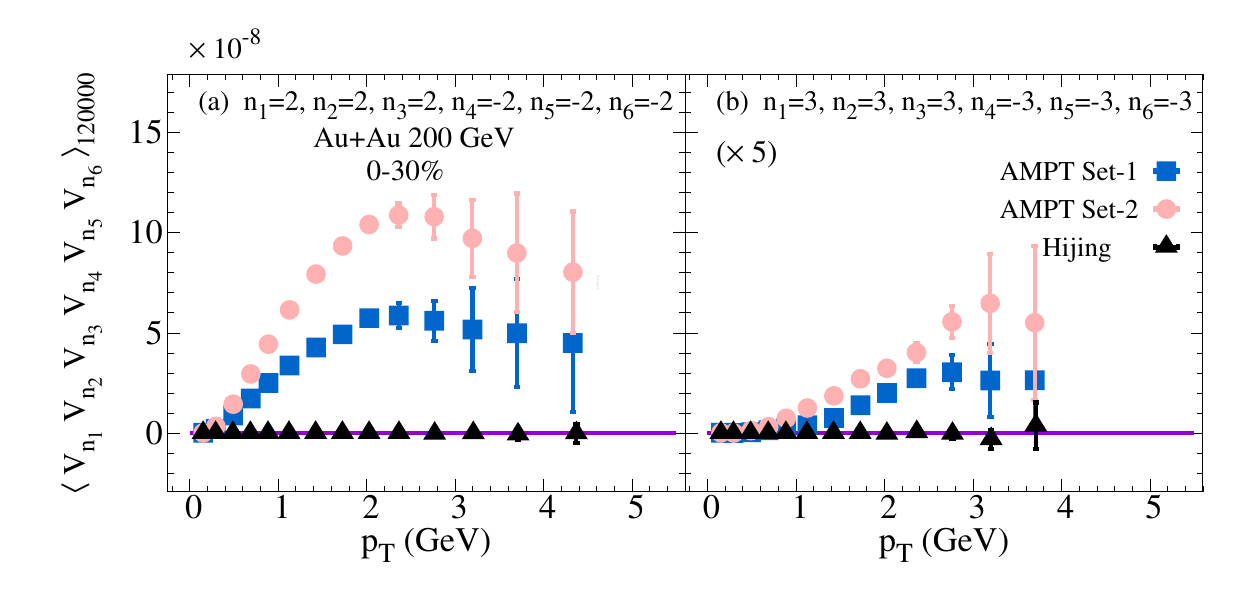}
\vskip -0.4cm
\caption{
Same as in Fig.~\ref{fig:1-2psc} but for symmetric same harmonic correlations $\langle V_{2} V_{2} V_{2} V_{-2} V_{-2} V_{-2}\rangle_{120000}$ and $\langle V_{3} V_{3} V_{3} V_{-3} V_{-3} V_{-3}\rangle_{120000}$ using the traditional cumulant method.
\label{fig:5-2psc}}
\vskip -0.1cm
\end{figure}
%--------------------------------------------------------------------
In contrast to the correlations depicted in Figs.~\ref{fig:2-2psc} and \ref{fig:3-2psc}, the four- and six-particle same harmonic symmetric correlations with two POIs presented in Figs.~\ref{fig:4-2psc} and \ref{fig:5-2psc} are anticipated to encompass a contribution from the event-plane angle fluctuation effect. Consequently, comparing ratios between these two cases can offer insights into the $p_{T}$ dependence of the event-plane angle fluctuation effect~\cite{ALICE:2022dtx}. This effect on the four- and six-particle correlations will be subject to further discussion in forthcoming investigations. Additionally, the calculations presented in Figs.~\ref{fig:2-2psc} and \ref{fig:3-2psc} exhibit a decreasing trend with $\sigma_{pp}$, emphasizing their sensitivity to the final state effects elucidated by the AMPT model.
%--------------------------------------------------------------------
%--------------------------------------------------------------------
\begin{figure}[!h] 
\includegraphics[width=0.90 \linewidth, angle=-0,keepaspectratio=true,clip=true]{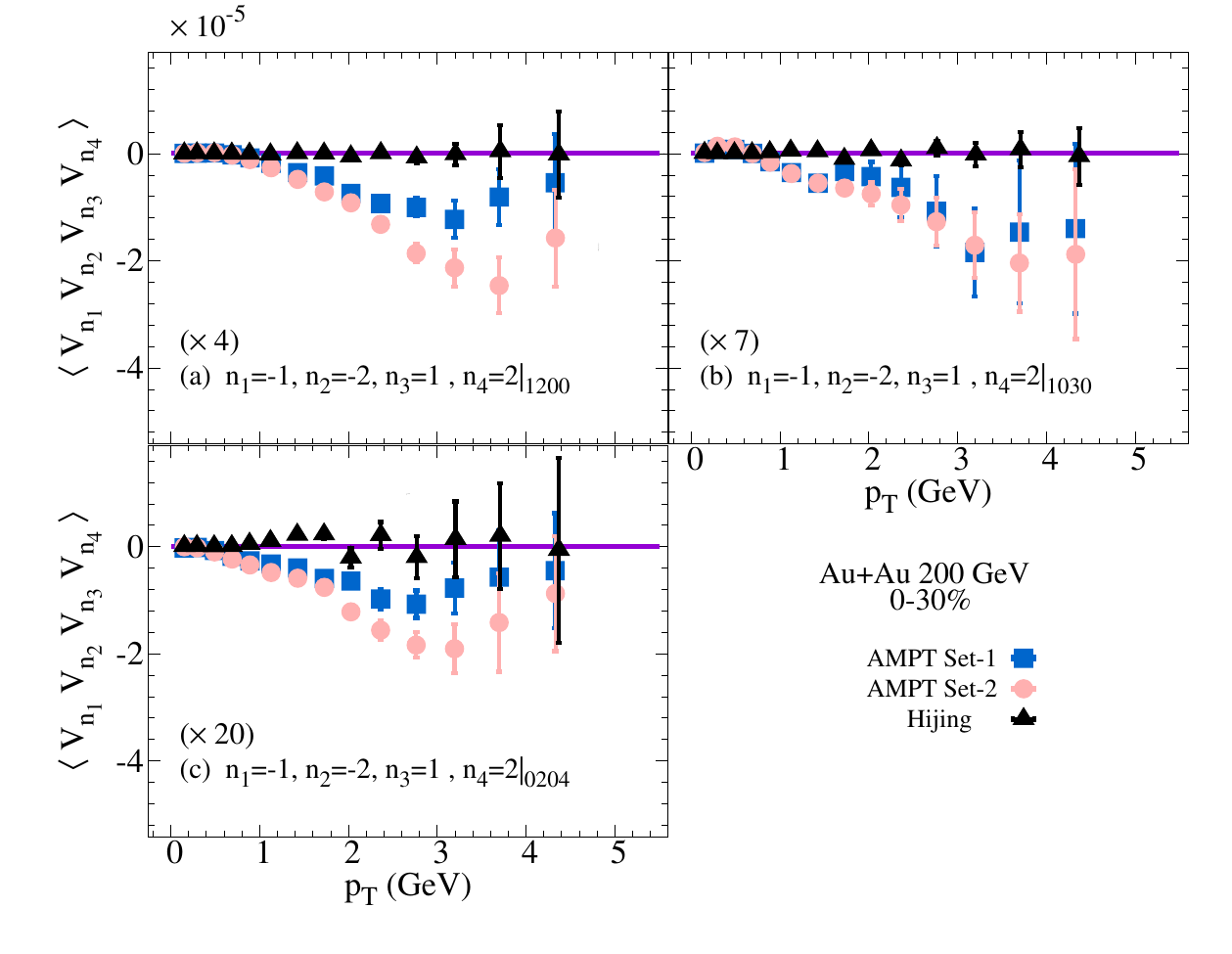}
\vskip -0.4cm
\caption{
Same as in Fig.~\ref{fig:1-2psc} but for symmetric mixed harmonic correlations $\langle V_{1} V_{2} V_{-1} V_{-2} \rangle_{1200}$, $\langle V_{1} V_{2} V_{-1} V_{-2}\rangle_{1030}$, and $\langle V_{1} V_{2} V_{-1} V_{-2}\rangle_{0204}$.
\label{fig:6-2psc}}
\vskip -0.1cm
\end{figure}
%--------------------------------------------------------------------
\begin{figure}[!h] 
\includegraphics[width=0.90 \linewidth, angle=-0,keepaspectratio=true,clip=true]{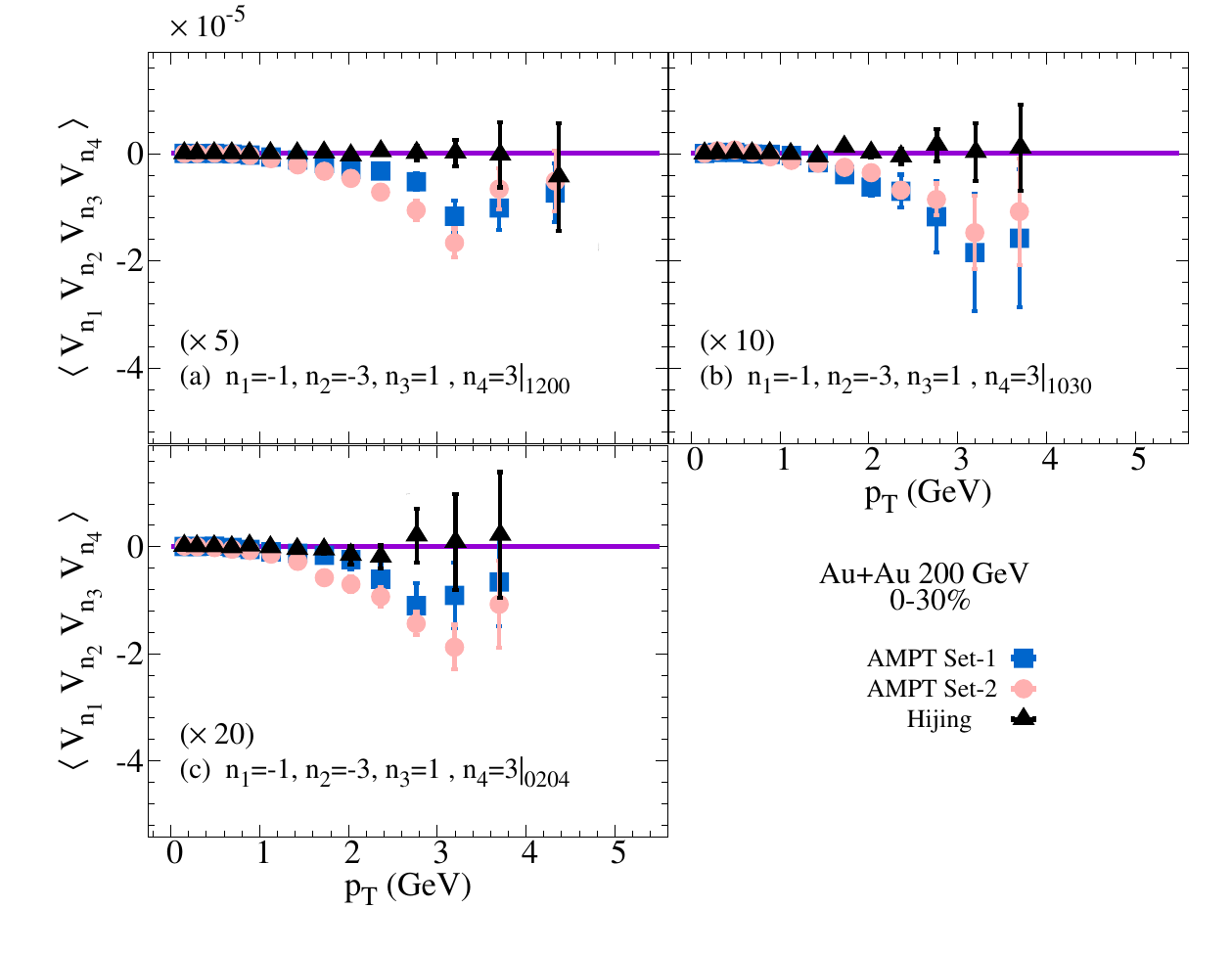}
\vskip -0.4cm
\caption{
Same as in Fig.~\ref{fig:1-2psc} but for symmetric mixed harmonic correlations $\langle V_{1} V_{3} V_{-1} V_{-3}\rangle_{1200}$, $\langle V_{1} V_{3} V_{-1} V_{-3}\rangle_{1030}$, and $\langle V_{1} V_{3} V_{-1} V_{-3}\rangle_{0204}$.
\label{fig:7-2psc}}
\vskip -0.1cm
\end{figure}
%--------------------------------------------------------------------
\begin{figure}[!h] 
\includegraphics[width=0.90 \linewidth, angle=-0,keepaspectratio=true,clip=true]{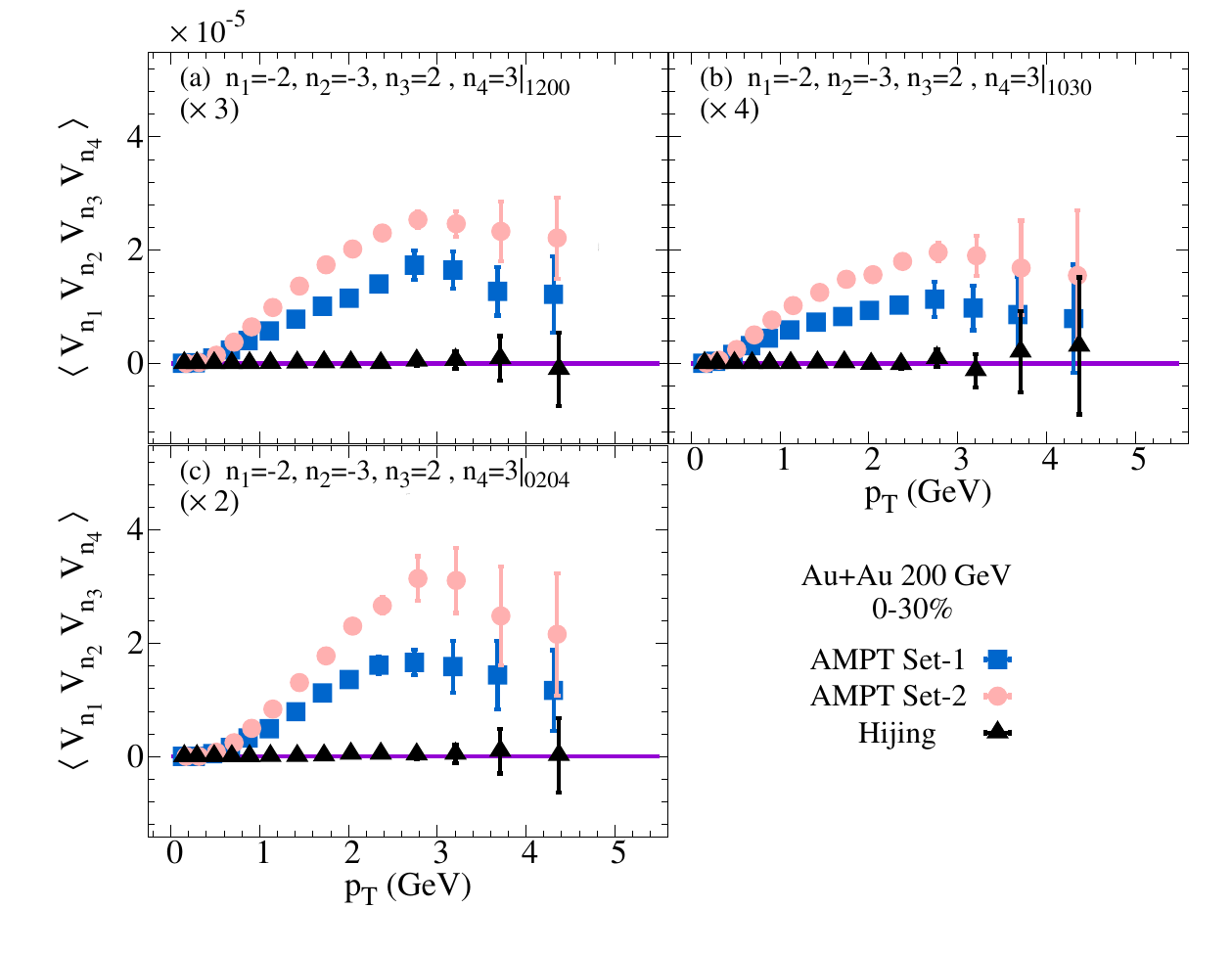}
\vskip -0.4cm
\caption{
Same as in Fig.~\ref{fig:1-2psc} but for symmetric mixed harmonic correlations $\langle V_{2} V_{3} V_{-2} V_{-3}\rangle_{1200}$, $\langle V_{2} V_{3} V_{-2} V_{-3}\rangle_{1030}$, and $\langle V_{2} V_{3} V_{-2} V_{-3}\rangle_{0204}$.
\label{fig:8-2psc}}
\vskip -0.1cm
\end{figure}
%--------------------------------------------------------------------
\begin{figure}[!h] 
\includegraphics[width=0.90 \linewidth, angle=-0,keepaspectratio=true,clip=true]{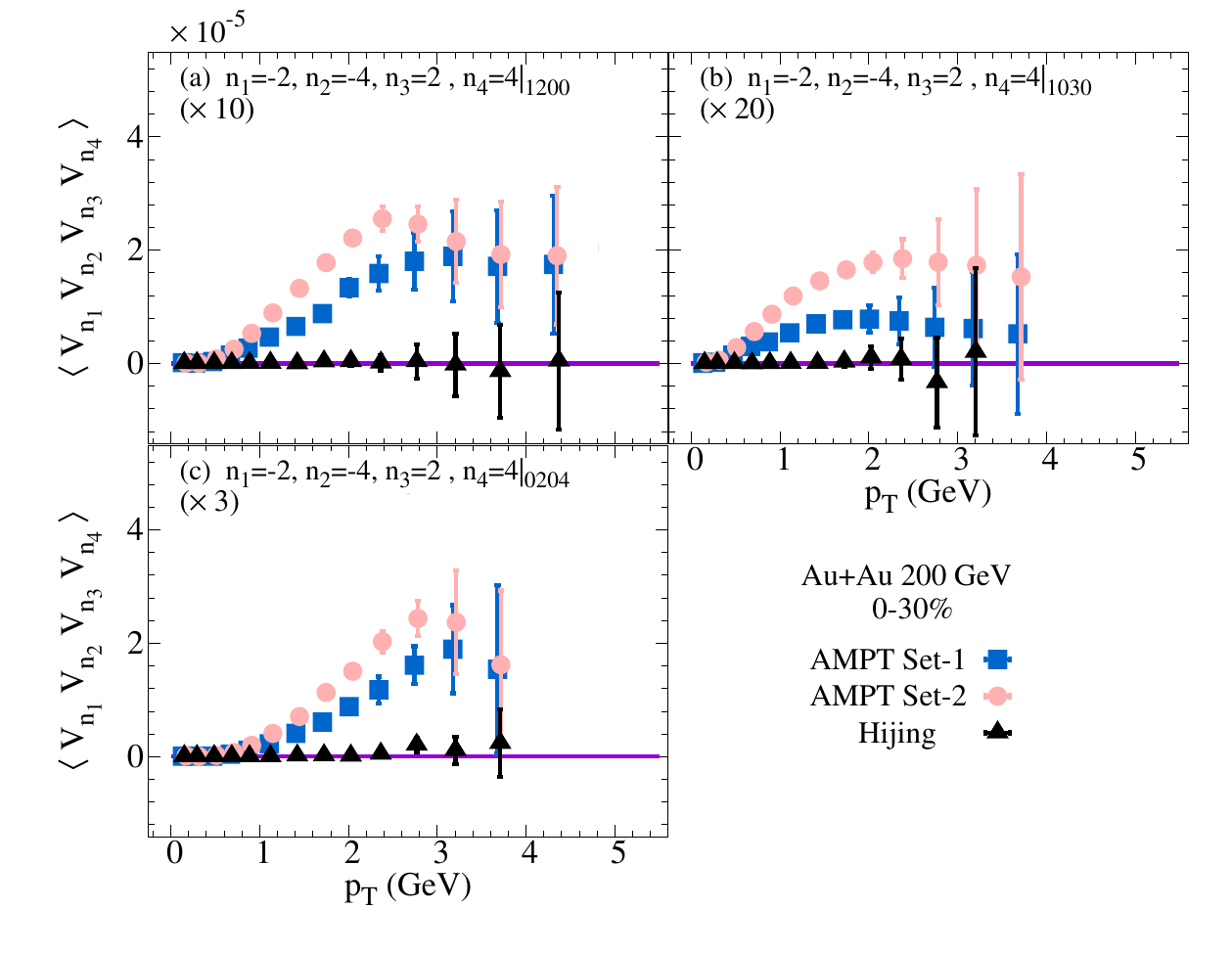}
\vskip -0.4cm
\caption{
Same as in Fig.~\ref{fig:1-2psc} but for symmetric mixed harmonic correlations $\langle V_{2} V_{4} V_{-2} V_{-4}\rangle_{1200}$, $\langle V_{2} V_{4} V_{-2} V_{-4}\rangle_{1030}$, and $\langle V_{2} V_{4} V_{-2} V_{-4}\rangle_{0204}$.
\label{fig:9-2psc}}
\vskip -0.1cm
\end{figure}
%--------------------------------------------------------------------
\begin{figure}[!h] 
\includegraphics[width=0.90 \linewidth, angle=-0,keepaspectratio=true,clip=true]{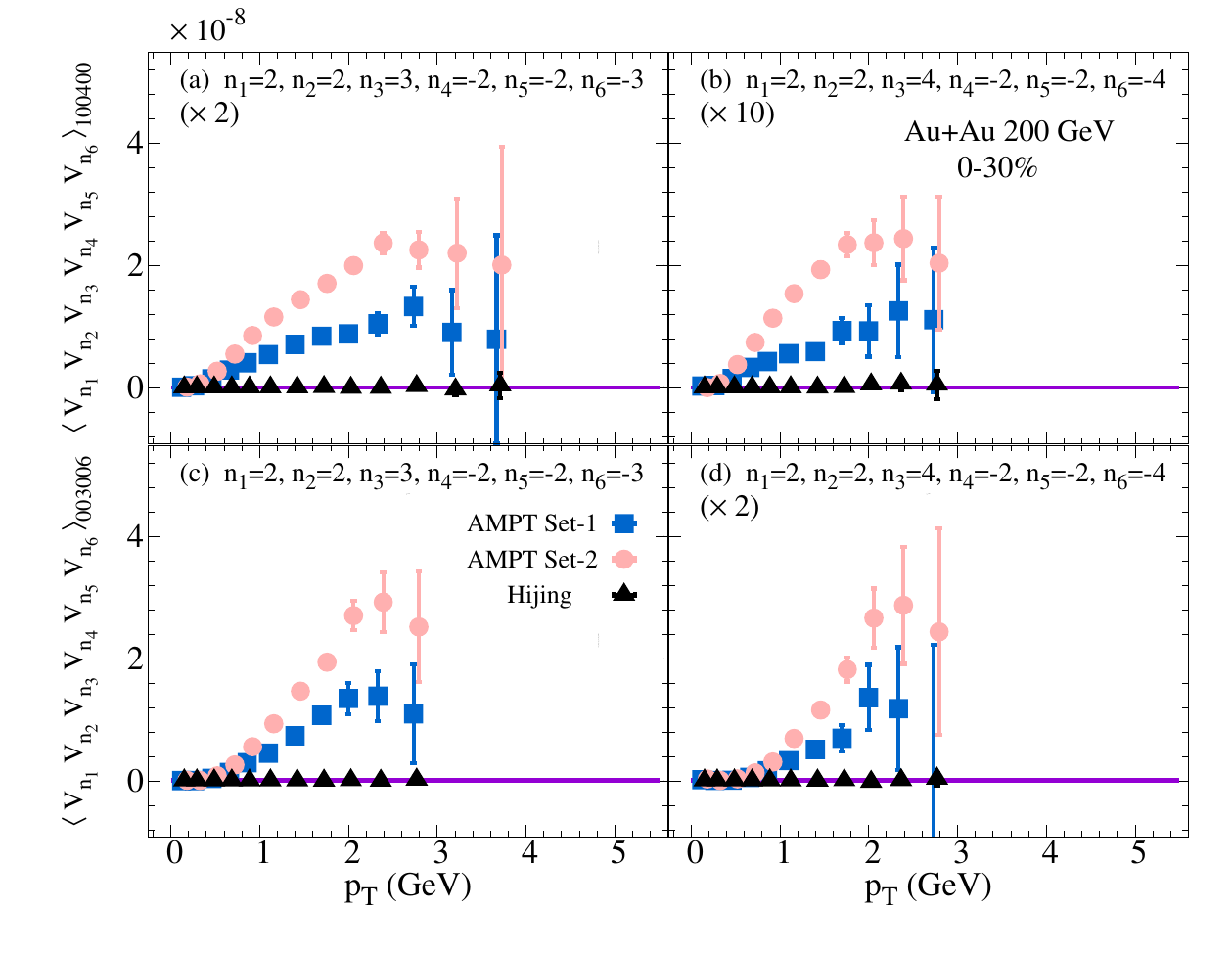}
\vskip -0.4cm
\caption{
Same as in Fig.~\ref{fig:1-2psc} but for symmetric mixed harmonic correlations $\langle V_{2} V_{2} V_{3} V_{-2} V_{-2} V_{-3}\rangle_{100400}$, $\langle V_{2} V_{2} V_{4} V_{-2} V_{-2} V_{-4}\rangle_{100400}$,  $\langle V_{2} V_{2} V_{3} V_{-2} V_{-2} V_{-3}\rangle_{003006}$, and $\langle V_{2} V_{2} V_{4} V_{-2} V_{-2} V_{-4}\rangle_{003006}$ using the traditional cumulant method.
\label{fig:10-2psc}}
\vskip -0.1cm
\end{figure}
%--------------------------------------------------------------------
\begin{figure}[!h] 
\includegraphics[width=0.90 \linewidth, angle=-0,keepaspectratio=true,clip=true]{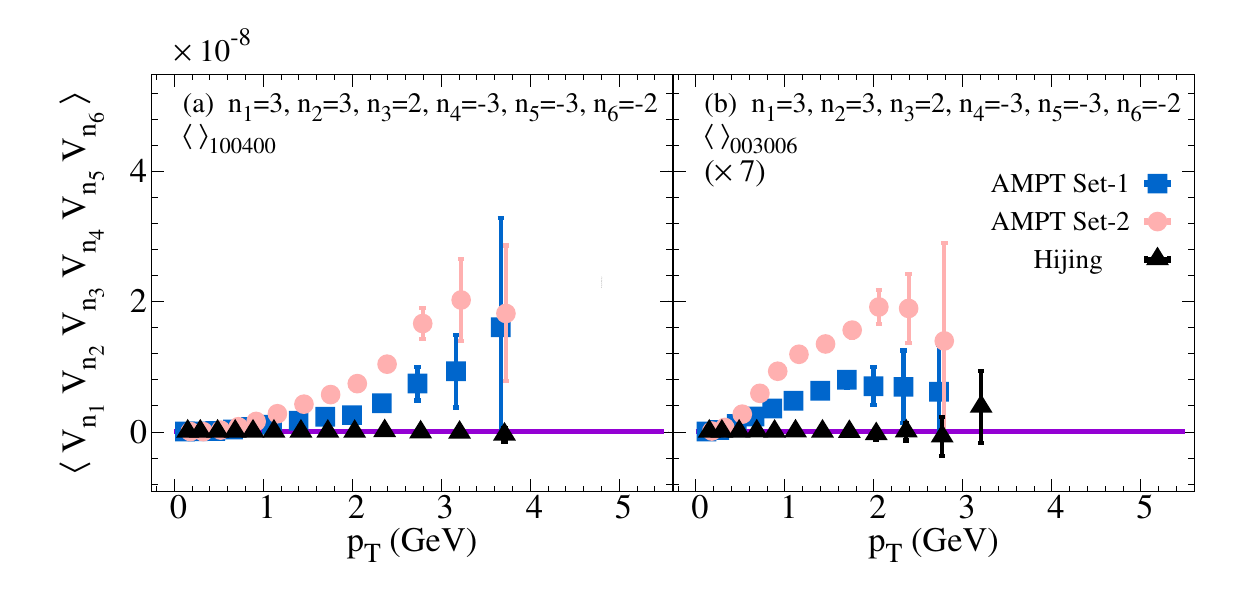}
\vskip -0.4cm
\caption{
Same as in Fig.~\ref{fig:1-2psc} but for symmetric mixed harmonic correlations $\langle V_{3} V_{3} V_{2} V_{-3} V_{-3} V_{-2}\rangle_{100400}$  and $\langle V_{3} V_{3} V_{2} V_{-3} V_{-3} V_{-2}\rangle_{003006}$ using the traditional cumulant method.
\label{fig:11-2psc}}
\vskip -0.1cm
\end{figure}
%--------------------------------------------------------------------
\begin{figure}[!h] 
\includegraphics[width=0.90 \linewidth, angle=-0,keepaspectratio=true,clip=true]{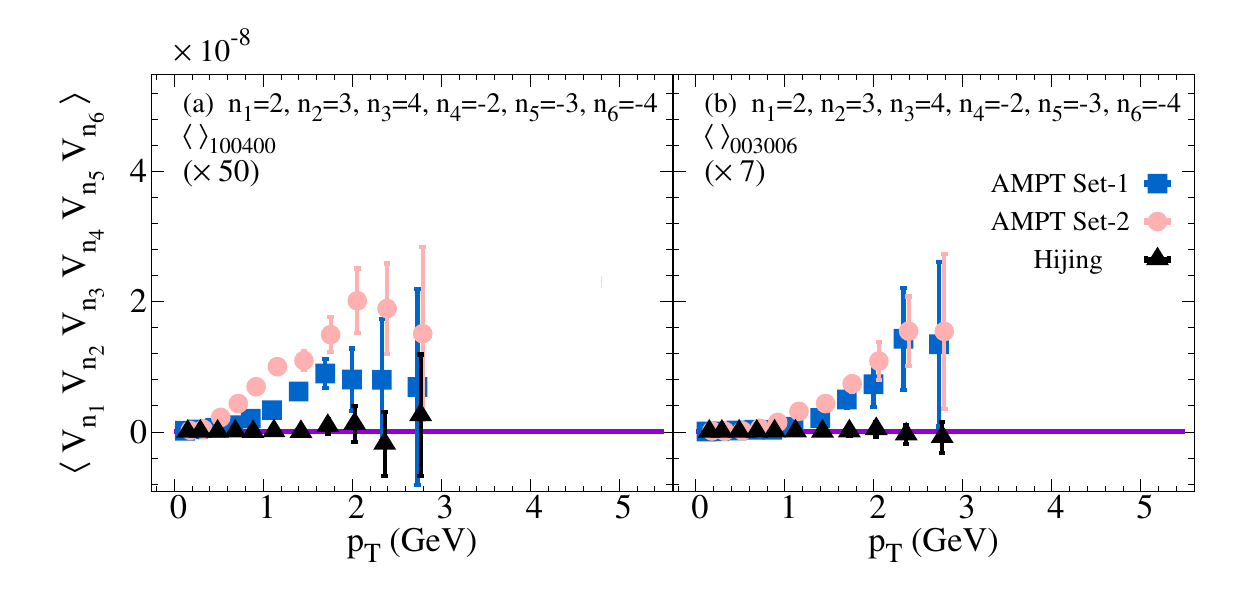}
\vskip -0.4cm
\caption{
Same as in Fig.~\ref{fig:1-2psc} but for symmetric mixed harmonic correlations $\langle V_{2} V_{3} V_{4} V_{-2} V_{-3} V_{-4}\rangle_{100400}$  and $\langle V_{2} V_{3} V_{4} V_{-2} V_{-3} V_{-4}\rangle_{003006}$ using the traditional cumulant method.
\label{fig:12-2psc}}
\vskip -0.1cm
\end{figure}

%--------------------------------------------------------------------
The HIJING and AMPT model calculations of the mixed harmonics four- and six-particle symmetric correlations with two POIs are illustrated in Figs.~\ref{fig:6-2psc}--\ref{fig:12-2psc}, showcasing their dependence on transverse momentum ($p_{T}$) for Au+Au collisions at 200 GeV. As previously mentioned, these observables not only reflect the correlations between different flow harmonics but also the $p_{T}$ dependence of the event-plane angle fluctuations. The mixed harmonics four-particle correlations presented in panel (a) of Figs.~\ref{fig:6-2psc}--\ref{fig:9-2psc} encompass a contribution from the event-plane angle fluctuations ($\cos(n\psi_{n}(p_{T}) + m\psi_{m}(p_{T}) - n\psi_{n} - m\psi_{m})$). In contrast, the remaining four- and six-particle correlations (with two particles of interest) are expected to be unaffected by the event-plane angle fluctuation effect. Thus, comparing ratios between these two cases can provide insights into the $p_{T}$ dependence of the event-plane angle fluctuation effect~\cite{ALICE:2022dtx}. Also, the calculations depicted in Figs.~\ref{fig:6-2psc}--\ref{fig:12-2psc} display a decreasing trend with $\sigma_{pp}$, highlighting their sensitivity to the final state effects given by the AMPT model.
%--------------------------------------------------------------------

%--------------------------------------------------------------------
The HIJING model calculations shown for the two-, four-, and six-particle $n > 1$ symmetric correlations in Figs.~\ref{fig:1-2psc}--\ref{fig:12-2psc} reveal small values that closely align with zero within the associated uncertainties. These findings suggest that the current two-, four-, and six-particle symmetric correlations with two POIs exhibit minimal non-flow contributions.

%\clearpage
%--------------------------------------------------------------------
%--------------------------------------------------------------------
%--------------------------------------------------------------------
\subsection{Asymmetric correlations with two POIs}
%--------------------------------------------------------------------
\begin{figure}[ht] 
\includegraphics[width=0.9 \linewidth, angle=-0,keepaspectratio=true,clip=true]{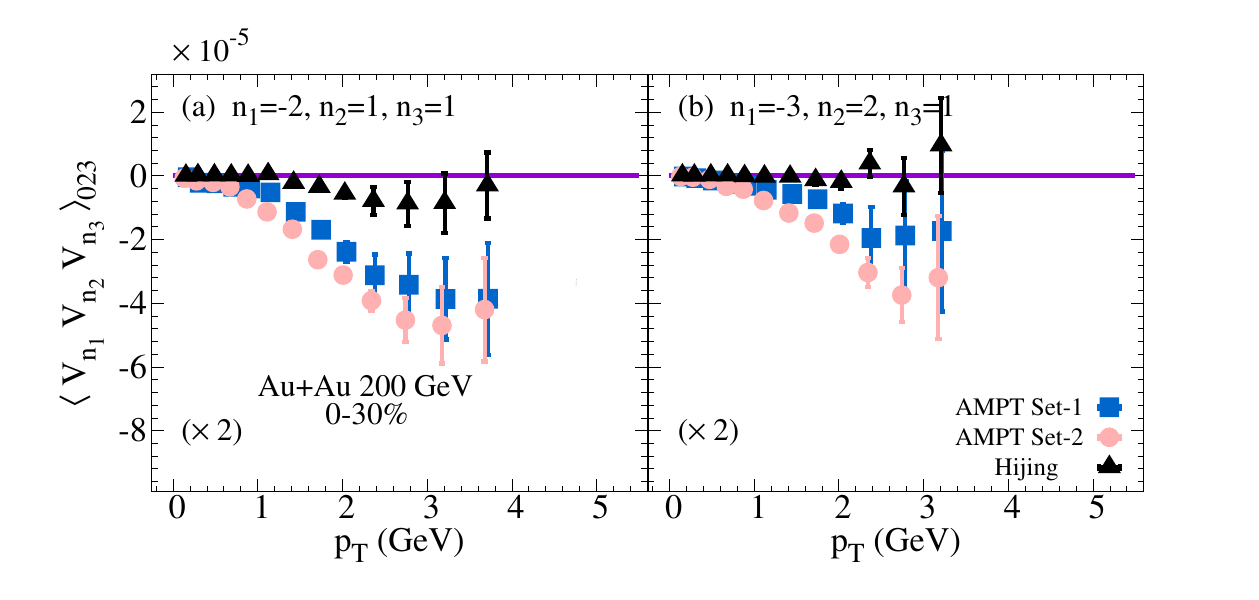}
\vskip -0.4cm
\caption{
The $p_{T}$ dependence of the three-particle asymmetric correlations $\langle V_{-2} V_{1} V_{1}\rangle_{023}$ and $\langle V_{-3} V_{1} V_{2}\rangle_{023}$ with two POSs using the two-subevents method for Au+Au at $\sqrt{\textit{s}_{NN}}~=$ 200~GeV from the AMPT and HIJING models.
\label{fig:1-2pasc}}
\vskip -0.1cm
\end{figure}
%--------------------------------------------------------------------
\begin{figure}[ht] 
\includegraphics[width=0.9 \linewidth, angle=-0,keepaspectratio=true,clip=true]{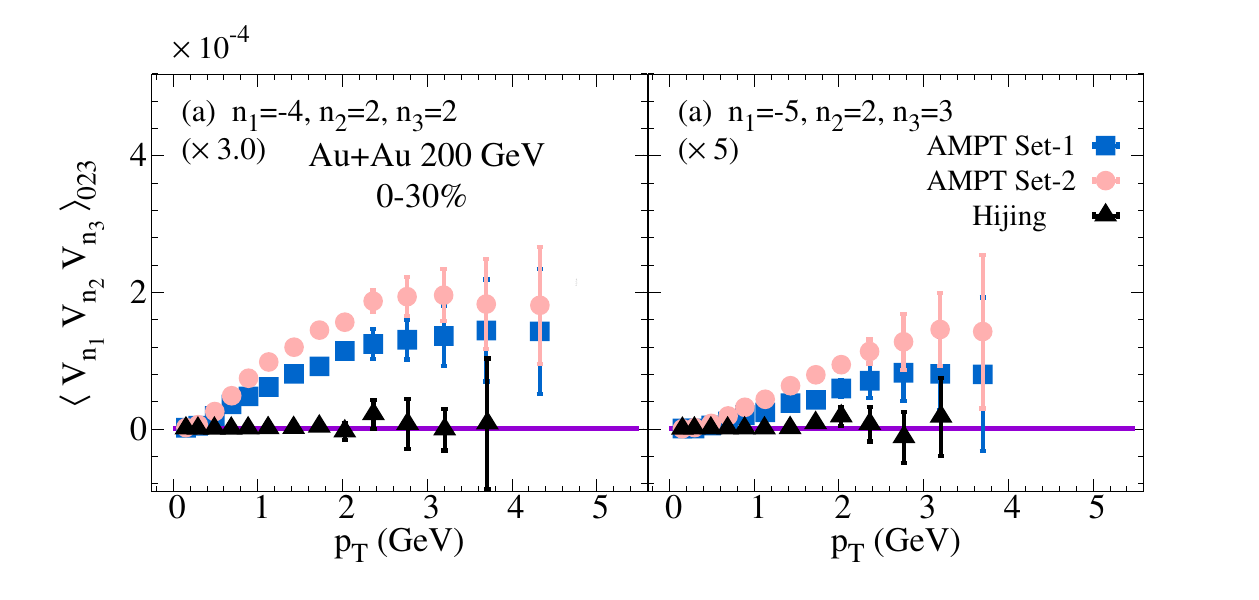}
\vskip -0.4cm
\caption{
Same as in Fig.~\ref{fig:1-2pasc} but for asymmetric correlations $\langle V_{-2} V_{2} V_{2}\rangle_{023}$ and $\langle V_{-5} V_{2} V_{3}\rangle_{023}$ .
\label{fig:2-2pasc}}
\vskip -0.1cm
\end{figure}
%--------------------------------------------------------------------
\begin{figure}[ht] 
\includegraphics[width=0.9 \linewidth, angle=-0,keepaspectratio=true,clip=true]{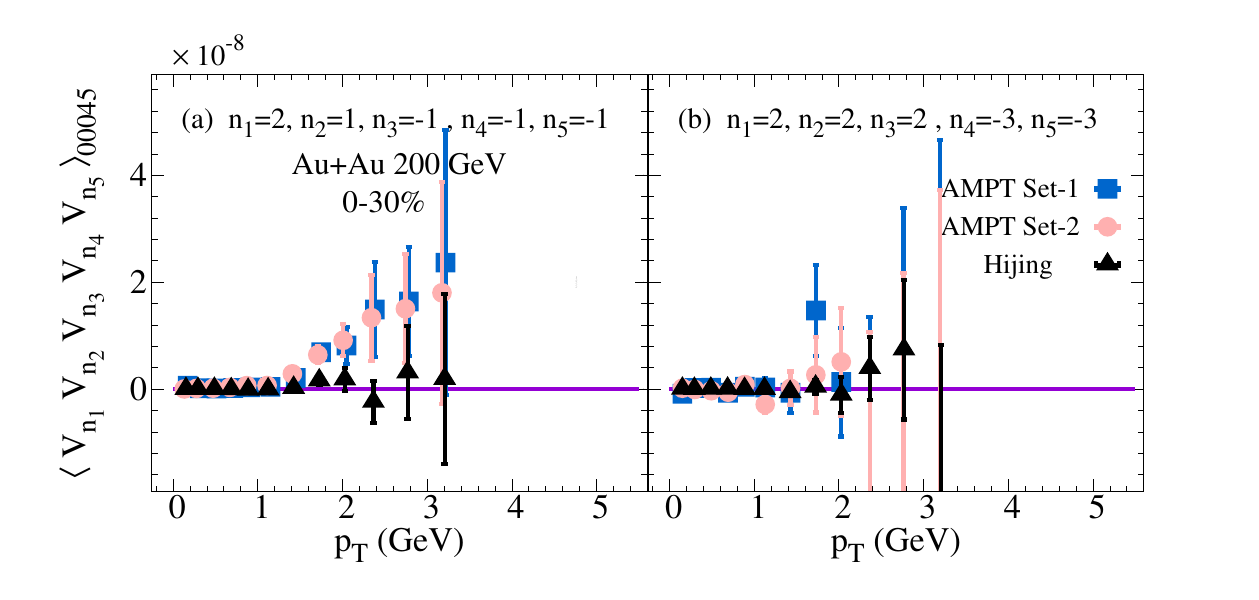}
\vskip -0.4cm
\caption{
Same as in Fig.~\ref{fig:1-2pasc} but for asymmetric correlations $\langle V_{-2} V_{1} V_{-1} V_{1} V_{1}\rangle_{00045}$ and $\langle V_{-2} V_{2} V_{2} V_{-3} V_{-3}\rangle_{00045}$.
\label{fig:3-2pasc}}
\vskip -0.1cm
\end{figure}
%--------------------------------------------------------------------
\begin{figure}[ht] 
\includegraphics[width=0.9\linewidth, angle=-0,keepaspectratio=true,clip=true]{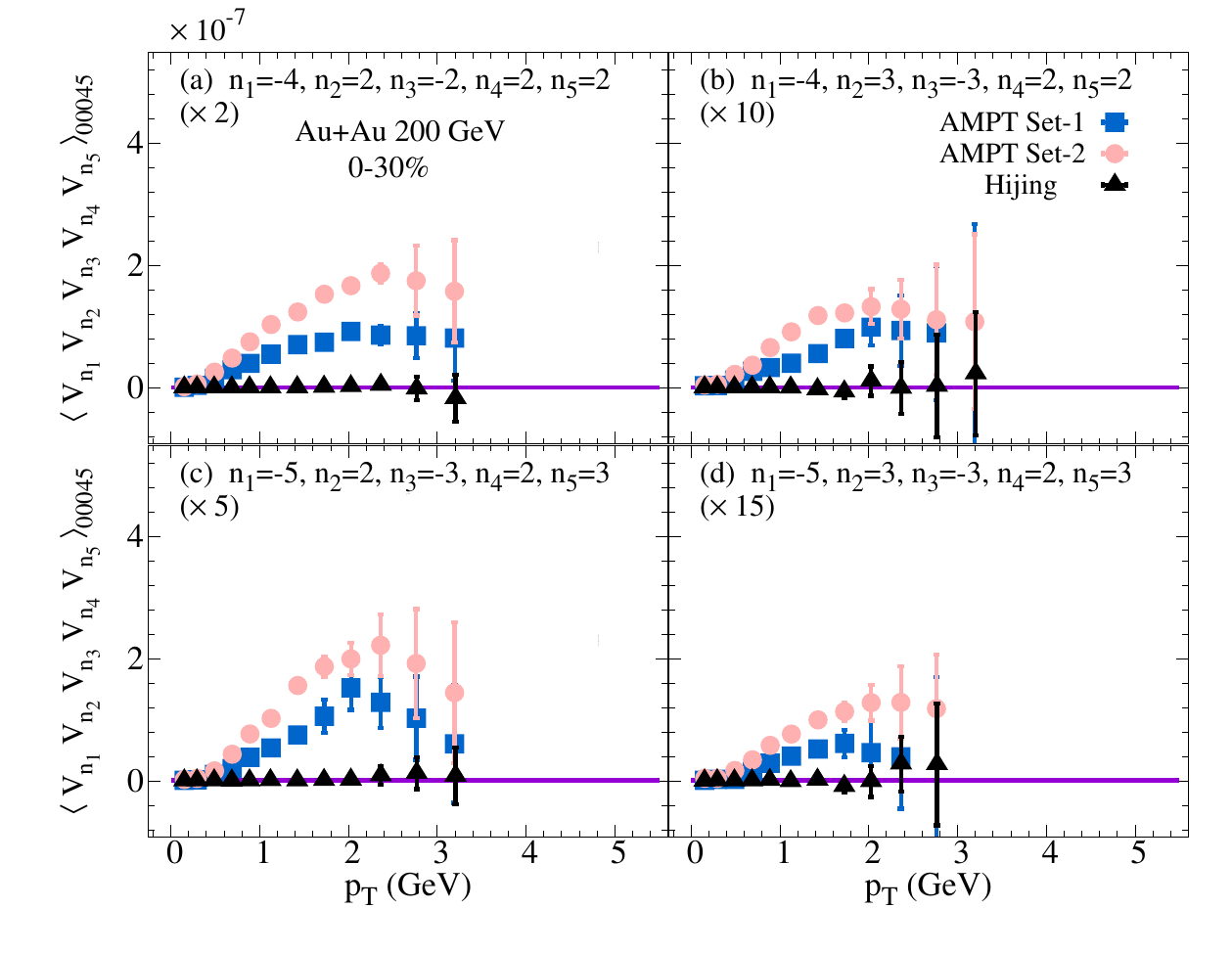}
\vskip -0.4cm
\caption{
Same as in Fig.~\ref{fig:1-2pasc} but for asymmetric correlations $\langle V_{-4} V_{2} V_{-2} V_{2} V_{2}\rangle_{00045}$, $\langle V_{-4} V_{3} V_{-3} V_{2} V_{2}\rangle_{00045}$, $\langle V_{-5} V_{2} V_{-2} V_{2} V_{3}\rangle_{00045}$, and $\langle V_{-4} V_{3} V_{-3} V_{2} V_{3}\rangle_{00045}$.
\label{fig:4-2pasc}}
\vskip -0.1cm
\end{figure}
%--------------------------------------------------------------------

%--------------------------------------------------------------------
Figures~\ref{fig:1-2pasc}--\ref{fig:4-2pasc} showcase the three- and five-particle asymmetric correlations with two POIs plotted vs. $p_{T}$ for Au+Au collisions at 200 GeV, as simulated by the HIJING and AMPT models. These asymmetric correlations with two POIs provide insights into the $p_{T}$ dependence of flow angle correlations when one or more of the flow angles depend on $p_T$. The calculations reveal a noticeable dependence on $p_{T}$ alongside a decrease in correlation strength with the $\sigma_{pp}$ values, indicating the sensitivity of the three- and five-particle asymmetric correlations to final state effects as captured by the AMPT model. Furthermore, the asymmetric correlations $\langle V_{2} V_{2} V_{2} V_{-3} V_{-3} \rangle_{00045}$ in panel (b) of Fig.~\ref{fig:3-2pasc} give values around zero as a function of $p_{T}$, consistent with the anticipated weak correlations between $\psi_{2}$ and $\psi_3$. Additionally, the HIJING model calculations in Figs.~\ref{fig:1-2pasc}--\ref{fig:4-2pasc} yield values consistent with zero within the uncertainties, suggesting minimal (if any) non-flow contribution to the presented three- and five-particle asymmetric correlations.

%\clearpage
%--------------------------------------------------------------------
%--------------------------------------------------------------------
%--------------------------------------------------------------------
\section{Summary and outlook}\label{sec:5}
%--------------------------------------------------------------------
This study delves into the $p_{T}$ dependencies of single and double POIs symmetric and asymmetric correlations for Au+Au collisions at 200 GeV. Utilizing the AMPT model, I examined the sensitivity of these correlations to final state effects, offering insights into their potential to constrain such effects. The HIJING model, serving as a benchmark for non-flow correlations, underscored the impact of non-flow effects on interpreting the $p_T$ dependant SC and ASC data. This work indicates: (i) Using the traditional correlation method for the five- and six-particle correlations and the two-subevents correlation method for the two-, three-, and four-particle correlations can reduce the non-flow effect on such calculations. (ii)  The SC and ASC indicated a sensitivity to the final stat effect given by varying the $\sigma_{pp}$ in the AMPT model. 

In summary, this research underscores the utility of SC and ASC with one and two POIs in constraining $p_T$ dependence of the final state effects. Moreover, it advocates for comprehensive experimental measurements of differential SC and ASC across diverse beam energies and system sizes to serve as additional constraints for theoretical models.
%-----------------------------------------------
\section*{Acknowledgments}
%-----------------------------------------------
The author thanks Somadutta Bhatta and Tanner Mengel for the valuable discussions and for pointing out essential references.
This work was supported in part by funding from the Division of Nuclear Physics of the U.S. Department of Energy under Grant No. DE-FG02-96ER40982.
%--------------------------------------------------------------------
%--------------------------------------------------------------------
%--------------------------------------------------------------------
%\clearpage

\appendix
%--------------------------------------------------------------------
%--------------------------------------------------------------------
\begin{widetext}
%--------------------------------------------------------------------
%--------------------------------------------------------------------
%--------------------------------------------------------------------
\section{k-Particle Correlations}
\label{App:A}

The events generated were analyzed using the two- and multi-particle correlations given via the use of the traditional and subevents correlations methods~\cite{Jia:2017hbm, Huo:2017nms, Zhang:2018lls, Magdy:2020bhd}.

\subsection{The traditional particle correlations;}

The observables discussed in this work using the traditional particle correlations method can be given in terms of the flow vectors as;
%--------------------------------------------------------------------
{\tiny
\begin{eqnarray}\label{eq:A-1}
Q(p,n,k) &=&  \sum^{M_{p}}_{j=1} \omega^{k}_{j} e^{in\varphi_{j}},  \nonumber \\
Q(Q,n,k) &=&  \sum^{M_{Q}}_{j=1} \omega^{k}_{j} e^{in\varphi_{j}},  \nonumber \\
Q(q,n,k) &=&  \sum^{M_{q}}_{j=1} \omega^{k}_{j} e^{in\varphi_{j}},  \nonumber \\
\end{eqnarray}
}
where $\phi_i$ is the azimuthal angle of the $\mathit{i}^{th}$ particle. The order $n$ can take positive or negative values, for $n$=0, $Q(x,0,k)$ gives the event multiplicity. The  $p$ represents particles of interest, $Q$ refers to particles of reference, and $q$ refers to particles that can be indicated as particles of reference and particles of interest.
%--------------------------------------------------------------------

The five-particle correlations using one POI can be given as,
{\tiny
\begin{eqnarray}\label{eq:A-6}
\langle n1,n2,n3,n4,n5 \rangle_{10000}&=& A1(1) + A2(1)  + A3(1)  + A4(1)  + A5(1)  + A6(1)  + A7(1),  \\
A1(1)  &=& Q(p,n1,1)~Q(Q,n2,1)~Q(Q,n3,1)~Q(Q,n4,1)~Q(Q,n5,1),  \nonumber \\
\end{eqnarray}
\begin{eqnarray}
A2(1)  &=&-Q(q,n1+n2,2)~Q(Q,n3,1)~Q(Q,n4,1)~Q(Q,n5,1)-Q(q,n1+n3,2)~Q(Q,n2,1)~Q(Q,n4,1)~Q(Q,n5,1)  \nonumber \\
       &-& Q(q,n1+n4,2)~Q(Q,n2,1)~Q(Q,n3,1)~Q(Q,n5,1)-Q(q,n1+n5,2)~Q(Q,n2,1)~Q(Q,n3,1)~Q(Q,n4,1)  \nonumber \\
       &-& Q(Q,n2+n3,2)~Q(p,n1,1)~Q(Q,n4,1)~Q(Q,n5,1)-Q(Q,n2+n4,2)~Q(p,n1,1)~Q(Q,n3,1)~Q(Q,n5,1)  \nonumber \\
       &-& Q(Q,n2+n5,2)~Q(p,n1,1)~Q(Q,n3,1)~Q(Q,n4,1)-Q(Q,n3+n4,2)~Q(p,n1,1)~Q(Q,n2,1)~Q(Q,n5,1)  \nonumber \\
       &-& Q(Q,n3+n5,2)~Q(p,n1,1)~Q(Q,n2,1)~Q(Q,n4,1)-Q(Q,n4+n5,2)~Q(p,n1,1)~Q(Q,n2,1)~Q(Q,n3,1)  \nonumber \\
\end{eqnarray}
\begin{eqnarray}
A3(1)  &=&2.~Q(q,n1+n2+n3,3)~Q(Q,n4,1)~Q(Q,n5,1)+2.~Q(q,n1+n2+n4,3)~Q(Q,n3,1)~Q(Q,n5,1)  \nonumber \\
       &+&2.~Q(q,n1+n2+n5,3)~Q(Q,n3,1)~Q(Q,n4,1)+2.~Q(q,n1+n3+n4,3)~Q(Q,n2,1)~Q(Q,n5,1)  \nonumber \\
       &+&2.~Q(q,n1+n3+n5,3)~Q(Q,n2,1)~Q(Q,n4,1)+2.~Q(q,n1+n4+n5,3)~Q(Q,n2,1)~Q(Q,n3,1)  \nonumber \\
       &+&2.~Q(Q,n2+n3+n4,3)~Q(p,n1,1)~Q(Q,n5,1)+2.~Q(Q,n2+n3+n5,3)~Q(p,n1,1)~Q(Q,n4,1)  \nonumber \\
       &+&2.~Q(Q,n2+n4+n5,3)~Q(p,n1,1)~Q(Q,n3,1)+2.~Q(Q,n3+n4+n5,3)~Q(p,n1,1)~Q(Q,n2,1)  \nonumber \\
\end{eqnarray}
\begin{eqnarray}
A4(1)  &=&Q(q,n1+n2,2)~Q(Q,n3+n4,2)~Q(Q,n5,1)+Q(q,n1+n2,2)~Q(Q,n3+n5,2)~Q(Q,n4,1) \nonumber \\
       &+&Q(q,n1+n2,2)~Q(Q,n4+n5,2)~Q(Q,n3,1)+Q(q,n1+n3,2)~Q(Q,n2+n4,2)~Q(Q,n5,1) \nonumber \\
       &+&Q(q,n1+n3,2)~Q(Q,n2+n5,2)~Q(Q,n4,1)+Q(q,n1+n3,2)~Q(Q,n4+n5,2)~Q(Q,n2,1) \nonumber \\
       &+&Q(q,n1+n4,2)~Q(Q,n2+n3,2)~Q(Q,n5,1)+Q(q,n1+n4,2)~Q(Q,n2+n5,2)~Q(Q,n3,1) \nonumber \\
       &+&Q(q,n1+n4,2)~Q(Q,n3+n5,2)~Q(Q,n2,1)+Q(q,n1+n5,2)~Q(Q,n2+n3,2)~Q(Q,n4,1) \nonumber \\
       &+&Q(q,n1+n5,2)~Q(Q,n2+n4,2)~Q(Q,n3,1)+Q(q,n1+n5,2)~Q(Q,n3+n4,2)~Q(Q,n2,1) \nonumber \\
       &+&Q(Q,n2+n5,2)~Q(Q,n3+n4,2)~Q(p,n1,1)+Q(Q,n2+n3,2)~Q(Q,n4+n5,2)~Q(p,n1,1) \nonumber \\
       &+&Q(Q,n2+n4,2)~Q(Q,n3+n5,2)~Q(p,n1,1),  \nonumber \\
\end{eqnarray}
\begin{eqnarray}
A5(1)  &=&-2.~Q(q,n1+n2+n3,3)~Q(Q,n4+n5,2)-2.~Q(q,n1+n2+n4,3)~Q(Q,n3+n5,2) \nonumber \\
       &-& 2.~Q(q,n1+n2+n5,3)~Q(Q,n3+n4,2)-2.~Q(q,n1+n3+n4,3)~Q(Q,n2+n5,2) \nonumber \\
       &-& 2.~Q(q,n1+n3+n5,3)~Q(Q,n2+n4,2)-2.~Q(q,n1+n4+n5,3)~Q(Q,n2+n3,2) \nonumber \\
       &-& 2.~Q(Q,n2+n3+n4,3)~Q(q,n1+n5,2)-2.~Q(Q,n2+n3+n5,3)~Q(q,n1+n4,2) \nonumber \\
       &-& 2.~Q(Q,n2+n4+n5,3)~Q(q,n1+n3,2)-2.~Q(Q,n3+n4+n5,3)~Q(q,n1+n2,2), \nonumber \\
\end{eqnarray}
\begin{eqnarray}
A6(1)  &=&-6.~Q(q,n1+n2+n3+n4,4)~Q(Q,n5,1)-6.~Q(q,n1+n2+n3+n5,4)~Q(Q,n4,1) \nonumber \\
       &-& 6.~Q(q,n1+n2+n4+n5,4)~Q(Q,n3,1)-6.~Q(q,n1+n3+n4+n5,4)~Q(Q,n2,1) \nonumber \\
       &-& 6.~Q(Q,n2+n3+n4+n5,4)~Q(p,n1,1), \nonumber \\
\end{eqnarray}
\begin{eqnarray}
A7(1)  &=& 24.~Q(q,n1+n2+n3+n4+n5,5).  \nonumber \\
\end{eqnarray}
}

In addition, the five-particle correlations using two POIs can be given as,
{\tiny
\begin{eqnarray}\label{eq:A-7}
\langle n1,n2,n3,n4,n5 \rangle_{12000} &=& A1(2) + A2(2)  + A3(2)  + A4(2)  + A5(2)  + A6(2)  + A7(2),\\
A1(2) &=& Q(p,n1,1)~Q(p,n2,1)~Q(Q,n3,1)~Q(Q,n4,1)~Q(Q,n5,1),\nonumber \\
\end{eqnarray}
\begin{eqnarray}
A2(2) &=&-Q(p,n1+n2,2)~Q(Q,n3,1)~Q(Q,n4,1)~Q(Q,n5,1) - Q(q,n1+n3,2)~Q(p,n2,1)~Q(Q,n4,1)~Q(Q,n5,1)\nonumber \\
   &-& Q(q,n1+n4,2)~Q(p,n2,1)~Q(Q,n3,1)~Q(Q,n5,1) - Q(q,n1+n5,2)~Q(p,n2,1)~Q(Q,n3,1)~Q(Q,n4,1)\nonumber \\
   &-& Q(q,n2+n3,2)~Q(p,n1,1)~Q(Q,n4,1)~Q(Q,n5,1) - Q(q,n2+n4,2)~Q(p,n1,1)~Q(Q,n3,1)~Q(Q,n5,1)\nonumber \\
   &-& Q(q,n2+n5,2)~Q(p,n1,1)~Q(Q,n3,1)~Q(Q,n4,1) - Q(Q,n3+n4,2)~Q(p,n1,1)~Q(p,n2,1)~Q(Q,n5,1)\nonumber \\
   &-& Q(Q,n3+n5,2)~Q(p,n1,1)~Q(p,n2,1)~Q(Q,n4,1) - Q(Q,n4+n5,2)~Q(p,n1,1)~Q(p,n2,1)~Q(Q,n3,1),\nonumber \\
\end{eqnarray}
\begin{eqnarray}
A3(2) &=& 2.~Q(q,n1+n2+n3,3)~Q(Q,n4,1)~Q(Q,n5,1) + 2.~Q(q,n1+n2+n4,3)~Q(Q,n3,1)~Q(Q,n5,1)\nonumber \\
   &+& 2.~Q(q,n1+n2+n5,3)~Q(Q,n3,1)~Q(Q,n4,1) + 2.~Q(q,n1+n3+n4,3)~Q(p,n2,1)~Q(Q,n5,1)\nonumber \\
   &+& 2.~Q(q,n1+n3+n5,3)~Q(p,n2,1)~Q(Q,n4,1) + 2.~Q(q,n1+n4+n5,3)~Q(p,n2,1)~Q(Q,n3,1)\nonumber \\
   &+& 2.~Q(q,n2+n3+n4,3)~Q(p,n1,1)~Q(Q,n5,1) + 2.~Q(q,n2+n3+n5,3)~Q(p,n1,1)~Q(Q,n4,1)\nonumber \\
   &+& 2.~Q(q,n2+n4+n5,3)~Q(p,n1,1)~Q(Q,n3,1) + 2.~Q(Q,n3+n4+n5,3)~Q(p,n1,1)~Q(p,n2,1),\nonumber \\
\end{eqnarray}
\begin{eqnarray}
A4(2) &=& Q(p,n1+n2,2)~Q(Q,n3+n4,2)~Q(Q,n5,1) + Q(p,n1+n2,2)~Q(Q,n3+n5,2)~Q(Q,n4,1)\nonumber \\
   &+& Q(p,n1+n2,2)~Q(Q,n4+n5,2)~Q(Q,n3,1) + Q(q,n1+n3,2)~Q(q,n2+n4,2)~Q(Q,n5,1)\nonumber \\
   &+& Q(q,n1+n3,2)~Q(q,n2+n5,2)~Q(Q,n4,1) + Q(q,n1+n3,2)~Q(Q,n4+n5,2)~Q(p,n2,1)\nonumber \\
   &+& Q(q,n1+n4,2)~Q(q,n2+n3,2)~Q(Q,n5,1) + Q(q,n1+n4,2)~Q(q,n2+n5,2)~Q(Q,n3,1)\nonumber \\
   &+& Q(q,n1+n4,2)~Q(Q,n3+n5,2)~Q(p,n2,1) + Q(q,n1+n5,2)~Q(q,n2+n3,2)~Q(Q,n4,1)\nonumber \\
   &+& Q(q,n1+n5,2)~Q(q,n2+n4,2)~Q(Q,n3,1) + Q(q,n1+n5,2)~Q(Q,n3+n4,2)~Q(p,n2,1)\nonumber \\
   &+& Q(q,n2+n5,2)~Q(Q,n3+n4,2)~Q(p,n1,1) + Q(q,n2+n3,2)~Q(Q,n4+n5,2)~Q(p,n1,1)\nonumber \\
   &+& Q(q,n2+n4,2)~Q(Q,n3+n5,2)~Q(p,n1,1),\nonumber \\
\end{eqnarray}
\begin{eqnarray}
A5(2) &=&-2.~Q(q,n1+n2+n3,3)~Q(Q,n4+n5,2) - 2.~Q(q,n1+n2+n4,3)~Q(Q,n3+n5,2)\nonumber \\
   &-& 2.~Q(q,n1+n2+n5,3)~Q(Q,n3+n4,2) - 2.~Q(q,n1+n3+n4,3)~Q(q,n2+n5,2)\nonumber \\
   &-& 2.~Q(q,n1+n3+n5,3)~Q(q,n2+n4,2) - 2.~Q(q,n1+n4+n5,3)~Q(q,n2+n3,2)\nonumber \\
   &-& 2.~Q(q,n2+n3+n4,3)~Q(q,n1+n5,2) - 2.~Q(q,n2+n3+n5,3)~Q(q,n1+n4,2)\nonumber \\
   &-& 2.~Q(q,n2+n4+n5,3)~Q(q,n1+n3,2) - 2.~Q(Q,n3+n4+n5,3)~Q(p,n1+n2,2),\nonumber \\
\end{eqnarray}
\begin{eqnarray}
A6(2) &=&-6.~Q(q,n1+n2+n3+n4,4)~Q(Q,n5,1) - 6.~Q(q,n1+n2+n3+n5,4)~Q(Q,n4,1)\nonumber \\
   &-& 6.~Q(q,n1+n2+n4+n5,4)~Q(Q,n3,1) - 6.~Q(q,n1+n3+n4+n5,4)~Q(p,n2,1)\nonumber \\
   &-& 6.~Q(q,n2+n3+n4+n5,4)~Q(p,n1,1),\nonumber \\
\end{eqnarray}
\begin{eqnarray}
A7(2) &=& 24.~Q(q,n1+n2+n3+n4+n5,5). \nonumber \\
\end{eqnarray}

}

Then, the event average five particle correlations will be given as:
{\tiny
\begin{eqnarray}
\langle V_{n_1} V_{n_2} V_{n_3} V_{n_4} V_{n_5} \rangle_{P} &=& \langle \langle n1,n2,n3,n4,n5 \rangle_{P}\rangle/\langle \langle 0,0,0,0,0 \rangle_{P}\rangle.
\end{eqnarray}
}

The six-particle correlations using one POI can be given as,
{\tiny
\begin{eqnarray}\label{eq:A-8}
\langle n1,n2,n3,n4,n5,n6 \rangle_{100000} &=& B1(1) + B2(1)  + B3(1)  + B4(1)  + B5(1)  + B6(1)  + B7(1) + A8(1) + A9(1) + A10(1) + A11(1), \\
B1(1)&=& Q(p,n1,1)~Q(Q,n2,1)~Q(Q,n3,1)~Q(Q,n4,1)~Q(Q,n5,1)~Q(Q,n6,1),\nonumber \\
\end{eqnarray}
\begin{eqnarray}
B2(1)&=&-Q(q,n1+n2,2)~Q(Q,n3,1)~Q(Q,n4,1)~Q(Q,n5,1)~Q(Q,n6,1) - Q(q,n1+n3,2)~Q(Q,n2,1)~Q(Q,n4,1)~Q(Q,n5,1)~Q(Q,n6,1) \nonumber \\
     &-& Q(q,n1+n4,2)~Q(Q,n2,1)~Q(Q,n3,1)~Q(Q,n5,1)~Q(Q,n6,1) - Q(q,n1+n5,2)~Q(Q,n2,1)~Q(Q,n3,1)~Q(Q,n4,1)~Q(Q,n6,1) \nonumber \\
     &-& Q(q,n1+n6,2)~Q(Q,n2,1)~Q(Q,n3,1)~Q(Q,n4,1)~Q(Q,n5,1) - Q(Q,n2+n3,2)~Q(p,n1,1)~Q(Q,n4,1)~Q(Q,n5,1)~Q(Q,n6,1) \nonumber \\
     &-& Q(Q,n2+n4,2)~Q(p,n1,1)~Q(Q,n3,1)~Q(Q,n5,1)~Q(Q,n6,1) - Q(Q,n2+n5,2)~Q(p,n1,1)~Q(Q,n3,1)~Q(Q,n4,1)~Q(Q,n6,1) \nonumber \\
     &-& Q(Q,n2+n6,2)~Q(p,n1,1)~Q(Q,n3,1)~Q(Q,n4,1)~Q(Q,n5,1) - Q(Q,n3+n4,2)~Q(p,n1,1)~Q(Q,n2,1)~Q(Q,n5,1)~Q(Q,n6,1) \nonumber \\
     &-& Q(Q,n3+n5,2)~Q(p,n1,1)~Q(Q,n2,1)~Q(Q,n4,1)~Q(Q,n6,1) - Q(Q,n3+n6,2)~Q(p,n1,1)~Q(Q,n2,1)~Q(Q,n4,1)~Q(Q,n5,1) \nonumber \\
     &-& Q(Q,n4+n5,2)~Q(p,n1,1)~Q(Q,n2,1)~Q(Q,n3,1)~Q(Q,n6,1) - Q(Q,n4+n6,2)~Q(p,n1,1)~Q(Q,n2,1)~Q(Q,n3,1)~Q(Q,n5,1) \nonumber \\
     &-& Q(Q,n5+n6,2)~Q(p,n1,1)~Q(Q,n2,1)~Q(Q,n3,1)~Q(Q,n4,1), \nonumber \\
\end{eqnarray}
\begin{eqnarray}
B3(1)&=& 2.~Q(q,n1+n2+n3,3)~Q(Q,n4,1)~Q(Q,n5,1)~Q(Q,n6,1) + 2.~Q(q,n1+n2+n4,3)~Q(Q,n3,1)~Q(Q,n5,1)~Q(Q,n6,1) \nonumber \\
		&+& 2.~Q(q,n1+n2+n5,3)~Q(Q,n3,1)~Q(Q,n4,1)~Q(Q,n6,1) + 2.~Q(q,n1+n2+n6,3)~Q(Q,n3,1)~Q(Q,n4,1)~Q(Q,n5,1) \nonumber \\
		&+& 2.~Q(q,n1+n3+n4,3)~Q(Q,n2,1)~Q(Q,n5,1)~Q(Q,n6,1) + 2.~Q(q,n1+n3+n5,3)~Q(Q,n2,1)~Q(Q,n4,1)~Q(Q,n6,1) \nonumber \\
		&+& 2.~Q(q,n1+n3+n6,3)~Q(Q,n2,1)~Q(Q,n4,1)~Q(Q,n5,1) + 2.~Q(q,n1+n4+n5,3)~Q(Q,n2,1)~Q(Q,n3,1)~Q(Q,n6,1) \nonumber \\
		&+& 2.~Q(q,n1+n4+n6,3)~Q(Q,n2,1)~Q(Q,n3,1)~Q(Q,n5,1) + 2.~Q(q,n1+n5+n6,3)~Q(Q,n2,1)~Q(Q,n3,1)~Q(Q,n4,1) \nonumber \\
		&+& 2.~Q(Q,n2+n3+n4,3)~Q(p,n1,1)~Q(Q,n5,1)~Q(Q,n6,1) + 2.~Q(Q,n2+n3+n5,3)~Q(p,n1,1)~Q(Q,n4,1)~Q(Q,n6,1) \nonumber \\
		&+& 2.~Q(Q,n2+n3+n6,3)~Q(p,n1,1)~Q(Q,n4,1)~Q(Q,n5,1) + 2.~Q(Q,n2+n4+n5,3)~Q(p,n1,1)~Q(Q,n3,1)~Q(Q,n6,1) \nonumber \\
		&+& 2.~Q(Q,n2+n4+n6,3)~Q(p,n1,1)~Q(Q,n3,1)~Q(Q,n5,1) + 2.~Q(Q,n2+n5+n6,3)~Q(p,n1,1)~Q(Q,n3,1)~Q(Q,n4,1) \nonumber \\
		&+& 2.~Q(Q,n3+n4+n5,3)~Q(p,n1,1)~Q(Q,n2,1)~Q(Q,n6,1) + 2.~Q(Q,n3+n4+n6,3)~Q(p,n1,1)~Q(Q,n2,1)~Q(Q,n5,1) \nonumber \\
		&+& 2.~Q(Q,n3+n5+n6,3)~Q(p,n1,1)~Q(Q,n2,1)~Q(Q,n4,1) + 2.~Q(Q,n4+n5+n6,3)~Q(p,n1,1)~Q(Q,n2,1)~Q(Q,n3,1),\nonumber \\
\end{eqnarray}
\begin{eqnarray}
B4(1)&=& Q(q,n1+n2,2)~Q(Q,n3+n4,2)~Q(Q,n5,1)~Q(Q,n6,1) + Q(q,n1+n2,2)~Q(Q,n3+n5,2)~Q(Q,n4,1)~Q(Q,n6,1) \nonumber \\
		&+& Q(q,n1+n2,2)~Q(Q,n3+n6,2)~Q(Q,n4,1)~Q(Q,n5,1) + Q(q,n1+n2,2)~Q(Q,n4+n5,2)~Q(Q,n3,1)~Q(Q,n6,1) \nonumber \\
		&+& Q(q,n1+n2,2)~Q(Q,n4+n6,2)~Q(Q,n3,1)~Q(Q,n5,1) + Q(q,n1+n2,2)~Q(Q,n5+n6,2)~Q(Q,n3,1)~Q(Q,n4,1) \nonumber \\
		&+& Q(q,n1+n3,2)~Q(Q,n2+n4,2)~Q(Q,n5,1)~Q(Q,n6,1) + Q(q,n1+n3,2)~Q(Q,n2+n5,2)~Q(Q,n4,1)~Q(Q,n6,1) \nonumber \\
		&+& Q(q,n1+n3,2)~Q(Q,n2+n6,2)~Q(Q,n4,1)~Q(Q,n5,1) + Q(q,n1+n3,2)~Q(Q,n4+n5,2)~Q(Q,n2,1)~Q(Q,n6,1) \nonumber \\
		&+& Q(q,n1+n3,2)~Q(Q,n4+n6,2)~Q(Q,n2,1)~Q(Q,n5,1) + Q(q,n1+n3,2)~Q(Q,n5+n6,2)~Q(Q,n2,1)~Q(Q,n4,1) \nonumber \\
		&+& Q(q,n1+n4,2)~Q(Q,n2+n3,2)~Q(Q,n5,1)~Q(Q,n6,1) + Q(q,n1+n4,2)~Q(Q,n2+n5,2)~Q(Q,n3,1)~Q(Q,n6,1) \nonumber \\
		&+& Q(q,n1+n4,2)~Q(Q,n2+n6,2)~Q(Q,n3,1)~Q(Q,n5,1) + Q(q,n1+n4,2)~Q(Q,n3+n5,2)~Q(Q,n2,1)~Q(Q,n6,1) \nonumber \\
		&+& Q(q,n1+n4,2)~Q(Q,n3+n6,2)~Q(Q,n2,1)~Q(Q,n5,1) + Q(q,n1+n4,2)~Q(Q,n5+n6,2)~Q(Q,n2,1)~Q(Q,n3,1) \nonumber \\
		&+& Q(q,n1+n5,2)~Q(Q,n2+n3,2)~Q(Q,n4,1)~Q(Q,n6,1) + Q(q,n1+n5,2)~Q(Q,n2+n4,2)~Q(Q,n3,1)~Q(Q,n6,1) \nonumber \\
		&+& Q(q,n1+n5,2)~Q(Q,n2+n6,2)~Q(Q,n3,1)~Q(Q,n4,1) + Q(q,n1+n5,2)~Q(Q,n3+n4,2)~Q(Q,n2,1)~Q(Q,n6,1) \nonumber \\
		&+& Q(q,n1+n5,2)~Q(Q,n3+n6,2)~Q(Q,n2,1)~Q(Q,n4,1) + Q(q,n1+n5,2)~Q(Q,n4+n6,2)~Q(Q,n2,1)~Q(Q,n3,1) \nonumber \\
		&+& Q(q,n1+n6,2)~Q(Q,n2+n3,2)~Q(Q,n4,1)~Q(Q,n5,1) + Q(q,n1+n6,2)~Q(Q,n2+n4,2)~Q(Q,n3,1)~Q(Q,n5,1) \nonumber \\
		&+& Q(q,n1+n6,2)~Q(Q,n2+n5,2)~Q(Q,n3,1)~Q(Q,n4,1) + Q(q,n1+n6,2)~Q(Q,n3+n4,2)~Q(Q,n2,1)~Q(Q,n5,1) \nonumber \\
		&+& Q(q,n1+n6,2)~Q(Q,n3+n5,2)~Q(Q,n2,1)~Q(Q,n4,1) + Q(q,n1+n6,2)~Q(Q,n4+n5,2)~Q(Q,n2,1)~Q(Q,n3,1) \nonumber \\
		&+& Q(Q,n2+n3,2)~Q(Q,n4+n5,2)~Q(p,n1,1)~Q(Q,n6,1) + Q(Q,n2+n3,2)~Q(Q,n4+n6,2)~Q(p,n1,1)~Q(Q,n5,1) \nonumber \\
		&+& Q(Q,n2+n3,2)~Q(Q,n5+n6,2)~Q(p,n1,1)~Q(Q,n4,1) + Q(Q,n2+n4,2)~Q(Q,n3+n5,2)~Q(p,n1,1)~Q(Q,n6,1) \nonumber \\
		&+& Q(Q,n2+n4,2)~Q(Q,n3+n6,2)~Q(p,n1,1)~Q(Q,n5,1) + Q(Q,n2+n4,2)~Q(Q,n5+n6,2)~Q(p,n1,1)~Q(Q,n3,1) \nonumber \\
		&+& Q(Q,n2+n5,2)~Q(Q,n3+n4,2)~Q(p,n1,1)~Q(Q,n6,1) + Q(Q,n2+n5,2)~Q(Q,n3+n6,2)~Q(p,n1,1)~Q(Q,n4,1) \nonumber \\
		&+& Q(Q,n2+n5,2)~Q(Q,n4+n6,2)~Q(p,n1,1)~Q(Q,n3,1) + Q(Q,n3+n4,2)~Q(Q,n2+n6,2)~Q(p,n1,1)~Q(Q,n5,1) \nonumber \\
		&+& Q(Q,n3+n4,2)~Q(Q,n5+n6,2)~Q(p,n1,1)~Q(Q,n2,1) + Q(Q,n3+n5,2)~Q(Q,n2+n6,2)~Q(p,n1,1)~Q(Q,n4,1) \nonumber \\
		&+& Q(Q,n3+n5,2)~Q(Q,n4+n6,2)~Q(p,n1,1)~Q(Q,n2,1) + Q(Q,n4+n5,2)~Q(Q,n2+n6,2)~Q(p,n1,1)~Q(Q,n3,1) \nonumber \\
		&+& Q(Q,n4+n5,2)~Q(Q,n3+n6,2)~Q(p,n1,1)~Q(Q,n2,1),\nonumber \\
\end{eqnarray}
\begin{eqnarray}
B5(1)&=&-2.~Q(q,n1+n2+n3,3)~Q(Q,n4+n5,2)~Q(Q,n6,1) - 2.~Q(q,n1+n2+n3,3)~Q(Q,n4+n6,2)~Q(Q,n5,1) \nonumber \\
		&-& 2.~Q(q,n1+n2+n3,3)~Q(Q,n5+n6,2)~Q(Q,n4,1) - 2.~Q(q,n1+n2+n4,3)~Q(Q,n3+n5,2)~Q(Q,n6,1) \nonumber \\
		&-& 2.~Q(q,n1+n2+n4,3)~Q(Q,n3+n6,2)~Q(Q,n5,1) - 2.~Q(q,n1+n2+n4,3)~Q(Q,n5+n6,2)~Q(Q,n3,1) \nonumber \\
		&-& 2.~Q(q,n1+n2+n5,3)~Q(Q,n3+n4,2)~Q(Q,n6,1) - 2.~Q(q,n1+n2+n5,3)~Q(Q,n3+n6,2)~Q(Q,n4,1) \nonumber \\
		&-& 2.~Q(q,n1+n2+n5,3)~Q(Q,n4+n6,2)~Q(Q,n3,1) - 2.~Q(q,n1+n2+n6,3)~Q(Q,n3+n4,2)~Q(Q,n5,1) \nonumber \\
		&-& 2.~Q(q,n1+n2+n6,3)~Q(Q,n3+n5,2)~Q(Q,n4,1) - 2.~Q(q,n1+n2+n6,3)~Q(Q,n4+n5,2)~Q(Q,n3,1) \nonumber \\
		&-& 2.~Q(q,n1+n3+n4,3)~Q(Q,n2+n5,2)~Q(Q,n6,1) - 2.~Q(q,n1+n3+n4,3)~Q(Q,n2+n6,2)~Q(Q,n5,1) \nonumber \\
		&-& 2.~Q(q,n1+n3+n4,3)~Q(Q,n5+n6,2)~Q(Q,n2,1) - 2.~Q(q,n1+n3+n5,3)~Q(Q,n2+n4,2)~Q(Q,n6,1) \nonumber \\
		&-& 2.~Q(q,n1+n3+n5,3)~Q(Q,n2+n6,2)~Q(Q,n4,1) - 2.~Q(q,n1+n3+n5,3)~Q(Q,n4+n6,2)~Q(Q,n2,1) \nonumber \\
		&-& 2.~Q(q,n1+n3+n6,3)~Q(Q,n2+n4,2)~Q(Q,n5,1) - 2.~Q(q,n1+n3+n6,3)~Q(Q,n2+n5,2)~Q(Q,n4,1) \nonumber \\
		&-& 2.~Q(q,n1+n3+n6,3)~Q(Q,n4+n5,2)~Q(Q,n2,1) - 2.~Q(q,n1+n4+n5,3)~Q(Q,n2+n3,2)~Q(Q,n6,1) \nonumber \\
		&-& 2.~Q(q,n1+n4+n5,3)~Q(Q,n2+n6,2)~Q(Q,n3,1) - 2.~Q(q,n1+n4+n5,3)~Q(Q,n3+n6,2)~Q(Q,n2,1) \nonumber \\
		&-& 2.~Q(q,n1+n4+n6,3)~Q(Q,n2+n3,2)~Q(Q,n5,1) - 2.~Q(q,n1+n4+n6,3)~Q(Q,n2+n5,2)~Q(Q,n3,1) \nonumber \\
		&-& 2.~Q(q,n1+n4+n6,3)~Q(Q,n3+n5,2)~Q(Q,n2,1) - 2.~Q(q,n1+n5+n6,3)~Q(Q,n2+n3,2)~Q(Q,n4,1) \nonumber \\
		&-& 2.~Q(q,n1+n5+n6,3)~Q(Q,n2+n4,2)~Q(Q,n3,1) - 2.~Q(q,n1+n5+n6,3)~Q(Q,n3+n4,2)~Q(Q,n2,1) \nonumber \\
		&-& 2.~Q(Q,n2+n3+n4,3)~Q(q,n1+n5,2)~Q(Q,n6,1) - 2.~Q(Q,n2+n3+n4,3)~Q(q,n1+n6,2)~Q(Q,n5,1) \nonumber \\
		&-& 2.~Q(Q,n2+n3+n4,3)~Q(Q,n5+n6,2)~Q(p,n1,1) - 2.~Q(Q,n2+n3+n5,3)~Q(q,n1+n4,2)~Q(Q,n6,1) \nonumber \\
		&-& 2.~Q(Q,n2+n3+n5,3)~Q(q,n1+n6,2)~Q(Q,n4,1) - 2.~Q(Q,n2+n3+n5,3)~Q(Q,n4+n6,2)~Q(p,n1,1) \nonumber \\
		&-& 2.~Q(Q,n2+n3+n6,3)~Q(q,n1+n4,2)~Q(Q,n5,1) - 2.~Q(Q,n2+n3+n6,3)~Q(q,n1+n5,2)~Q(Q,n4,1) \nonumber \\
		&-& 2.~Q(Q,n2+n3+n6,3)~Q(Q,n4+n5,2)~Q(p,n1,1) - 2.~Q(Q,n2+n4+n5,3)~Q(q,n1+n3,2)~Q(Q,n6,1) \nonumber \\
		&-& 2.~Q(Q,n2+n4+n5,3)~Q(q,n1+n6,2)~Q(Q,n3,1) - 2.~Q(Q,n2+n4+n5,3)~Q(Q,n3+n6,2)~Q(p,n1,1) \nonumber \\
		&-& 2.~Q(Q,n2+n4+n6,3)~Q(q,n1+n3,2)~Q(Q,n5,1) - 2.~Q(Q,n2+n4+n6,3)~Q(q,n1+n5,2)~Q(Q,n3,1) \nonumber \\
		&-& 2.~Q(Q,n2+n4+n6,3)~Q(Q,n3+n5,2)~Q(p,n1,1) - 2.~Q(Q,n2+n5+n6,3)~Q(q,n1+n3,2)~Q(Q,n4,1) \nonumber \\
		&-& 2.~Q(Q,n2+n5+n6,3)~Q(q,n1+n4,2)~Q(Q,n3,1) - 2.~Q(Q,n2+n5+n6,3)~Q(Q,n3+n4,2)~Q(p,n1,1) \nonumber \\
		&-& 2.~Q(Q,n3+n4+n5,3)~Q(q,n1+n2,2)~Q(Q,n6,1) - 2.~Q(Q,n3+n4+n5,3)~Q(q,n1+n6,2)~Q(Q,n2,1) \nonumber \\
		&-& 2.~Q(Q,n3+n4+n5,3)~Q(Q,n2+n6,2)~Q(p,n1,1) - 2.~Q(Q,n3+n4+n6,3)~Q(q,n1+n2,2)~Q(Q,n5,1) \nonumber \\
		&-& 2.~Q(Q,n3+n4+n6,3)~Q(q,n1+n5,2)~Q(Q,n2,1) - 2.~Q(Q,n3+n4+n6,3)~Q(Q,n2+n5,2)~Q(p,n1,1) \nonumber \\
		&-& 2.~Q(Q,n3+n5+n6,3)~Q(q,n1+n2,2)~Q(Q,n4,1) - 2.~Q(Q,n3+n5+n6,3)~Q(q,n1+n4,2)~Q(Q,n2,1) \nonumber \\
		&-& 2.~Q(Q,n3+n5+n6,3)~Q(Q,n2+n4,2)~Q(p,n1,1) - 2.~Q(Q,n4+n5+n6,3)~Q(q,n1+n2,2)~Q(Q,n3,1) \nonumber \\
		&-& 2.~Q(Q,n4+n5+n6,3)~Q(q,n1+n3,2)~Q(Q,n2,1) - 2.~Q(Q,n4+n5+n6,3)~Q(Q,n2+n3,2)~Q(p,n1,1),\nonumber \\
\end{eqnarray}
\begin{eqnarray}
B6(1)&=&-6.~Q(q,n1+n2+n3+n4,4)~Q(Q,n5,1)~Q(Q,n6,1) - 6.~Q(q,n1+n2+n3+n5,4)~Q(Q,n4,1)~Q(Q,n6,1) \nonumber \\
		&-& 6.~Q(q,n1+n2+n4+n5,4)~Q(Q,n3,1)~Q(Q,n6,1) - 6.~Q(q,n1+n3+n4+n5,4)~Q(Q,n2,1)~Q(Q,n6,1) \nonumber \\
		&-& 6.~Q(q,n1+n2+n3+n6,4)~Q(Q,n4,1)~Q(Q,n5,1) - 6.~Q(q,n1+n2+n4+n6,4)~Q(Q,n3,1)~Q(Q,n5,1) \nonumber \\
		&-& 6.~Q(q,n1+n3+n4+n6,4)~Q(Q,n2,1)~Q(Q,n5,1) - 6.~Q(q,n1+n2+n5+n6,4)~Q(Q,n3,1)~Q(Q,n4,1) \nonumber \\
		&-& 6.~Q(q,n1+n3+n5+n6,4)~Q(Q,n2,1)~Q(Q,n4,1) - 6.~Q(q,n1+n4+n5+n6,4)~Q(Q,n2,1)~Q(Q,n3,1) \nonumber \\
		&-& 6.~Q(Q,n2+n3+n4+n5,4)~Q(p,n1,1)~Q(Q,n6,1) - 6.~Q(Q,n2+n3+n4+n6,4)~Q(p,n1,1)~Q(Q,n5,1) \nonumber \\
		&-& 6.~Q(Q,n2+n3+n5+n6,4)~Q(p,n1,1)~Q(Q,n4,1) - 6.~Q(Q,n2+n4+n5+n6,4)~Q(p,n1,1)~Q(Q,n3,1) \nonumber \\
		&-& 6.~Q(Q,n3+n4+n5+n6,4)~Q(p,n1,1)~Q(Q,n2,1), \nonumber \\
\end{eqnarray}
\begin{eqnarray}
B7(1)&=&-Q(q,n1+n2,2)~Q(Q,n3+n4,2)~Q(Q,n5+n6,2) - Q(q,n1+n2,2)~Q(Q,n3+n5,2)~Q(Q,n4+n6,2) \nonumber \\
    &-& Q(q,n1+n2,2)~Q(Q,n4+n5,2)~Q(Q,n3+n6,2) - Q(q,n1+n3,2)~Q(Q,n4+n5,2)~Q(Q,n2+n6,2) \nonumber \\
    &-& Q(q,n1+n3,2)~Q(Q,n2+n5,2)~Q(Q,n4+n6,2) - Q(q,n1+n3,2)~Q(Q,n2+n4,2)~Q(Q,n5+n6,2) \nonumber \\
    &-& Q(q,n1+n4,2)~Q(Q,n3+n5,2)~Q(Q,n2+n6,2) - Q(q,n1+n4,2)~Q(Q,n2+n5,2)~Q(Q,n3+n6,2) \nonumber \\
    &-& Q(q,n1+n4,2)~Q(Q,n2+n3,2)~Q(Q,n5+n6,2) - Q(q,n1+n5,2)~Q(Q,n3+n4,2)~Q(Q,n2+n6,2) \nonumber \\
    &-& Q(q,n1+n5,2)~Q(Q,n2+n4,2)~Q(Q,n3+n6,2) - Q(q,n1+n5,2)~Q(Q,n2+n3,2)~Q(Q,n4+n6,2) \nonumber \\
    &-& Q(q,n1+n6,2)~Q(Q,n3+n4,2)~Q(Q,n2+n5,2) - Q(q,n1+n6,2)~Q(Q,n2+n4,2)~Q(Q,n3+n5,2) \nonumber \\
    &-& Q(q,n1+n6,2)~Q(Q,n2+n3,2)~Q(Q,n4+n5,2), \nonumber \\  
\end{eqnarray}
\begin{eqnarray}
B8(1)&=& 4.~Q(q,n1+n2+n3,3)~Q(Q,n4+n5+n6,3) + 4.~Q(q,n1+n2+n4,3)~Q(Q,n3+n5+n6,3) \nonumber \\
    &+& 4.~Q(q,n1+n2+n5,3)~Q(Q,n3+n4+n6,3) + 4.~Q(q,n1+n3+n4,3)~Q(Q,n2+n5+n6,3) \nonumber \\
    &+& 4.~Q(q,n1+n3+n5,3)~Q(Q,n2+n4+n6,3) + 4.~Q(q,n1+n4+n5,3)~Q(Q,n2+n3+n6,3) \nonumber \\
    &+& 4.~Q(Q,n2+n3+n4,3)~Q(q,n1+n5+n6,3) + 4.~Q(Q,n2+n3+n5,3)~Q(q,n1+n4+n6,3) \nonumber \\
    &+& 4.~Q(Q,n2+n4+n5,3)~Q(q,n1+n3+n6,3) + 4.~Q(Q,n3+n4+n5,3)~Q(q,n1+n2+n6,3),\nonumber \\
\end{eqnarray}
\begin{eqnarray}
B9(1)&=& 6.~Q(q,n1+n2+n3+n4,4)~Q(Q,n5+n6,2) + 6.~Q(q,n1+n2+n3+n5,4)~Q(Q,n4+n6,2) \nonumber \\
    &+& 6.~Q(q,n1+n2+n3+n6,4)~Q(Q,n4+n5,2) + 6.~Q(q,n1+n2+n4+n5,4)~Q(Q,n3+n6,2) \nonumber \\
    &+& 6.~Q(q,n1+n2+n4+n6,4)~Q(Q,n3+n5,2) + 6.~Q(q,n1+n2+n5+n6,4)~Q(Q,n3+n4,2) \nonumber \\
    &+& 6.~Q(q,n1+n3+n4+n5,4)~Q(Q,n2+n6,2) + 6.~Q(q,n1+n3+n4+n6,4)~Q(Q,n2+n5,2) \nonumber \\
    &+& 6.~Q(q,n1+n3+n5+n6,4)~Q(Q,n2+n4,2) + 6.~Q(q,n1+n4+n5+n6,4)~Q(Q,n2+n3,2) \nonumber \\
    &+& 6.~Q(Q,n2+n3+n4+n5,4)~Q(q,n1+n6,2) + 6.~Q(Q,n2+n3+n4+n6,4)~Q(q,n1+n5,2) \nonumber \\
    &+& 6.~Q(Q,n2+n3+n5+n6,4)~Q(q,n1+n4,2) + 6.~Q(Q,n2+n4+n5+n6,4)~Q(q,n1+n3,2) \nonumber \\
    &+& 6.~Q(Q,n3+n4+n5+n6,4)~Q(q,n1+n2,2), \nonumber \\
\end{eqnarray}
\begin{eqnarray}
B10(1)&=& 24.~Q(q,n1+n2+n3+n4+n5,5)~Q(Q,n6,1) + 24.~Q(q,n1+n2+n3+n4+n6,5)~Q(Q,n5,1) \nonumber \\
     &+& 24.~Q(q,n1+n2+n3+n5+n6,5)~Q(Q,n4,1) + 24.~Q(q,n1+n2+n4+n5+n6,5)~Q(Q,n3,1) \nonumber \\
     &+& 24.~Q(q,n1+n3+n4+n5+n6,5)~Q(Q,n2,1) + 24.~Q(Q,n2+n3+n4+n5+n6,5)~Q(p,n1,1),\nonumber \\
\end{eqnarray}
\begin{eqnarray}
B11(1)&=&-120.~Q(q,n1+n2+n3+n4+n5+n6,6).\nonumber \\
\end{eqnarray}

}

In addition, the six-particle correlations using two POIs can be given as,
{\tiny
\begin{eqnarray}\label{eq:A-9}
\langle n1,n2,n3,n4,n5,n6 \rangle_{120000} &=& B1(2) + B2(2)  + B3(2)  + B4(2)  + B5(2)  + B6(2)  + B7(2) + A8(2) + A9(2) + A10(2) + A11(2), \\
B1(2)&=&  Q(p,n1,1)~Q(p,n2,1)~Q(Q,n3,1)~Q(Q,n4,1)~Q(Q,n5,1)~Q(Q,n6,1),\nonumber \\
\end{eqnarray}
\begin{eqnarray}
B2(2)&=&-Q(p,n1+n2,2)~Q(Q,n3,1)~Q(Q,n4,1)~Q(Q,n5,1)~Q(Q,n6,1) - Q(q,n1+n3,2)~Q(p,n2,1)~Q(Q,n4,1)~Q(Q,n5,1)~Q(Q,n6,1)\nonumber \\
     &-& Q(q,n1+n4,2)~Q(p,n2,1)~Q(Q,n3,1)~Q(Q,n5,1)~Q(Q,n6,1) - Q(q,n1+n5,2)~Q(p,n2,1)~Q(Q,n3,1)~Q(Q,n4,1)~Q(Q,n6,1)\nonumber \\
     &-& Q(q,n1+n6,2)~Q(p,n2,1)~Q(Q,n3,1)~Q(Q,n4,1)~Q(Q,n5,1) - Q(q,n2+n3,2)~Q(p,n1,1)~Q(Q,n4,1)~Q(Q,n5,1)~Q(Q,n6,1)\nonumber \\
     &-& Q(q,n2+n4,2)~Q(p,n1,1)~Q(Q,n3,1)~Q(Q,n5,1)~Q(Q,n6,1) - Q(q,n2+n5,2)~Q(p,n1,1)~Q(Q,n3,1)~Q(Q,n4,1)~Q(Q,n6,1)\nonumber \\
     &-& Q(q,n2+n6,2)~Q(p,n1,1)~Q(Q,n3,1)~Q(Q,n4,1)~Q(Q,n5,1) - Q(Q,n3+n4,2)~Q(p,n1,1)~Q(p,n2,1)~Q(Q,n5,1)~Q(Q,n6,1)\nonumber \\
     &-& Q(Q,n3+n5,2)~Q(p,n1,1)~Q(p,n2,1)~Q(Q,n4,1)~Q(Q,n6,1) - Q(Q,n3+n6,2)~Q(p,n1,1)~Q(p,n2,1)~Q(Q,n4,1)~Q(Q,n5,1)\nonumber \\
     &-& Q(Q,n4+n5,2)~Q(p,n1,1)~Q(p,n2,1)~Q(Q,n3,1)~Q(Q,n6,1) - Q(Q,n4+n6,2)~Q(p,n1,1)~Q(p,n2,1)~Q(Q,n3,1)~Q(Q,n5,1)\nonumber \\
     &-& Q(Q,n5+n6,2)~Q(p,n1,1)~Q(p,n2,1)~Q(Q,n3,1)~Q(Q,n4,1),\nonumber \\
\end{eqnarray}
\begin{eqnarray}
B3(2)&=& 2.~Q(q,n1+n2+n3,3)~Q(Q,n4,1)~Q(Q,n5,1)~Q(Q,n6,1) + 2.~Q(q,n1+n2+n4,3)~Q(Q,n3,1)~Q(Q,n5,1)~Q(Q,n6,1)\nonumber \\
     &+& 2.~Q(q,n1+n2+n5,3)~Q(Q,n3,1)~Q(Q,n4,1)~Q(Q,n6,1) + 2.~Q(q,n1+n2+n6,3)~Q(Q,n3,1)~Q(Q,n4,1)~Q(Q,n5,1)\nonumber \\
     &+& 2.~Q(q,n1+n3+n4,3)~Q(p,n2,1)~Q(Q,n5,1)~Q(Q,n6,1) + 2.~Q(q,n1+n3+n5,3)~Q(p,n2,1)~Q(Q,n4,1)~Q(Q,n6,1)\nonumber \\
     &+& 2.~Q(q,n1+n3+n6,3)~Q(p,n2,1)~Q(Q,n4,1)~Q(Q,n5,1) + 2.~Q(q,n1+n4+n5,3)~Q(p,n2,1)~Q(Q,n3,1)~Q(Q,n6,1)\nonumber \\
     &+& 2.~Q(q,n1+n4+n6,3)~Q(p,n2,1)~Q(Q,n3,1)~Q(Q,n5,1) + 2.~Q(q,n1+n5+n6,3)~Q(p,n2,1)~Q(Q,n3,1)~Q(Q,n4,1)\nonumber \\
     &+& 2.~Q(q,n2+n3+n4,3)~Q(p,n1,1)~Q(Q,n5,1)~Q(Q,n6,1) + 2.~Q(q,n2+n3+n5,3)~Q(p,n1,1)~Q(Q,n4,1)~Q(Q,n6,1)\nonumber \\
     &+& 2.~Q(q,n2+n3+n6,3)~Q(p,n1,1)~Q(Q,n4,1)~Q(Q,n5,1) + 2.~Q(q,n2+n4+n5,3)~Q(p,n1,1)~Q(Q,n3,1)~Q(Q,n6,1)\nonumber \\
     &+& 2.~Q(q,n2+n4+n6,3)~Q(p,n1,1)~Q(Q,n3,1)~Q(Q,n5,1) + 2.~Q(q,n2+n5+n6,3)~Q(p,n1,1)~Q(Q,n3,1)~Q(Q,n4,1)\nonumber \\
     &+& 2.~Q(Q,n3+n4+n5,3)~Q(p,n1,1)~Q(p,n2,1)~Q(Q,n6,1) + 2.~Q(Q,n3+n4+n6,3)~Q(p,n1,1)~Q(p,n2,1)~Q(Q,n5,1)\nonumber \\
     &+& 2.~Q(Q,n3+n5+n6,3)~Q(p,n1,1)~Q(p,n2,1)~Q(Q,n4,1) + 2.~Q(Q,n4+n5+n6,3)~Q(p,n1,1)~Q(p,n2,1)~Q(Q,n3,1),\nonumber \\
\end{eqnarray}
\begin{eqnarray}
B4(2)&=& Q(p,n1+n2,2)~Q(Q,n3+n4,2)~Q(Q,n5,1)~Q(Q,n6,1) + Q(p,n1+n2,2)~Q(Q,n3+n5,2)~Q(Q,n4,1)~Q(Q,n6,1)\nonumber \\
     &+& Q(p,n1+n2,2)~Q(Q,n3+n6,2)~Q(Q,n4,1)~Q(Q,n5,1) + Q(p,n1+n2,2)~Q(Q,n4+n5,2)~Q(Q,n3,1)~Q(Q,n6,1)\nonumber \\
     &+& Q(p,n1+n2,2)~Q(Q,n4+n6,2)~Q(Q,n3,1)~Q(Q,n5,1) + Q(p,n1+n2,2)~Q(Q,n5+n6,2)~Q(Q,n3,1)~Q(Q,n4,1)\nonumber \\
     &+& Q(q,n1+n3,2)~Q(q,n2+n4,2)~Q(Q,n5,1)~Q(Q,n6,1) + Q(q,n1+n3,2)~Q(q,n2+n5,2)~Q(Q,n4,1)~Q(Q,n6,1)\nonumber \\
     &+& Q(q,n1+n3,2)~Q(q,n2+n6,2)~Q(Q,n4,1)~Q(Q,n5,1) + Q(q,n1+n3,2)~Q(Q,n4+n5,2)~Q(p,n2,1)~Q(Q,n6,1)\nonumber \\
     &+& Q(q,n1+n3,2)~Q(Q,n4+n6,2)~Q(p,n2,1)~Q(Q,n5,1) + Q(q,n1+n3,2)~Q(Q,n5+n6,2)~Q(p,n2,1)~Q(Q,n4,1)\nonumber \\
     &+& Q(q,n1+n4,2)~Q(q,n2+n3,2)~Q(Q,n5,1)~Q(Q,n6,1) + Q(q,n1+n4,2)~Q(q,n2+n5,2)~Q(Q,n3,1)~Q(Q,n6,1)\nonumber \\
     &+& Q(q,n1+n4,2)~Q(q,n2+n6,2)~Q(Q,n3,1)~Q(Q,n5,1) + Q(q,n1+n4,2)~Q(Q,n3+n5,2)~Q(p,n2,1)~Q(Q,n6,1)\nonumber \\
     &+& Q(q,n1+n4,2)~Q(Q,n3+n6,2)~Q(p,n2,1)~Q(Q,n5,1) + Q(q,n1+n4,2)~Q(Q,n5+n6,2)~Q(p,n2,1)~Q(Q,n3,1)\nonumber \\
     &+& Q(q,n1+n5,2)~Q(q,n2+n3,2)~Q(Q,n4,1)~Q(Q,n6,1) + Q(q,n1+n5,2)~Q(q,n2+n4,2)~Q(Q,n3,1)~Q(Q,n6,1)\nonumber \\
     &+& Q(q,n1+n5,2)~Q(q,n2+n6,2)~Q(Q,n3,1)~Q(Q,n4,1) + Q(q,n1+n5,2)~Q(Q,n3+n4,2)~Q(p,n2,1)~Q(Q,n6,1)\nonumber \\
     &+& Q(q,n1+n5,2)~Q(Q,n3+n6,2)~Q(p,n2,1)~Q(Q,n4,1) + Q(q,n1+n5,2)~Q(Q,n4+n6,2)~Q(p,n2,1)~Q(Q,n3,1)\nonumber \\
     &+& Q(q,n1+n6,2)~Q(q,n2+n3,2)~Q(Q,n4,1)~Q(Q,n5,1) + Q(q,n1+n6,2)~Q(q,n2+n4,2)~Q(Q,n3,1)~Q(Q,n5,1)\nonumber \\
     &+& Q(q,n1+n6,2)~Q(q,n2+n5,2)~Q(Q,n3,1)~Q(Q,n4,1) + Q(q,n1+n6,2)~Q(Q,n3+n4,2)~Q(p,n2,1)~Q(Q,n5,1)\nonumber \\
     &+& Q(q,n1+n6,2)~Q(Q,n3+n5,2)~Q(p,n2,1)~Q(Q,n4,1) + Q(q,n1+n6,2)~Q(Q,n4+n5,2)~Q(p,n2,1)~Q(Q,n3,1)\nonumber \\
     &+& Q(q,n2+n3,2)~Q(Q,n4+n5,2)~Q(p,n1,1)~Q(Q,n6,1) + Q(q,n2+n3,2)~Q(Q,n4+n6,2)~Q(p,n1,1)~Q(Q,n5,1)\nonumber \\
     &+& Q(q,n2+n3,2)~Q(Q,n5+n6,2)~Q(p,n1,1)~Q(Q,n4,1) + Q(q,n2+n4,2)~Q(Q,n3+n5,2)~Q(p,n1,1)~Q(Q,n6,1)\nonumber \\
     &+& Q(q,n2+n4,2)~Q(Q,n3+n6,2)~Q(p,n1,1)~Q(Q,n5,1) + Q(q,n2+n4,2)~Q(Q,n5+n6,2)~Q(p,n1,1)~Q(Q,n3,1)\nonumber \\
     &+& Q(q,n2+n5,2)~Q(Q,n3+n4,2)~Q(p,n1,1)~Q(Q,n6,1) + Q(q,n2+n5,2)~Q(Q,n3+n6,2)~Q(p,n1,1)~Q(Q,n4,1)\nonumber \\
     &+& Q(q,n2+n5,2)~Q(Q,n4+n6,2)~Q(p,n1,1)~Q(Q,n3,1) + Q(Q,n3+n4,2)~Q(q,n2+n6,2)~Q(p,n1,1)~Q(Q,n5,1)\nonumber \\
     &+& Q(Q,n3+n4,2)~Q(Q,n5+n6,2)~Q(p,n1,1)~Q(p,n2,1) + Q(Q,n3+n5,2)~Q(q,n2+n6,2)~Q(p,n1,1)~Q(Q,n4,1)\nonumber \\
     &+& Q(Q,n3+n5,2)~Q(Q,n4+n6,2)~Q(p,n1,1)~Q(p,n2,1) + Q(Q,n4+n5,2)~Q(q,n2+n6,2)~Q(p,n1,1)~Q(Q,n3,1)\nonumber \\
     &+& Q(Q,n4+n5,2)~Q(Q,n3+n6,2)~Q(p,n1,1)~Q(p,n2,1),\nonumber \\
\end{eqnarray}
\begin{eqnarray}
B5(2)&=&-2.~Q(q,n1+n2+n3,3)~Q(Q,n4+n5,2)~Q(Q,n6,1) - 2.~Q(q,n1+n2+n3,3)~Q(Q,n4+n6,2)~Q(Q,n5,1)\nonumber \\
     &-& 2.~Q(q,n1+n2+n3,3)~Q(Q,n5+n6,2)~Q(Q,n4,1) - 2.~Q(q,n1+n2+n4,3)~Q(Q,n3+n5,2)~Q(Q,n6,1)\nonumber \\
     &-& 2.~Q(q,n1+n2+n4,3)~Q(Q,n3+n6,2)~Q(Q,n5,1) - 2.~Q(q,n1+n2+n4,3)~Q(Q,n5+n6,2)~Q(Q,n3,1)\nonumber \\
     &-& 2.~Q(q,n1+n2+n5,3)~Q(Q,n3+n4,2)~Q(Q,n6,1) - 2.~Q(q,n1+n2+n5,3)~Q(Q,n3+n6,2)~Q(Q,n4,1)\nonumber \\
     &-& 2.~Q(q,n1+n2+n5,3)~Q(Q,n4+n6,2)~Q(Q,n3,1) - 2.~Q(q,n1+n2+n6,3)~Q(Q,n3+n4,2)~Q(Q,n5,1)\nonumber \\
     &-& 2.~Q(q,n1+n2+n6,3)~Q(Q,n3+n5,2)~Q(Q,n4,1) - 2.~Q(q,n1+n2+n6,3)~Q(Q,n4+n5,2)~Q(Q,n3,1)\nonumber \\
     &-& 2.~Q(q,n1+n3+n4,3)~Q(q,n2+n5,2)~Q(Q,n6,1) - 2.~Q(q,n1+n3+n4,3)~Q(q,n2+n6,2)~Q(Q,n5,1)\nonumber \\
     &-& 2.~Q(q,n1+n3+n4,3)~Q(Q,n5+n6,2)~Q(p,n2,1) - 2.~Q(q,n1+n3+n5,3)~Q(q,n2+n4,2)~Q(Q,n6,1)\nonumber \\
     &-& 2.~Q(q,n1+n3+n5,3)~Q(q,n2+n6,2)~Q(Q,n4,1) - 2.~Q(q,n1+n3+n5,3)~Q(Q,n4+n6,2)~Q(p,n2,1)\nonumber \\
     &-& 2.~Q(q,n1+n3+n6,3)~Q(q,n2+n4,2)~Q(Q,n5,1) - 2.~Q(q,n1+n3+n6,3)~Q(q,n2+n5,2)~Q(Q,n4,1)\nonumber \\
     &-& 2.~Q(q,n1+n3+n6,3)~Q(Q,n4+n5,2)~Q(p,n2,1) - 2.~Q(q,n1+n4+n5,3)~Q(q,n2+n3,2)~Q(Q,n6,1)\nonumber \\
     &-& 2.~Q(q,n1+n4+n5,3)~Q(q,n2+n6,2)~Q(Q,n3,1) - 2.~Q(q,n1+n4+n5,3)~Q(Q,n3+n6,2)~Q(p,n2,1)\nonumber \\
     &-& 2.~Q(q,n1+n4+n6,3)~Q(q,n2+n3,2)~Q(Q,n5,1) - 2.~Q(q,n1+n4+n6,3)~Q(q,n2+n5,2)~Q(Q,n3,1)\nonumber \\
     &-& 2.~Q(q,n1+n4+n6,3)~Q(Q,n3+n5,2)~Q(p,n2,1) - 2.~Q(q,n1+n5+n6,3)~Q(q,n2+n3,2)~Q(Q,n4,1)\nonumber \\
     &-& 2.~Q(q,n1+n5+n6,3)~Q(q,n2+n4,2)~Q(Q,n3,1) - 2.~Q(q,n1+n5+n6,3)~Q(Q,n3+n4,2)~Q(p,n2,1)\nonumber \\
     &-& 2.~Q(q,n2+n3+n4,3)~Q(q,n1+n5,2)~Q(Q,n6,1) - 2.~Q(q,n2+n3+n4,3)~Q(q,n1+n6,2)~Q(Q,n5,1)\nonumber \\
     &-& 2.~Q(q,n2+n3+n4,3)~Q(Q,n5+n6,2)~Q(p,n1,1) - 2.~Q(q,n2+n3+n5,3)~Q(q,n1+n4,2)~Q(Q,n6,1)\nonumber \\
     &-& 2.~Q(q,n2+n3+n5,3)~Q(q,n1+n6,2)~Q(Q,n4,1) - 2.~Q(q,n2+n3+n5,3)~Q(Q,n4+n6,2)~Q(p,n1,1)\nonumber \\
     &-& 2.~Q(q,n2+n3+n6,3)~Q(q,n1+n4,2)~Q(Q,n5,1) - 2.~Q(q,n2+n3+n6,3)~Q(q,n1+n5,2)~Q(Q,n4,1)\nonumber \\
     &-& 2.~Q(q,n2+n3+n6,3)~Q(Q,n4+n5,2)~Q(p,n1,1) - 2.~Q(q,n2+n4+n5,3)~Q(q,n1+n3,2)~Q(Q,n6,1)\nonumber \\
     &-& 2.~Q(q,n2+n4+n5,3)~Q(q,n1+n6,2)~Q(Q,n3,1) - 2.~Q(q,n2+n4+n5,3)~Q(Q,n3+n6,2)~Q(p,n1,1)\nonumber \\
     &-& 2.~Q(q,n2+n4+n6,3)~Q(q,n1+n3,2)~Q(Q,n5,1) - 2.~Q(q,n2+n4+n6,3)~Q(q,n1+n5,2)~Q(Q,n3,1)\nonumber \\
     &-& 2.~Q(q,n2+n4+n6,3)~Q(Q,n3+n5,2)~Q(p,n1,1) - 2.~Q(q,n2+n5+n6,3)~Q(q,n1+n3,2)~Q(Q,n4,1)\nonumber \\
     &-& 2.~Q(q,n2+n5+n6,3)~Q(q,n1+n4,2)~Q(Q,n3,1) - 2.~Q(q,n2+n5+n6,3)~Q(Q,n3+n4,2)~Q(p,n1,1)\nonumber \\
     &-& 2.~Q(Q,n3+n4+n5,3)~Q(p,n1+n2,2)~Q(Q,n6,1) - 2.~Q(Q,n3+n4+n5,3)~Q(q,n1+n6,2)~Q(p,n2,1)\nonumber \\
     &-& 2.~Q(Q,n3+n4+n5,3)~Q(q,n2+n6,2)~Q(p,n1,1) - 2.~Q(Q,n3+n4+n6,3)~Q(p,n1+n2,2)~Q(Q,n5,1)\nonumber \\
     &-& 2.~Q(Q,n3+n4+n6,3)~Q(q,n1+n5,2)~Q(p,n2,1) - 2.~Q(Q,n3+n4+n6,3)~Q(q,n2+n5,2)~Q(p,n1,1)\nonumber \\
     &-& 2.~Q(Q,n3+n5+n6,3)~Q(p,n1+n2,2)~Q(Q,n4,1) - 2.~Q(Q,n3+n5+n6,3)~Q(q,n1+n4,2)~Q(p,n2,1)\nonumber \\
     &-& 2.~Q(Q,n3+n5+n6,3)~Q(q,n2+n4,2)~Q(p,n1,1) - 2.~Q(Q,n4+n5+n6,3)~Q(p,n1+n2,2)~Q(Q,n3,1)\nonumber \\
     &-& 2.~Q(Q,n4+n5+n6,3)~Q(q,n1+n3,2)~Q(p,n2,1) - 2.~Q(Q,n4+n5+n6,3)~Q(q,n2+n3,2)~Q(p,n1,1),\nonumber \\
\end{eqnarray}
\begin{eqnarray}
B6(2)&=&-6.~Q(q,n1+n2+n3+n4,4)~Q(Q,n5,1)~Q(Q,n6,1) - 6.~Q(q,n1+n2+n3+n5,4)~Q(Q,n4,1)~Q(Q,n6,1)\nonumber \\
     &-& 6.~Q(q,n1+n2+n4+n5,4)~Q(Q,n3,1)~Q(Q,n6,1) - 6.~Q(q,n1+n3+n4+n5,4)~Q(p,n2,1)~Q(Q,n6,1)\nonumber \\
     &-& 6.~Q(q,n1+n2+n3+n6,4)~Q(Q,n4,1)~Q(Q,n5,1) - 6.~Q(q,n1+n2+n4+n6,4)~Q(Q,n3,1)~Q(Q,n5,1)\nonumber \\
     &-& 6.~Q(q,n1+n3+n4+n6,4)~Q(p,n2,1)~Q(Q,n5,1) - 6.~Q(q,n1+n2+n5+n6,4)~Q(Q,n3,1)~Q(Q,n4,1)\nonumber \\
     &-& 6.~Q(q,n1+n3+n5+n6,4)~Q(p,n2,1)~Q(Q,n4,1) - 6.~Q(q,n1+n4+n5+n6,4)~Q(p,n2,1)~Q(Q,n3,1)\nonumber \\
     &-& 6.~Q(q,n2+n3+n4+n5,4)~Q(p,n1,1)~Q(Q,n6,1) - 6.~Q(q,n2+n3+n4+n6,4)~Q(p,n1,1)~Q(Q,n5,1)\nonumber \\
     &-& 6.~Q(q,n2+n3+n5+n6,4)~Q(p,n1,1)~Q(Q,n4,1) - 6.~Q(q,n2+n4+n5+n6,4)~Q(p,n1,1)~Q(Q,n3,1)\nonumber \\
     &-& 6.~Q(Q,n3+n4+n5+n6,4)~Q(p,n1,1)~Q(p,n2,1),\nonumber \\
\end{eqnarray}
\begin{eqnarray}
B7(2)&=&-Q(p,n1+n2,2)~Q(Q,n3+n4,2)~Q(Q,n5+n6,2) - Q(p,n1+n2,2)~Q(Q,n3+n5,2)~Q(Q,n4+n6,2)\nonumber \\
     &-& Q(p,n1+n2,2)~Q(Q,n4+n5,2)~Q(Q,n3+n6,2) - Q(q,n1+n3,2)~Q(Q,n4+n5,2)~Q(q,n2+n6,2)\nonumber \\
     &-& Q(q,n1+n3,2)~Q(q,n2+n5,2)~Q(Q,n4+n6,2) - Q(q,n1+n3,2)~Q(q,n2+n4,2)~Q(Q,n5+n6,2)\nonumber \\
     &-& Q(q,n1+n4,2)~Q(Q,n3+n5,2)~Q(q,n2+n6,2) - Q(q,n1+n4,2)~Q(q,n2+n5,2)~Q(Q,n3+n6,2)\nonumber \\
     &-& Q(q,n1+n4,2)~Q(q,n2+n3,2)~Q(Q,n5+n6,2) - Q(q,n1+n5,2)~Q(Q,n3+n4,2)~Q(q,n2+n6,2)\nonumber \\
     &-& Q(q,n1+n5,2)~Q(q,n2+n4,2)~Q(Q,n3+n6,2) - Q(q,n1+n5,2)~Q(q,n2+n3,2)~Q(Q,n4+n6,2)\nonumber \\
     &-& Q(q,n1+n6,2)~Q(Q,n3+n4,2)~Q(q,n2+n5,2) - Q(q,n1+n6,2)~Q(q,n2+n4,2)~Q(Q,n3+n5,2)\nonumber \\
     &-& Q(q,n1+n6,2)~Q(q,n2+n3,2)~Q(Q,n4+n5,2),\nonumber \\
\end{eqnarray}
\begin{eqnarray}
B8(2)&=& 4.~Q(q,n1+n2+n3,3)~Q(Q,n4+n5+n6,3) + 4.~Q(q,n1+n2+n4,3)~Q(Q,n3+n5+n6,3)\nonumber \\
     &+& 4.~Q(q,n1+n2+n5,3)~Q(Q,n3+n4+n6,3) + 4.~Q(q,n1+n3+n4,3)~Q(q,n2+n5+n6,3)\nonumber \\
     &+& 4.~Q(q,n1+n3+n5,3)~Q(q,n2+n4+n6,3) + 4.~Q(q,n1+n4+n5,3)~Q(q,n2+n3+n6,3)\nonumber \\
     &+& 4.~Q(q,n2+n3+n4,3)~Q(q,n1+n5+n6,3) + 4.~Q(q,n2+n3+n5,3)~Q(q,n1+n4+n6,3)\nonumber \\
     &+& 4.~Q(q,n2+n4+n5,3)~Q(q,n1+n3+n6,3) + 4.~Q(Q,n3+n4+n5,3)~Q(q,n1+n2+n6,3),\nonumber \\
\end{eqnarray}
\begin{eqnarray}
B9(2)&=& 6.~Q(q,n1+n2+n3+n4,4)~Q(Q,n5+n6,2) + 6.~Q(q,n1+n2+n3+n5,4)~Q(Q,n4+n6,2)\nonumber \\
     &+& 6.~Q(q,n1+n2+n3+n6,4)~Q(Q,n4+n5,2) + 6.~Q(q,n1+n2+n4+n5,4)~Q(Q,n3+n6,2)\nonumber \\
     &+& 6.~Q(q,n1+n2+n4+n6,4)~Q(Q,n3+n5,2) + 6.~Q(q,n1+n2+n5+n6,4)~Q(Q,n3+n4,2)\nonumber \\
     &+& 6.~Q(q,n1+n3+n4+n5,4)~Q(q,n2+n6,2) + 6.~Q(q,n1+n3+n4+n6,4)~Q(q,n2+n5,2)\nonumber \\
     &+& 6.~Q(q,n1+n3+n5+n6,4)~Q(q,n2+n4,2) + 6.~Q(q,n1+n4+n5+n6,4)~Q(q,n2+n3,2)\nonumber \\
     &+& 6.~Q(q,n2+n3+n4+n5,4)~Q(q,n1+n6,2) + 6.~Q(q,n2+n3+n4+n6,4)~Q(q,n1+n5,2)\nonumber \\
     &+& 6.~Q(q,n2+n3+n5+n6,4)~Q(q,n1+n4,2) + 6.~Q(q,n2+n4+n5+n6,4)~Q(q,n1+n3,2)\nonumber \\
     &+& 6.~Q(Q,n3+n4+n5+n6,4)~Q(p,n1+n2,2),\nonumber \\
\end{eqnarray}
\begin{eqnarray}
B10(2)&=& 24.~Q(q,n1+n2+n3+n4+n5,5)~Q(Q,n6,1) + 24.~Q(q,n1+n2+n3+n4+n6,5)~Q(Q,n5,1)\nonumber \\
     &+& 24.~Q(q,n1+n2+n3+n5+n6,5)~Q(Q,n4,1) + 24.~Q(q,n1+n2+n4+n5+n6,5)~Q(Q,n3,1)\nonumber \\
     &+& 24.~Q(q,n1+n3+n4+n5+n6,5)~Q(p,n2,1) + 24.~Q(q,n2+n3+n4+n5+n6,5)~Q(p,n1,1)\nonumber \\
\end{eqnarray}
\begin{eqnarray}
B11(2)&=& -120.~Q(q,n1+n2+n3+n4+n5+n6,6),\nonumber \\
\end{eqnarray}

}

Then, the event average six particle correlations will be given as:
{\tiny
\begin{eqnarray}
\langle V_{n_1} V_{n_2} V_{n_3} V_{n_4} V_{n_5} V_{n_6} \rangle_{P} &=& \langle \langle n1,n2,n3,n4,n5,n6\rangle_{P}\rangle/\langle \langle 0,0,0,0,0,0 \rangle_{P}\rangle.
\end{eqnarray}
}

\subsection{The two subevents particle correlations;}
The observables discussed in this work using the two subevents particle correlations method can be given in terms of the flow vectors;
%--------------------------------------------------------------------
{\tiny
\begin{eqnarray}\label{eq:A-1}
Q(p,X,n,k)              &=&  \sum^{M_{p}}_{i=1} \omega^{k}_{i} e^{in\varphi_{i, X}},  \nonumber \\
Q(Q,X,n,k)              &=&  \sum^{M_{Q}}_{i=1} \omega^{k}_{i} e^{in\varphi_{i, X}},  \nonumber \\
Q(q,X,n,k)              &=&  \sum^{M_{q}}_{i=1} \omega^{k}_{i} e^{in\varphi_{i}, X},  \nonumber \\
\end{eqnarray}
}
where $X$ represent the subevent of the particle, $X$ $=$S1 or S2 for first and second subevents respectively.
%--------------------------------------------------------------------

The two-particle correlations using one and two POIs can be given as,
{\tiny
\begin{eqnarray}
\langle n1,n2\rangle_{1,10} &=& QS(p,S1,n1,1)~QS(Q,S2,n2,1),\\
\langle n1,n2\rangle_{2,12} &=& QS(p,S1,n1,1)~QS(p,S2,n2,1),
\end{eqnarray}
}
and the event average two-particle correlations will be given as:
{\tiny
\begin{eqnarray}
\langle V_{n_1} V_{n_2} \rangle_{P} &=& \langle \langle n1,n2\rangle_{P}\rangle/\langle \langle 0,0\rangle_{P}\rangle.
\end{eqnarray}
}

The three-particle correlations using one and two POIs can be given as,
{\tiny
\begin{eqnarray}
 \langle n1,n2,n3\rangle_{003}  &=& (QS(Q,S1,n1,1)~QS(Q,S1,n2,1) - QS(Q,S1,n1+n2,2) ) ~ QS(p,S2,n3,1), \\
 \langle n1,n2,n3\rangle_{120}  &=& (QS(p,S1,n1,1)~QS(p,S1,n2,1) - QS(p,S1,n1+n2,2) ) ~ QS(Q,S2,n3,1), \\
 \langle n1,n2,n3\rangle_{103}  &=& (QS(p,S1,n1,1)~QS(Q,S1,n2,1) - QS(q,S1,n1+n2,2) ) ~ QS(p,S2,n3,1),
\end{eqnarray}
}
and then the event average three-particle correlations will be given as:
{\tiny
\begin{eqnarray}
\langle V_{n_1} V_{n_2}  V_{n_3} \rangle_{P} &=& \langle \langle n1,n2,n3\rangle_{P}\rangle/\langle \langle 0,0,0\rangle_{P}\rangle.
\end{eqnarray}
}

The four-particle correlations using one and two POIs can be given as,
{\tiny
\begin{eqnarray}
\langle n1,n2,n3,n4\rangle_{1000} &=& QS(p,S1,n1   ,1)~QS(Q,S1,n2   ,1)~QS(Q,S2,n3   ,1)~QS(Q,S2,n4,1) \\
          &-& QS(p,S1,n1   ,1)~QS(Q,S1,n2   ,1)~QS(Q,S2,n3+n4,2)\nonumber \\
          &-& QS(Q,S2,n3   ,1)~QS(Q,S2,n4   ,1)~QS(q,S1,n1+n2,2)\nonumber \\
          &+& QS(q,S1,n1+n2,2)~QS(Q,S2,n3+n4,2)\nonumber
\end{eqnarray}
\begin{eqnarray}
\langle n1,n2,n3,n4\rangle_{1200}  &=&  QS(p,S1,n1   ,1)~QS(p,S1,n2   ,1)~QS(Q,S2,n3   ,1)~QS(Q,S2,n4,1) \\
               &-& QS(p,S1,n1   ,1)~QS(p,S1,n2   ,1)~QS(Q,S2,n3+n4,2)\nonumber \\
               &-& QS(Q,S2,n3   ,1)~QS(Q,S2,n4   ,1)~QS(p,S1,n1+n2,2)\nonumber \\
               &+& QS(p,S1,n1+n2,2)~QS(Q,S2,n3+n4,2) \nonumber
\end{eqnarray}
\begin{eqnarray}
\langle n1,n2,n3,n4\rangle_{1030}  &=& QS(p,S1,n1   ,1)~QS(Q,S1,n2   ,1)~QS(p,S2,n3   ,1)~QS(Q,S2,n4,1) \\
               &-& QS(p,S1,n1   ,1)~QS(Q,S1,n2   ,1)~QS(q,S2,n3+n4,2) \nonumber \\
               &-& QS(p,S2,n3   ,1)~QS(Q,S2,n4   ,1)~QS(q,S1,n1+n2,2) \nonumber \\
               &+& QS(q,S1,n1+n2,2)~QS(q,S2,n3+n4,2) \nonumber
\end{eqnarray}
}

and then the event average four-particle correlations will be given as:
{\tiny
\begin{eqnarray}
\langle V_{n_1} V_{n_2} V_{n_3} V_{n_4} \rangle_{P} &=& \langle \langle n1,n2,n3,n4\rangle_{P}\rangle/\langle \langle 0,0,0,0\rangle_{P}\rangle.
\end{eqnarray}
}

%--------------------------------------------------------------------
%--------------------------------------------------------------------
\end{widetext}
%--------------------------------------------------------------------
%--------------------------------------------------------------------
%--------------------------------------------------------------------

%--------------------------------------------------------------------
%--------------------------------------------------------------------
%--------------------------------------------------------------------
%\bibliographystyle{aipauth4-1}
\bibliography{ref} 
%--------------------------------------------------------------------
\end{document}